\documentclass[twocolumn]{openjournal}
\usepackage{natbib}
\usepackage{graphicx}
\usepackage{graphicx,amsmath,amssymb,amstext}
\usepackage{amsbsy,amsfonts,amsthm,color}
\usepackage{caption}
\usepackage{subcaption}
\usepackage[dvipsnames]{xcolor}
\usepackage{graphicx}
\usepackage{lipsum}
\usepackage{bm}
\usepackage{cancel}
\usepackage{xspace}
\usepackage{appendix}

\usepackage{tikz}
\usetikzlibrary{shapes,arrows,shadows,fit}
\usetikzlibrary{positioning}
\usetikzlibrary{bayesnet}
\definecolor{ourgreen}{HTML}{06D6A0}
\definecolor{ourblue}{HTML}{118AB2}
\usepackage{enumitem} 

\usepackage{hyperref}
\hypersetup{colorlinks,linkcolor=ourblue,citecolor=ourblue,urlcolor=ourblue}

\definecolor{ork}{rgb}{0.9,0.1,0.3}
\definecolor{grbl}{rgb}{0.3,0.6,0.7}
\definecolor{bleu}{rgb}{0,0.5,0.5}

\usepackage{soul}

\usepackage{xcolor}

\newcommand{\revision}[1]{{\textcolor{black}{#1}}}
\newcommand{\geu}{iPTF16geu\xspace}
\newcommand{\pbh}{$f_{\rm dc}$\xspace}
\newcommand{\as}{$''$\xspace}

\newcommand{\wfcuvis}{WFC3/UVIS\xspace}
\newcommand{\wfcir}{WFC3/IR\xspace}

\usepackage{comment}

\begin{document}

\journalinfo{The Open Journal of Astrophysics}
\shorttitle{Microlensing of strongly lensed supernova Zwicky \& iPTF16geu}

\title{Microlensing of lensed supernovae Zwicky \& iPTF16geu: constraints \\ on the lens galaxy mass slope and dark compact object fraction \vspace{-1cm}}

\author{Nikki Arendse$^{1}$}
\author{Edvard Mörtsell$^{1}$}
\author{Luke Weisenbach$^{2}$}
\author{Erin Hayes$^{3}$}
\author{Stephen Thorp$^{1,3}$}
\author{Suhail Dhawan$^{3,4}$}
\author{Ariel Goobar$^{1}$}
\author{Sacha Guerrini$^{5}$}
\author{Jacob Osman Hjortlund$^{1}$}
\author{Joel Johansson$^{1}$}
\author{Cameron Lemon$^{1}$}
\author{Abdullah Al Zaif$^{3}$}
 
\affiliation{}
\affiliation{$^1$ Oskar Klein Centre, Department of Physics, Stockholm University, SE-106 91 Stockholm, Sweden}
\affiliation{$^2$ Institute of Cosmology and Gravitation, University of Portsmouth, Burnaby Road, Portsmouth, PO1 3FX, UK}
\affiliation{$^3$ Institute of Astronomy and Kavli Institute for Cosmology, University of Cambridge, Madingley Road, Cambridge CB3 0HA, UK}
\affiliation{$^4$ School of Physics and Astronomy, University of Birmingham, Birmingham, UK}
\affiliation{$^{5}$ Universit\'e Paris Cit\'e, Universit\'e Paris-Saclay, CEA, CNRS, AIM, 91191, Gif-sur-Yvette, France}

\email{nikki.arendse@proton.me}

\begin{abstract}

To date, only two strongly lensed type Ia supernovae (SNIa) have been discovered with an isolated galaxy acting as the lens: \geu and SN Zwicky. The observed image fluxes for both lens systems were inconsistent with predictions from a smooth macro lens model. A potential explanation for the anomalous flux ratios is microlensing: additional (de)magnification caused by stars and other compact objects in the lens galaxy. 
In this work, we combine observations of \geu and SN Zwicky with simulated microlensing magnification maps, leveraging their standardizable candle properties to constrain the lens galaxy mass slope, $\eta$, and the fraction of dark compact objects, \pbh. The resulting mass slopes are $\eta = 1.70 \pm 0.07$ for \geu and $\eta = 1.81 \pm 0.10$ for SN Zwicky. Our results indicate no evidence for a population of dark compact objects, placing upper limits at the $95\%$ confidence level of \pbh $< 0.25$ for \geu and \pbh $< 0.47$ for SN Zwicky \revision{(for compact objects with masses above $ 0.02 M_{\odot}$)}. Assuming a constant fraction of dark compact objects for both lensed SNe, we obtain \pbh $< 0.19$. These results highlight the potential of strongly lensed SNIa to probe the innermost parts of lens galaxies and learn about compact matter.


\end{abstract}

\maketitle

\begin{figure*}
	\centering
{\includegraphics[width=\textwidth,clip=true]{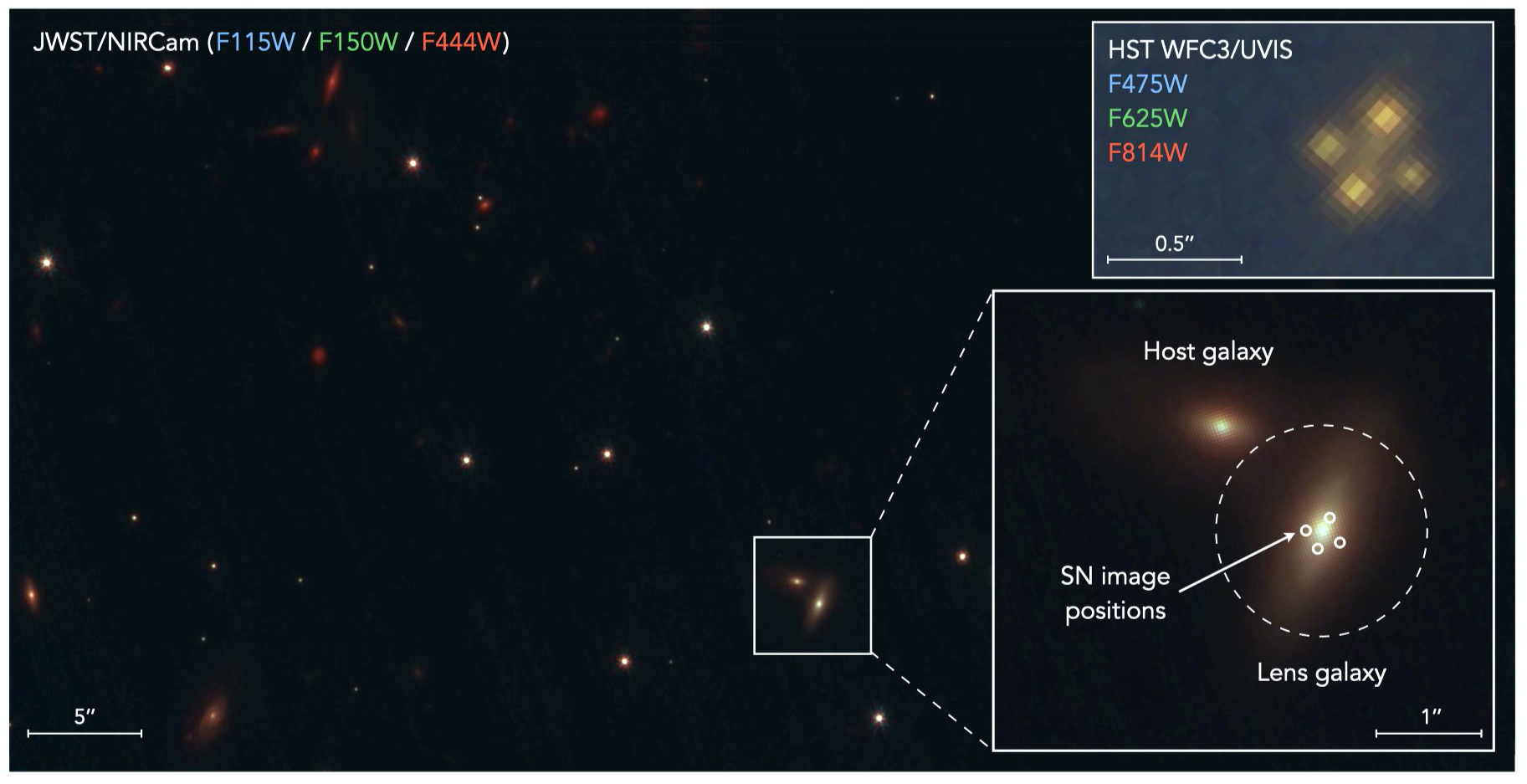}}
	\caption{\small 
    \textit{JWST} image of the field of SN Zwicky post-explosion, showing its lens and host galaxies.  
    The upper right inset shows SN Zwicky when it was active in earlier \textit{HST} observations.
    The lower right inset provides a zoomed-in view of the \textit{JWST} field, showing the lens galaxy and the host galaxy, which is not strongly lensed. The SN image positions are derived from the \textit{HST} observations and projected onto the observations as white circles. The dashed circle with a $1''$ radius indicates the aperture within which the stellar mass measurement of the lens galaxy has been conducted.  \\}
	\label{fig:jwst_zwicky}
\end{figure*}

\begin{figure*}
	\centering
{\includegraphics[width=\textwidth,clip=true]{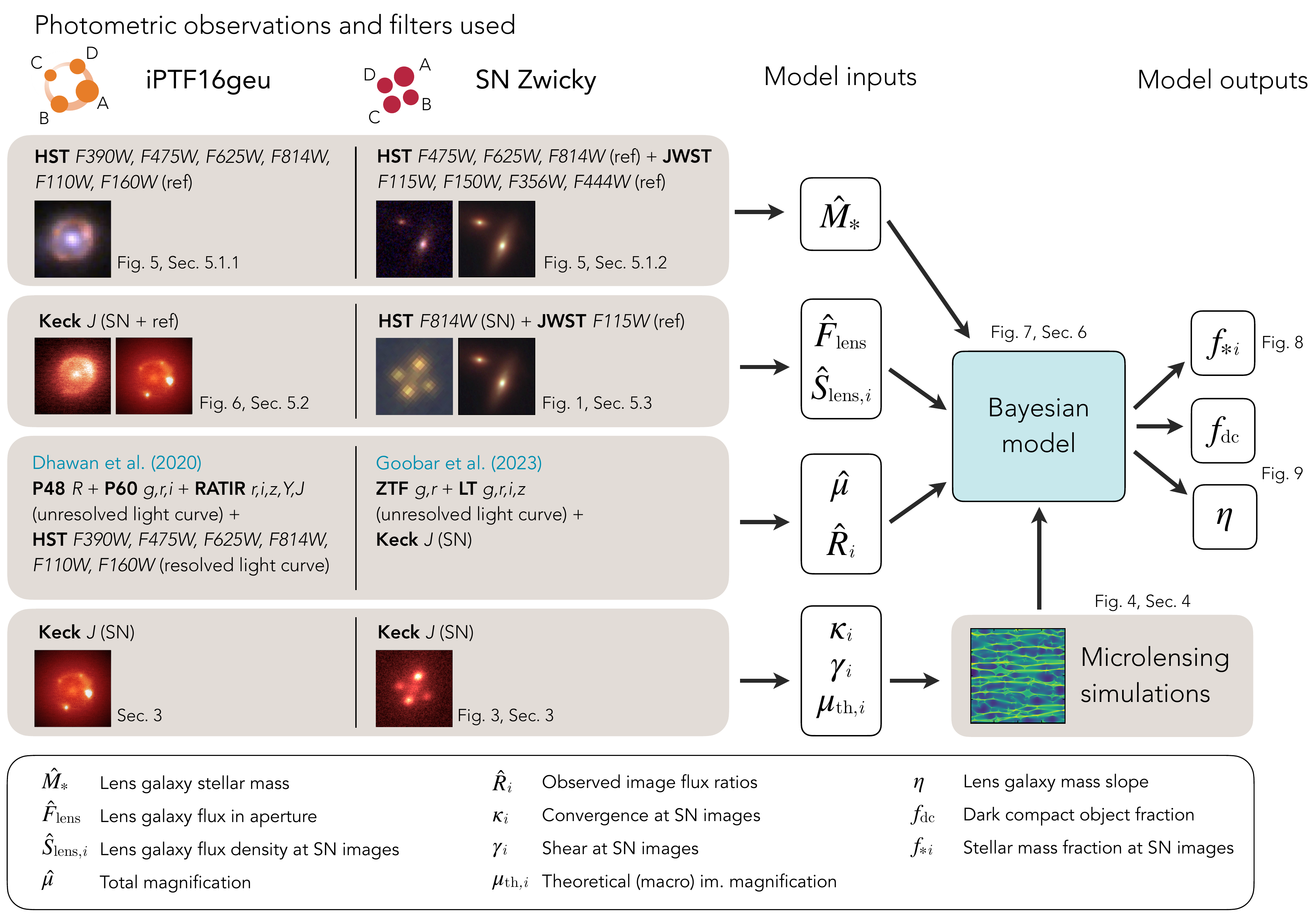}}
	\caption{\small 
    Flowchart of the observations and inferred quantities of this work. The left shaded panels show the specific observations and filters that go into modelling the derived quantities in the middle column, which in turn serve as input for the microlensing simulations and our Bayesian model. The final outputs, shown on the right, are the lens galaxy mass slope, $\eta$, fraction of dark compact objects, \pbh, and the stellar mass fraction at the SN image positions, $f_{*,i}$.  \\}
	\label{fig:flowchart}
\end{figure*} 

\section{Introduction}\label{Sec:Intro}

Strong gravitational lensing occurs when light from a distant light source is deflected by the gravitational potential of a massive galaxy or galaxy cluster along the line of sight, resulting in multiple images of the event. When the light source is a transient event, such as a supernova (SN), we can measure the time delays between the appearance of the multiple images. Combined with a model for the gravitational potential of the lens galaxy and line-of-sight structures, this constrains the present-day expansion rate of the Universe, the Hubble constant \revision{$H_0$} \citep{refsdal1964hubble, treu2016cosmography, Suyu2024, Birrer2024}. Additionally, lensed SNe enable detailed studies of lensing galaxies and high-redshift SNe \citep{cano2018spectroscopic, Johansson2021, Dhawan2024}.

\medskip
While strongly lensed SNe are very useful, they are also extremely rare and challenging to discover. To date, only eight multiply-imaged lensed SNe have been found. Six of them have been discovered behind galaxy clusters: `SN Refsdal' \citep{kelly2015multiple}, `SN Requiem' \citep{rodney2021gravitationally}, `AT2022riv' \citep{Kelly2022riv}, `C22’ \citep{Chen2022_glSNAbell}, `SN H0pe' \citep{Frye2024}, and `SN Encore' \citep{Pierel2023_Encore}. Additionally, two lensed SNe were lensed by isolated elliptical galaxies: `iPTF16geu' \citep{goobar2017iptf16geu} and `SN Zwicky' \citep{Goobar2023_SNZwicky}, shown in Fig.~\ref{fig:jwst_zwicky}. Both of these galaxy-scale lensed SNe were type Ia SNe (SNIa) that were recognised as being gravitationally lensed from their brightness relative to their redshifts. This could be inferred thanks to their standardizable-candle nature, and because they were bright enough to obtain a spectroscopic redshift. \geu was found with the intermediate Palomar Transient Factory (iPTF), a time-domain survey that ran from 2013 to 2017 \citep{Kulkarni2013_iPTF}. From 2018 onwards, a new camera with a larger field of view was installed on the telescope and the survey continued as the Zwicky Transient Facility \citep[ZTF;][]{bellm2019ZTF, Graham2019_ZTF}. After four years of ZTF observations, SN Zwicky was discovered. Simulation studies show that a handful more, unidentified, lensed SNe are expected to be present in the ZTF data below the spectroscopic follow-up limit \citep{Sagues2024}. Archival studies have been conducted to search for these overlooked lensed SNe \citep{Magee2023, Townsend2024}, and in \citet{Townsend2024}, two potential lensed SNe are identified. The number of lensed SN detections is expected to increase orders of magnitudes with the upcoming Vera Rubin Observatory \citep{wojtak2019magnified, goldstein2019rates, oguri2010gravitationally, SainzdeMurieta2023, Arendse2024}.

\medskip
\geu and SN Zwicky share several characteristics: they are both highly magnified ($67.8 \pm 2.9$ and $23.7 \pm 3.2$ times, respectively) and have short time delays of the order of a day \citep{more2017interpreting, Dhawan2020_16geu, Goobar2023_SNZwicky}. Additionally, they are unusually compact lens systems: \geu has an Einstein radius of 0.29\as and SN Zwicky of only 0.17\as. Compared to the population of known lensed galaxies and lensed quasars, both are outliers with small Einstein radii and faint lens galaxies, and they were only identified because they magnified SNIa standard candles \citep[see][ and Fig.~4 in \citealt*{Goobar2023_SNZwicky}]{Lemon2024}.
Another interesting shared property is that the flux ratios of their lensed images, after corrections for differential dust extinction in the lensing galaxy, are inconsistent with predicted flux ratios from a smooth macro lens model \citep{more2017interpreting, Mortsell2020, Goobar2023_SNZwicky, Pierel2023}. A potential explanation for the observed fluxes is additional (de)magnification from stars, stellar remnants and other compact objects in the lens galaxy; known as \textit{microlensing} \citep{Chang1979}. Several papers have analysed SN Zwicky and found evidence that microlensing is needed to explain the observed flux ratios \citep{Goobar2023_SNZwicky, Pierel2023, Larison2024}.
Lensing at such small angular scales ($\sim 10^{-6}$ \as) results in unresolvable micro-images that sum to a total magnification which differs from the macro magnification of the lens galaxy \citep{Young1981, Paczynski1986, Vernardos2024}. This process perturbs the magnification of each image independently, resulting in flux ratios that are discrepant from the macro model predictions. Microlensing is often talked about in terms of complicating the use of lensed SNIa for time-delay cosmography \citep{dobler2006microlensing, Yahalomi2017, foxley2018standardization, Weisenbach2021, Weisenbach_2024}. However, the combination of microlensing and lensed SNe Ia offers a unique opportunity:
since SNe Ia are standardizable candles with predictable brightness for each lensed image, we are able to precisely detect the magnification contribution from microlensing. 

\medskip
Microlensing allows us to probe the abundance of compact objects at the image positions, even if they are dark or too faint to observe. This makes microlensing a unique probe to distinguish between the smooth and compact matter components of the lens galaxy. 
By comparing the fraction of compact objects inferred from microlensing to the stellar mass fraction determined from photometry, we can determine if there are signs of an additional population of dark compact objects, such as primordial black holes (PBHs). PBHs are hypothetical black holes that formed in the early Universe through direct gravitational collapse \citep[e.g.][]{Khlopov2010, Carr2016, Escriva2022}. Since they are not formed through stellar collapse, their masses are not limited to the narrow mass range of stellar black holes.
PBHs are candidates for part of the dark matter in the Universe, as well as potential seeds for supermassive black holes. In particular, gravitational wave observations from the Laser Interferometer Gravitational-Wave Observatory (LIGO) suggest that PBHs of intermediate mass ($10 M_{\odot} < M < 200M_{\odot}$) may constitute a substantial part of the dark matter in the Universe \citep{Clesse2015, Kashlinsky2016, Hawkins2020}, which we could detect through microlensing observations.

\medskip
Microlensing flux ratio studies have been conducted extensively for strongly lensed quasars \citep[e.g ][]{Chang1979, Wambsganss2006, Schmidt2010, Vernardos2019, Vernardos2024}, of which several studies have constrained the abundance of compact objects through microlensing \citep[e.g. ][]{Pooley2009, Pooley2012, Schechter2004, Schechter2014, Mediavilla2009, Mediavilla2017, Esteban2020, Medavilla2024}. Most of these studies conclude that the compact objects causing microlensing can be explained sufficiently by the normal stellar population, 
and the fraction of dark compact objects is constrained to be below $10 \%$ for masses above $0.002 M_{\odot}$ \citep{Gutierrez2023, Medavilla2024}.
For galaxy-scale lensed SNe, the field of microlensing studies has only recently opened up, with \geu the first object that allowed such investigation. In \citet{Mortsell2020}, microlensing was included in the analysis of the image magnifications to obtain the lens galaxy mass slope, $\eta \sim 1.8$. \citet{Diego2022} studied the observed flux ratios and light curves of \geu and concluded that microlensing effects from stars suffice to explain the data.
With the discovery of SN Zwicky, a second galaxy-scale lensed SN has become available to study microlensing.
In this work, we perform a micro and macro lensing analysis of photometric observations of SN Zwicky and \geu, and develop a Bayesian model to constrain the lens galaxy mass slopes and compact objects fractions. By combining these with stellar mass fractions measured from photometry data, we estimate how much of the mass can be in the form of dark compact objects. 

\medskip
We describe the observations used in this work in section.~\ref{Sec:Observations}, the macrolens model in section~\ref{Sec:Macromodel}, and our microlensing simulations in section~\ref{Sec:MicroSims}. 
In section~\ref{Sec:Mass_light}, we describe the lens galaxy stellar mass and light analyses, and in section~\ref{Sec:Model} we outline our Bayesian model. Section~\ref{Sec:Results} shows our results, and section~\ref{Sec:Discussion} gives our discussion and conclusions.
A flowchart of the data we use, inferred quantities, and the input and output of the Bayesian model is presented in Fig.~\ref{fig:flowchart}. 

\section{Observations}\label{Sec:Observations}

Here, we summarise the observations of \geu and SN Zwicky  used in our analysis. More elaborate descriptions can be found in \cite{goobar2017iptf16geu, Dhawan2020_16geu, Goobar2023_SNZwicky, Pierel2023}.

\subsection{iPTF16geu}

SN \geu was discovered in 2016 with iPTF. 
Spectroscopic identification was carried out with Spectral Energy Distribution Machine \citep[SEDM;][]{Blago2018, Rigault2019}  at the Palomar 60-inch telescope on 2 October 2016, and iPTF16geu was found to be consistent with a SNIa at $z = 0.4$.
For the purposes of this work, we focus on the high-resolution imaging data from the Keck~II telescope on Mauna Kea and the \textit{Hubble Space Telescope} (\textit{HST}). The observations from the Keck telescope used in this work were conducted with the Near-IR Camera 2 \citep[NIRC2;][]{2022Fremling} in the $J$-band, using laser guided adaptive optics (LGS-AO). Details of the reduction can be found in \citet{goobar2017iptf16geu,Dhawan2020_16geu}. The \textit{HST} data were taken with the Wide Field Camera~3,
using the UV (\wfcuvis) and near-IR (\wfcir) channels.  There were two epochs of observations in the $F390W$, $F475W$ filters and eight epochs in each of the $F625W$, $F814W$, $F110W$, and $F160W$ filters. After the SN faded, we obtained reference observations in each of the filters on 2018 November 10.
The reference images both for the ground-based and \textit{HST} data are a crucial benefit of lensed SNe, since we can model the lensed host galaxy without the presence of the point sources.

\subsection{SN Zwicky}

On August 2022, SN Zwicky was discovered with ZTF. The object was bright enough for an automatic classification with SEDM, where a spectrum taken on the 21st of August showed it to be a SNIa at a redshift of $z = 0.35$, confirming a brightness several magnitudes brighter than a typical SNIa at that redshift.
The multiple images of SN Zwicky were first resolved on 2022 September 15 with the Keck telescope in the $J$-band, using adaptive optics with the NIRC2 camera \citep{2022Fremling}. On September 21, the LensWatch collaboration\footnote{\href{https://www.lenswatch.org/home}{https://www.lenswatch.org/home}} observed SN Zwicky using \textit{HST} with the optical filters $F475W$, $F625W$ and $F814W$ (WFC3/UVIS Camera) and the NIR $F160W$ filter (WFC3/IR) \citep{2022Pierel, Pierel2023}. To allow for further studies of the lens and host galaxies, reference images after SN Zwicky faded were obtained with \textit{HST} in the same filters \citep{Larison2024} and also with the \textit{James Webb Space Telescope (JWST)}\footnote{Program ID: GO-2905} in the filters $F115W$, $F150W$, $F356W$, and $F444W$, as shown in a composite image in Fig.~\ref{fig:jwst_zwicky}.

\begin{figure*}
\centering
{\includegraphics[width=\textwidth,clip=true]{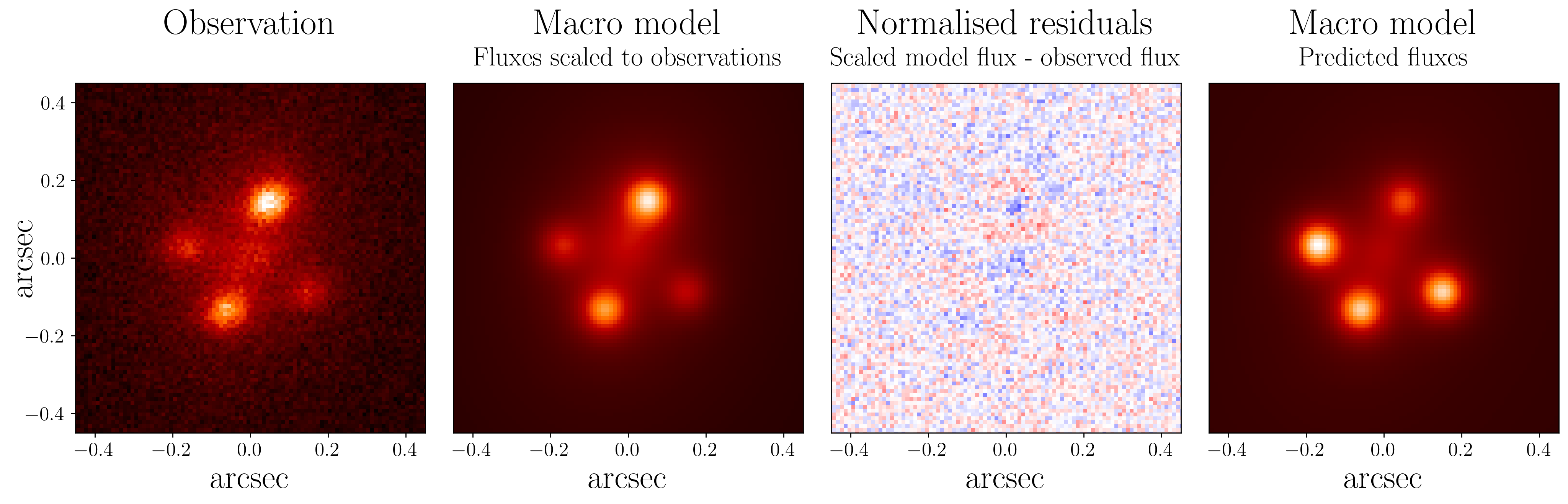}}
\caption{\small The macrolens model fit and residuals to Keck $J$-band observations of SN Zwicky, using a PEMD model with a mass slope of $\eta = 1.8$. From left to right, the panels show: 1.) Keck $J$-band observations of SN Zwicky. 2.) The best-fit macro lens model with the SN image fluxes scaled to match the observed fluxes. Only the SN image positions are used to fit the macro lens model, because the fluxes are affected by microlensing. 3.) Normalised residuals between panels 2 and 1, demonstrating an excellent fit when SN fluxes are scaled. 4.) The predicted fluxes from the best-fit macro model, which deviate from the observed fluxes. This discrepancy highlights the effects of microlensing, which is not captured by the macro model.}
\label{fig:lensmodel_zwicky}
\end{figure*}

\begin{table*}[t]
\caption{\small The macro lens modelling predictions for the convergence ($\kappa$), shear ($\gamma$), and magnifications ($\mu$) at the image positions of \geu and SN Zwicky, for different lens galaxy mass slopes $\eta$. `A', `B', `C', and `D' correspond to the lensed SN images, and $\Sigma = \textrm{A} + \textrm{B} + \textrm{C} + \textrm{D}$, which is used to denote the total magnification. The predicted macro magnifications are discrepant with the observed magnifications, pointing to the presence of additional microlensing magnification. The $\kappa$ and $\gamma$ values serve as input for the microlensing simulations, together with the compact object fraction.}
\centering
\begin{tabular}{l|lllll|lllll|lllll}
\multicolumn{7}{l}{\textbf{\geu}} &  \\
$\eta$ (slope) & \multicolumn{4}{l}{$\kappa$ (convergence)} & & \multicolumn{4}{l}{$\gamma$ (shear)} & & \multicolumn{5}{l}{$\mu$ (macro magnification)} \\
 & A   & B   & C  & D &  & A   & B   & C  & D & & A & B & C & D  & $\Sigma$   \\
\hline
2.10                        & 0.53         & 0.36         & 0.55        & 0.39    &    & 0.63       & 0.46       & 0.65       & 0.49  & &  5.79 & 5.35 & 4.78 & 7.91 & 23.8  \\
2.05                        & 0.55         & 0.39         & 0.57        & 0.42     &   & 0.60       & 0.44       & 0.62       & 0.47  & & 6.81 & 6.11 & 5.63 & 9.04 &  27.6   \\
2.00                        & 0.57         & 0.43         & 0.58        & 0.45     &   & 0.57       & 0.43       & 0.58       & 0.45  & & 7.69 & 6.79 & 6.39 & 10.0 & 30.9    \\
1.95                        & 0.58         & 0.46         & 0.59        & 0.48    &    & 0.53       & 0.41       & 0.54       & 0.43 & &  8.85 & 7.67 & 7.34 & 11.3 & 35.2    \\
1.90                        & 0.60         & 0.49         & 0.62        & 0.51     &   & 0.50       & 0.39       & 0.51       & 0.41 &  & 10.3 & 8.74 & 8.52 & 12.9 & 40.5    \\
1.85                        & 0.62         & 0.52         & 0.64        & 0.54     &   & 0.47       & 0.37       & 0.48       & 0.38  &  & 11.1 & 9.44 & 9.33 & 14.0 & 43.9   \\
1.80                        & 0.65         & 0.54         & 0.66        & 0.56     &   & 0.45       & 0.34       & 0.46       & 0.36  &  & 13.1 & 10.9 & 11.0 & 16.2 & 51.2   \\
1.75                        & 0.66         & 0.57         & 0.67        & 0.59     &   & 0.42       & 0.33       & 0.42       & 0.34  &  & 15.4 & 12.7 & 13.0 & 18.9 & 60.0   \\
1.70                        & 0.69         & 0.60         & 0.70        & 0.62    &    & 0.39       & 0.31       & 0.39       & 0.32   &  & 18.2 & 14.9 & 15.4 & 22.2 & 70.7  \\
1.65                        & 0.71         & 0.63         & 0.72        & 0.65    &    & 0.36       & 0.29       & 0.36       & 0.29   &  & 23.4 & 18.7 & 19.7 & 27.9 & 89.7  \\
1.60                        & 0.73         & 0.66         & 0.74        & 0.68    &    & 0.33       & 0.26       & 0.33       & 0.27  &   & 27.4 & 21.8 & 23.2 & 32.6 & 105  \\
1.55                        & 0.75         & 0.69         & 0.76        & 0.70    &    & 0.30       & 0.24       & 0.31       & 0.25   &  & 33.6 & 26.5 & 28.5 & 39.7 & 128  \\
1.50                        & 0.77         & 0.72         & 0.78        & 0.73     &   & 0.27       & 0.22       & 0.28       & 0.23  &   & 42.1 & 33.0 & 35.7 & 49.3 & 160  \\
1.45                        & 0.80         & 0.75         & 0.80        & 0.76   &     & 0.25       & 0.20       & 0.25       & 0.20   &  & 53.4 & 41.4 & 45.4 & 62.2 & 202  \\
1.40                        & 0.82         & 0.77         & 0.82        & 0.79   &     & 0.22       & 0.18       & 0.22       & 0.18   &  & 67.2 & 52.0 & 56.3 & 76.8 & 252 
\end{tabular} 

\vspace{0.2cm}

\begin{tabular}{l|lllll|lllll|lllll}
\multicolumn{7}{l}{\textbf{SN Zwicky}} &  \\
$\eta$ (slope) & \multicolumn{4}{l}{$\kappa$ (convergence)} & & \multicolumn{4}{l}{$\gamma$ (shear)} & & \multicolumn{5}{l}{$\mu$ (macro magnification)} \\
 & A   & B   & C  & D &  & A   & B   & C  & D & & A & B & C & D  & $\Sigma$   \\
\hline
2.10                        & 0.74         & 0.31         & 0.62        & 0.32    &    & 0.85       & 0.41       & 0.72       & 0.42  &  & 1.55 & 3.33 & 2.68  & 3.60 &  11.2 \\
2.05                        & 0.73         & 0.35         & 0.62        & 0.36    &    & 0.78       & 0.40       & 0.67       & 0.41 &  & 1.87 & 3.78 & 3.23 & 4.08 & 13.0    \\
2.00                        & 0.72         & 0.38         & 0.63        & 0.39    &    & 0.72       & 0.38       & 0.63       & 0.39   &   & 2.24 & 4.28 & 3.87 & 4.63 & 15.0 \\
1.95                        & 0.72         & 0.42         & 0.64        & 0.43    &    & 0.67       & 0.37       & 0.59       & 0.38  &  & 2.69 & 4.89 & 4.63 & 5.29 & 17.5   \\
1.90                        & 0.73         & 0.45         & 0.65        & 0.46    &    & 0.62       & 0.35       & 0.55       & 0.36  &   & 3.21 & 5.60 & 5.52 & 6.06 & 20.4   \\
1.85                        & 0.73         & 0.48         & 0.66        & 0.49    &    & 0.58       & 0.34       & 0.52       & 0.34  &  & 3.85 & 6.47 & 6.60 & 7.01 & 23.9    \\
1.80                        & 0.74         & 0.52         & 0.67        & 0.52    &    & 0.53       & 0.32       & 0.48       & 0.32  &  & 4.65 & 7.55 & 7.93 & 8.18 & 28.3   \\
1.75                        & 0.75         & 0.55         & 0.69        & 0.55     &   & 0.49       & 0.30       & 0.45       & 0.31  &   & 5.58 & 8.79 & 9.47 & 9.53 & 33.4  \\
1.70                        & 0.76         & 0.58         & 0.71        & 0.59   &     & 0.45       & 0.28       & 0.42       & 0.29  &  & 6.78 & 10.4 & 11.5 & 11.3 & 40.0   \\
1.65                        & 0.78         & 0.61         & 0.72        & 0.62   &     & 0.41       & 0.27       & 0.39       & 0.27  &  & 8.28 & 12.4 & 14.0 & 13.5 &  48.2  \\
1.60                        & 0.79         & 0.64         & 0.74        & 0.65    &    & 0.38       & 0.25       & 0.35       & 0.25 &  & 10.2 & 15.0 & 17.2 & 16.3 & 58.7    \\
1.55                        & 0.80         & 0.67         & 0.76        & 0.68   &     & 0.34       & 0.23       & 0.32       & 0.23  &  & 12.7 & 18.3 & 21.2 & 19.9 & 72.1   \\
1.50                        & 0.82         & 0.70         & 0.78        & 0.71   &     & 0.31       & 0.21       & 0.29       & 0.21 &  & 16.2 & 22.8 & 26.9 & 24.8 & 90.7    \\
1.45                        & 0.84         & 0.73         & 0.80        & 0.74    &    & 0.27       & 0.19       & 0.26       & 0.19  &  & 20.7 & 28.6 & 34.2 & 31.2 & 115    \\
1.40                        & 0.85         & 0.76         & 0.82        & 0.77    &    & 0.24       & 0.17       & 0.23       & 0.17  &  & 27.4 & 37.2 & 44.9 & 40.6 & 150 
\end{tabular}
\label{Table:kappa_gamma_mu}
\end{table*}


\section{Macro lens modelling}\label{Sec:Macromodel}

We describe the definitions of our mass and light models in section~\ref{subsect:model_def}, our fitting methodology in section~\ref{subsect:fitting_meth}, and explain how we constrain the galaxy mass slope from standard candle magnifications in section~\ref{subsect:slope_mu}. 

\subsection{Mass and light model definitions}
\label{subsect:model_def}

We model the mass distribution in the lens galaxies of SN Zwicky and iPTF16geu as a power-law elliptical mass distribution (PEMD):
\begin{align}
    \kappa(x, y) = \frac{3 - \eta}{2} \left(\frac{\theta_E}{\sqrt{q x^2 + y^2/q}} \right)^{\eta - 1},
    \label{eq:pemd}
\end{align}
where $\kappa$ is the projected surface mass density (convergence), $q$ denotes the projected axis ratio of the lens, $\theta_E$ is the Einstein radius, and $\eta$ corresponds to the logarithmic lens mass slope. The coordinates ($x, y$) are centred on the lens galaxy mass centre and are rotated by the lens orientation angle $\phi$ so that the $x$-axis aligns with the major axis of the lens.

\medskip
We model the light of the lens and host galaxies in the form of elliptical Sérsic profiles:
\begin{equation}
I(R) = I_{\rm e}\exp{\left\{-b_n\left[\left(\frac{R}{R_{\rm e}}\right)^{1/n}-1\right]\right\}},
\label{eq:sersic}
\end{equation}
where $I_{\rm e}$ is the intensity at the half-light radius $R_{\rm e}$, $b_n=1.9992 n -0.3271$ \citep{birrer2018lenstronomy} and
\begin{equation}
R\equiv\sqrt{x^2 + y^2/q_{\textrm{s}}^2},
\end{equation}
with $q_{\textrm{s}}$ the axis ratio of the Sérsic profile and ($x, y$) the coordinates that are centred on the lens galaxy light centre and rotated by the orientation angle $\phi_{\textrm{s}}$. The multiple lensed images of the SNe are modelled as point sources, where the Point Spread Function (PSF) was obtained by fitting a Moffat profile to stars in the field, where possible (e.g. \textit{HST} and \textit{JWST}). In observations with no suitable stars (e.g. Keck), the SN images themselves were used for the PSF modelling.

\revision{\subsection{Testing angular complexity in the mass model}}
\label{subsect:multipoles}

\revision{To investigate whether deviations from a smooth elliptical mass distribution can account for the observed flux anomalies in the SN images, we tested the inclusion of multipole perturbations in the lens model. Recent work by \citet{Cohen2024} has shown that higher-order angular structure in the lens galaxy can also reproduce flux-ratio anomalies, potentially providing an alternative explanation to microlensing. 
Specifically, we considered both circular and elliptical fourth-order multipole perturbations to the convergence, following the formalism of \citet{Xu2013} as implemented in the open-source lens modelling software \texttt{Lenstronomy}\footnote{\href{https://lenstronomy.readthedocs.io/en/latest/}{https://lenstronomy.readthedocs.io/en/latest/}} \citep{birrer2018lenstronomy, Birrer2021LenstronomyII}. The fourth-order multipole was added as an additional mass component with two free parameters: the amplitude $a_4$ and the phase $\phi_4$. These multipole terms allow for departures from purely elliptical isodensity contours, introducing boxy or disky shapes. We applied this extended mass model to Keck $J$-band observations of both SN Zwicky and \geu, fixing the mass slope $\eta$ to our best-fit values. The flux ratios of the SN images were included in the likelihood function to assess whether the lens model could reproduce them. }

\medskip
\revision{We briefly describe our main findings here. Our results for \geu yield a poor fit, with a high reduced $\chi^2$ value. This demonstrates that even with the included angular complexity in the mass distribution, the lens model fails to reproduce the observed flux ratios.
For SN Zwicky, the multipole model matches the observed flux ratios more closely, but the resulting mass distribution is highly irregular and unphysical, with best-fit multipole amplitudes ($a_4 = -0.192^{+0.007}_{-0.005}$ for circular multipoles and $a_4 = -0.103^{+0.006}_{-0.005}$ for elliptical multipoles) inconsistent with realistic galaxy profiles (which are often of the order $10^{-2}$, see e.g.\ \citealp{Hao2006}). These unphysical multipole amplitudes are likely a consequence of the absence of a lensed host galaxy in the SN Zwicky observations, which would otherwise have constrained the model to more realistic values, as is the case for \geu. We conclude that angular complexity does not provide a plausible physical explanation for the observed flux ratios in either lensed SN system. These results strengthen the case that microlensing is required to explain the observed flux anomalies.}

\medskip
\revision{When the flux ratio constraints are excluded, adding multipole perturbations to the lens model decreases the reduced $\chi^2$ by only 0.008 for \geu and provides no improvement for SN Zwicky. In our subsequent analysis, we do not include angular complexity and proceed with the mass model described in Section~\ref{subsect:model_def}.} \\

\subsection{Fitting methodology}
\label{subsect:fitting_meth}

Using \revision{\texttt{Lenstronomy}}, we simultaneously reconstruct the lens mass model, SN image positions, and the light distributions of the lens and host galaxy. For both \geu and SN Zwicky, we perform lens modelling for several observations from different telescopes and in different filters, each with characteristics that require slightly different approaches to the modelling. In this section, we outline our overarching modelling approach which applies to most observations, and we highlight deviations from this for specific cases in their respective sections.

\medskip
Most of our lens models have the following $25$ non-linear parameters: 

\begin{itemize}
[leftmargin=*,itemsep=1em]
    \item \textbf{Lens galaxy mass model }(PEMD): Einstein radius ($\theta_{\textrm{E}}$), axis ratio ($q_{\textrm{lens}}$), orientation angle ($\phi_{\textrm{lens}}$), lens galaxy mass centre ($x_{\textrm{lens}}, y_{\textrm{lens}}$). Note that for most runs, the power-law slope of the mass profile, $\eta$, is not a free parameter, but fixed to specific values (as explained further in section~\ref{subsect:slope_mu}). 
    \item \textbf{Lens galaxy light model} (Sérsic): Sérsic radius ($R_{\textrm{s,lens}}$), Sérsic index ($n_{\textrm{s,lens}}$), axis ratio ($q_{\textrm{s,lens}}$), orientation angle ($\phi_{\textrm{s,lens}}$), lens galaxy light centre ($x_{\textrm{s,lens}}, y_{\textrm{s,lens}}$).
    \item \textbf{Host galaxy light model}: Sérsic radius ($R_{\textrm{s,host}}$), Sérsic index ($n_{\textrm{s,host}}$), axis ratio ($q_{\textrm{s,host}}$), orientation angle ($\phi_{\textrm{s,host}}$), lens galaxy light centre ($x_{\textrm{s,host}}, y_{\textrm{s,host}}$).
    \item \textbf{SN images:} positions of four point sources ($x_{\textrm{A}}, \, y_{\textrm{A}}, \, x_{\textrm{B}}, \, y_{\textrm{B}}, \, x_{\textrm{C}}, \, y_{\textrm{C}}, \, x_{\textrm{D}}, \, y_{\textrm{D}}$).
\end{itemize}


Note that since the host galaxy of SN Zwicky is located at a large offset ($\sim 1.4 ''$) from the lens galaxy centre, and it is not strongly lensed, the host galaxy light model is not always included in the lens modelling. 

\medskip
We iterate three times over fitting preliminary models with particle swarm optimization (PSO). For each of the three runs of PSO, we use 100, 100, and 10 particles and 100, 100, and 10 iterations, respectively. For the first run, we initialise the particles according to wide Gaussian priors around the expected parameter values. For the second and third runs, we reduce the standard deviations of the particle initial positions to 10\% and 1\% of their original values, respectively, around the best fit parameter values from the previous PSO iteration. 
We also iterate over the PSF model, using a built-in feature in \texttt{Lenstronomy} that minimises the residuals between the observed and reconstructed image around the SN positions \citep{Shajib2019}. Once a suitable PSF is obtained, we run a Markov chain Monte Carlo (MCMC) sampler to obtain parameter estimates with uncertainties. In particular, we use an affine invariant MCMC ensemble sampler from \texttt{python} package \texttt{emcee} \citep{ForemanMackey2013, Goodman2010}. We typically use 200 walkers with 500 burn-in steps that we discard and 1500 steps from which we obtain the final posterior on the model parameters. Because we have already used PSO to bring the fitter close to the posterior maximum, we initialize the walkers with a standard deviation of 2\% of that specified in the Gaussian priors.

\begin{figure*}
	\centering
{\includegraphics[width=\textwidth,clip=true]{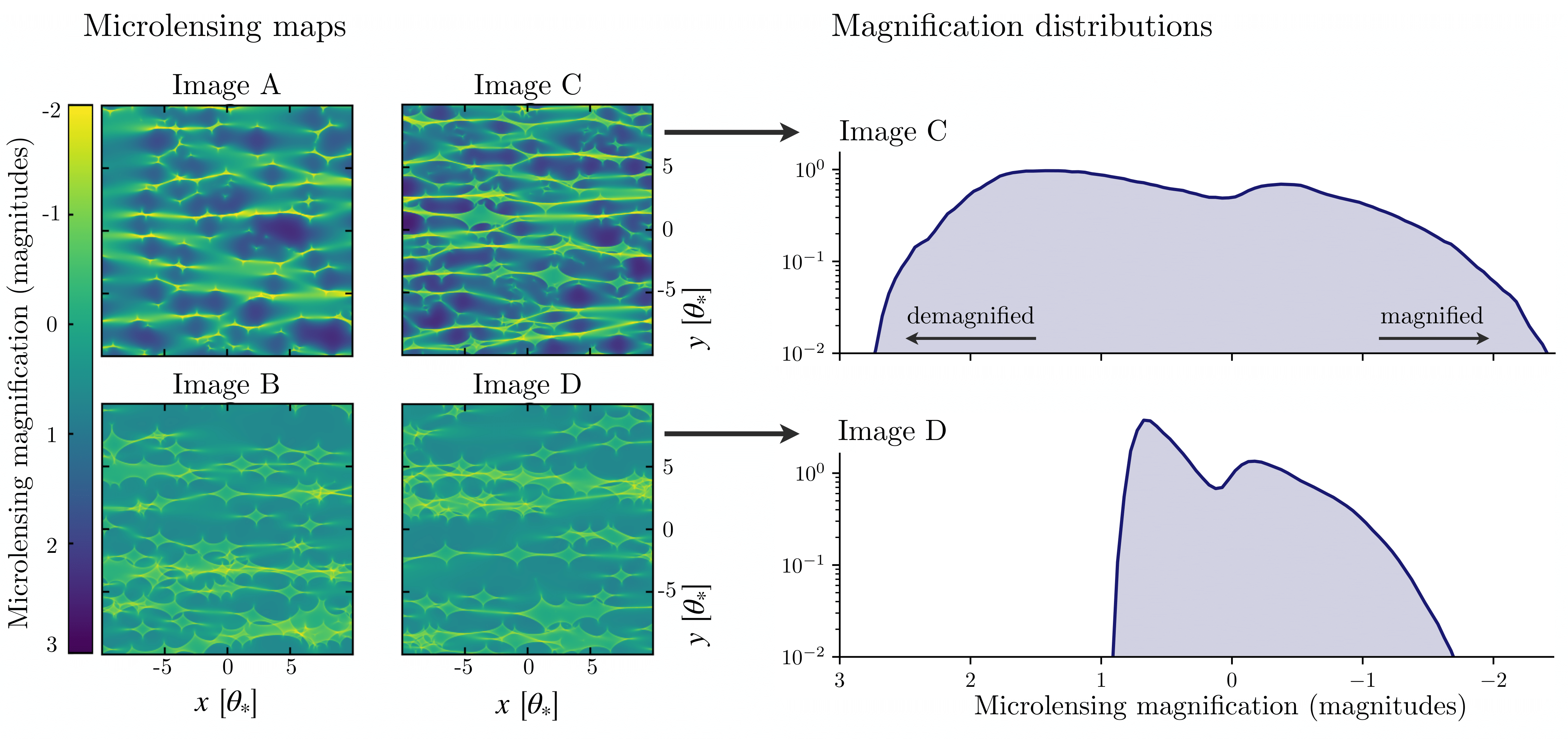}}
	\caption{\small Microlensing magnification maps and distributions for SN Zwicky, assuming a lens mass slope $\eta = 1.8$ and compact object fraction $f_c = 0.2$. The microlensing magnification is given in units of magnitudes, $\Delta m = -2.5\log_{10}(\mu_{\rm obs}/\mu_{\rm th})$.
    \textit{Left:} Microlensing magnification maps for images A, B, C, and D. The microlensing caustics show the areas of magnification and demagnification in the source plane due to compact objects in the lens galaxy. $\theta_*$ refers to the Einstein radius of an individual microlens. \textit{Right:} The magnification distributions for images C and D, obtained by convolving the SN size with the microlensing maps. 
    The microlensing magnification distributions for images A and C are similar and show a wide spread in magnifications, since they are both saddle points in the time-delay surface. Images B and D show similarly narrow distributions, because they represent minima in the time-delay surface.\\}	\label{fig:micromaps}
\end{figure*}

\subsection{Mass slope constraints from standard candle magnifications}
\label{subsect:slope_mu}

\medskip 
Constraining the mass slope $\eta$ of the lens galaxy from photometric observations alone is challenging, particularly for SN Zwicky, which occurred at an offset of 7 kpc from its host galaxy's centre and therefore lacks strongly lensed host galaxy arcs. This is not an uncommon scenario; around half of strongly lensed SNe is expected to be without a strongly lensed host \citep{SainzdeMurieta2024}.
An alternative approach for constraining the mass slope is to utilise the standard candle nature of SNIa, which provides information about the absolute magnification. The magnification of the macro lensing model, $\mu_{\rm th}$, depends strongly on the slope of the mass profile, with flatter profiles providing larger magnifications.
However, constraining the mass slope through the observed magnification can only be done when the additional magnification effects of microlensing are taken into consideration. Section~\ref{Sec:MicroSims} describes our approach to modelling microlensing effects through simulations, which require inputs for the convergence ($\kappa$) and shear ($\gamma$). 
We use observations from the Keck $J$-band to model the lens systems of \geu and SN Zwicky, and determine the $\kappa$ and $\gamma$ values for each SN image, for a range of mass slopes between $\eta = 1.4$ and $\eta = 2.1$ (step size 0.05). 
Our range of $\eta$ values was initially chosen to encompass the most common slope values from known lens galaxies, and eventually extended to lower $\eta$ values to capture the $95\%$ confidence level contours for \geu and SN Zwicky.
We also compute the predicted macro magnification $\mu_{\textrm{th}}$ for each SN image and slope value, which, in combination with the simulated microlensing magnification distributions, allows us to determine the best-fit slope values of \geu and SN Zwicky.
Table~\ref{Table:kappa_gamma_mu} shows the best-fit convergence, shear and magnification values per slope and per image for SN Zwicky and iPTF16geu, and Fig.~\ref{fig:lensmodel_zwicky} shows results and residuals of SN Zwicky's best-fit macrolens model for a slope of $\eta = 1.8$.


\section{Microlensing simulations}\label{Sec:MicroSims}

In order to incorporate the effects of microlensing at the SN image positions for both SN Zwicky and iPTF16geu, we simulate microlensing magnification maps. The three main parameters that govern the microlensing maps are the convergence (projected surface mass density), $\kappa$, the shear, $\gamma$, and the compact object fraction, $f_c$. From the macro lens model fits described in section~\ref{Sec:Macromodel}, we estimate the best-fit $\kappa$ and $\gamma$ values for each SN image and each mass slope (with $\eta$ ranging from 1.4 to 2.1; see Table~\ref{Table:kappa_gamma_mu}). For our range of $\eta$ values and for $f_c$ ranging from 0.1 to 1.0 in steps of 0.1, we create magnification maps for all four lensed SN images. An example of the resulting microlensing magnification maps and distributions for SN Zwicky with $\eta = 1.8$ and $f_c = 0.2$ is displayed in Fig.~\ref{fig:micromaps}. 

\medskip
The microlensing magnification maps are created using Inverse Polygon Mapping \citep{2006ApJ...653..942M, 2022ApJ...941...80J} implemented on a GPU for efficiency \revision{\citep{2025MNRAS.541..281W}}. The maps are 20 Einstein radii and 10,000 pixels per side. This is a large enough size to create a representative magnification distribution \citep{2013MNRAS.434..832V}, while the pixel scale corresponds to approximately 1 day of SN expansion. The maps are convolved with a uniform disk to approximate the size and profile of the expanding SN photosphere at observation, a sufficient approximation for our purposes \citep{pierel2019turning}. 
While more complex SN intensity models may formally be more accurate \citep{Goldstein2018microlensing, Huber2019LSSTcadence}, the main source parameter that impacts microlensing is the half-light radius of the source \citep{2005ApJ...628..594M, Vernardos2019}. 
For both \geu and SN Zwicky, the size of the photosphere at the time of observations is of the order of a light day.
We average 10 maps for the final microlensing magnification probability distributions in order to minimize scatter from simulating a finite source plane region.

\medskip
The maps are created using a single mass (in our case, $1M_\odot$) for all the microlenses, a common approximation as the magnification distribution is largely insensitive to a mass spectrum, except in extreme scenarios \citep{2004ApJ...613...77S}. We note that the precise value of the mass used only loosely effects results, as the SN is still only a fraction of an Einstein radius in size, and a smaller microlens mass is degenerate with a larger source size (higher SN expansion speed). The number of microlenses used to create each map depends on the macro-model parameters, but is large enough to ensure that the majority of microlensed flux is accounted for \citep{1986ApJ...306....2K}.


\section{Stellar mass and light inference}\label{Sec:Mass_light}

We derive estimates of the stellar mass fractions at the image positions for SN Zwicky and \geu, based on photometric observations taken when the SNe were active, and after they had faded away. To obtain the stellar mass fractions, we require estimates of the total stellar mass of the lens galaxies, along with the total flux and the flux density at the image positions, which allow us to scale the stellar mass to the image positions. Finally, we divide this by the total mass density at the image positions, obtained from the lens modelling, to get the stellar mass fractions at the SN image positions. This Section describes how we estimate the stellar masses and light models of the lens galaxies.


\begin{figure*}
	\centering
{\includegraphics[width=0.95\linewidth,clip=true]{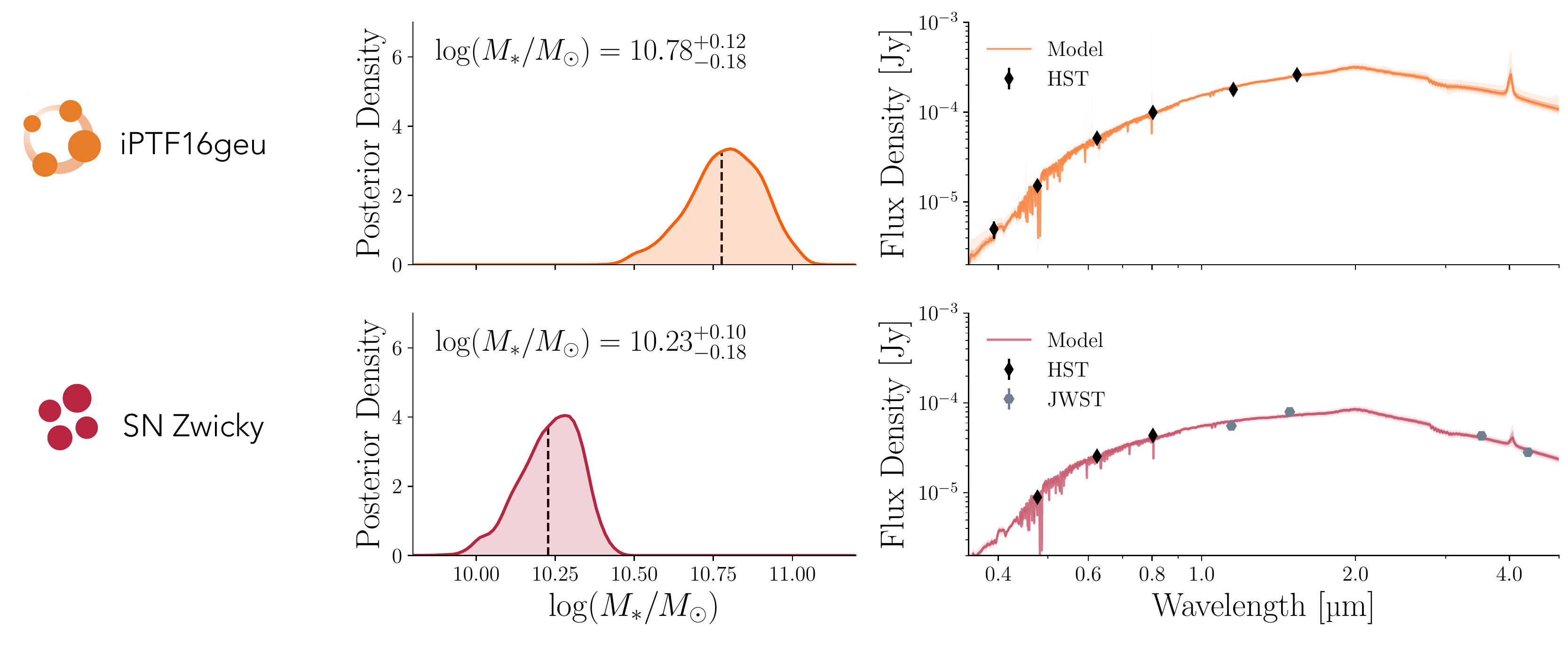}}
	\caption{\small Stellar mass estimates for \geu (top) and SN Zwicky (bottom) obtained with \texttt{prospector}. The left panels show the stellar mass posteriors, and the right panels the best-fit SEDs and photometric measurements used for the analysis. For \geu, the observations consist of six filters from \textit{HST}, and for SN Zwicky, three filters from \textit{HST} and four from \textit{JWST}. \\}
	\label{fig:SED_stellar_mass}
\end{figure*}

\subsection{Lens galaxy stellar mass estimates}\label{subseq:stellar_mass}

The lens galaxy stellar masses are inferred from a series of photometric observations in the visible and NIR, which are fitted with a spectral energy distribution (SED) model. We use \texttt{prospector} to fit an SED model based on flexible stellar population synthesis (FSPS; \citealp{Conroy2009, Conroy2010, Conroy2010ApJ}; via the \texttt{python-fsps} bindings; \citealp{Johnson2024}) to our photometric observations of the two lens galaxies. We infer the stellar mass ($M_*/M_\odot$) jointly with the stellar metallicity ($Z_*/Z_\odot$), optical depth of dust ($\tau_{5500}$), and two parameters describing the star formation history ($t_\text{age}$, $\tau_\text{SF}$). Although the photometry alone does not constrain the additional parameters, we leave them free to get a robust estimate of the stellar mass uncertainty. We use the MILES stellar template library \citep{SanchezBlazquez2006}, MIST isochrones \citep{Choi2016}, a \citet{Chabrier2003} initial mass function, and a parametric star formation history (SFH) that is linearly increasing at early times and exponentially decaying at late times \citep[see e.g.][]{Carnall2019, Johnson2021}. The SFH has the form
\begin{equation}
    \text{SFR}(t)\propto \frac{t_\text{age}-t}{\tau_\text{SF}} \times \exp\left(\frac{t-t_\text{age}}{\tau_\text{SF}}\right),
\end{equation}
where $t_\text{age}$ is the age of the galaxy, $0<t\leq t_\text{age}$ is lookback time, and $t_\text{age}-\tau_\text{SF}$ sets the peak of the SFH. We use the \citet{Calzetti2000} dust attenuation law, with fixed slope and optical depth $\tau_{5500}$ at 5500~\AA. Gas-phase metallicity is coupled to stellar metallicity, with nebular emission being provided by the \texttt{Cloudy} \citep{Ferland2013} model grids from \citet{Byler2017}. We correct for Milky-Way dust using the \citet{Gordon2023} extinction curve (implemented in \citealp{Gordon2024}, \revision{and relying on work from \citealp{Gordon2009, Fitzpatrick2019, Gordon2021, Decleir2022}}) with $R_V=3.1$, and $E(B-V)$ estimated from the \citet{Schlafly2011} dust map. We infer $E(B-V)=0.0725$~mag for \geu, and $E(B-V)=0.1558$~mag for SN Zwicky. We carry out our inference with the lens galaxy redshifts fixed.

\medskip
The posterior distribution over our five model parameters is sampled using a dynamic nested sampler \citep{Skilling2004, Skilling2006, Higson2019} implemented in \texttt{dynesty} \citep{Speagle2020, Koposov2024}, with a proposal strategy based on rejection sampling within ellipsoidal bounds \citep{Feroz2009}. Our priors are: $\log(M_*/M_\odot)\sim \mathcal{U}(8,12)$, $\log(Z_*/Z_\odot)\sim \mathcal{U}(-2.00,0.19)$, $\tau_{5500}\sim\mathcal{U}(0,2)$, $t_\text{age}/\text{Gyr}\sim \mathcal{U}(0,t_\text{Univ}(z))$, and $\log(\tau_\text{SF}/\text{Gyr})\sim\mathcal{U}(-1, \log(30))$. Fig.~\ref{fig:prospector_all_params} in the Appendix shows a corner plot of all the free parameters used in the stellar mass fits for \geu and SN Zwicky.

\begin{table*}[ht]
\caption{\small Photometry and lens fraction measurements for \geu{} and SN Zwicky, which serve as input for the stellar mass analysis with \texttt{prospector}. The lens fraction is the ratio between the lens galaxy flux and the total flux. Both the photometry and lens fractions are measured within an aperture with radius $1''$. For \geu, observations in six \textit{HST} filters are used. For SN Zwicky, \textit{HST} observations in the $F475W$, $F625W$, and $F814W$ bands, and \textit{JWST} observations in the $F115W$, $F150W$, $F356W$, and $F444W$ filters are used. We also quote the Milky Way extinction correction (in magnitudes) for each band, computed using the \citet{Gordon2023} curve. The photometric measurements are given in AB magnitudes, and have not been corrected for Milky-Way reddening or the lens fractions. }
\label{Table:16geu_zwicky_photometry}
\centering
\small
\begin{tabular}{l|llllll}
\multicolumn{7}{l}{\normalsize{\textbf{\makebox[2cm][l]{\geu{} lens galaxy photometry (\textit{HST/WFC3})}}}} \vspace{0.2cm} \\ 

              & $F390W$ & $F475W$ & $F625W$ & $F814W$  & $F110W$ & $F160W$ \\
\hline
Photometry    & $22.46\pm0.24$ & $21.08\pm0.12$ & $19.55\pm0.06$ & $18.68\pm0.05$ & $17.82\pm0.05$ & $17.27\pm0.05$              \\
Lens fraction & 0.980 & 0.880 & 0.780 & 0.720 & 0.620 & 0.560 \\
MW extinction & 0.320 & 0.267 & 0.191 & 0.130 & 0.068 & 0.042
\end{tabular}

\vspace{0.4cm}

\begin{tabular}{l | lllllll}
\multicolumn{8}{l}{\normalsize{\textbf{\makebox[2cm][l]{SN Zwicky lens galaxy photometry (\textit{HST/WFC3 + JWST/NIRCam})}}}} \vspace{0.2cm} \\
              & $F475W$ & $F625W$ & $F814W$  & $F115W$  & $F150W$  & $F356W$  & $F444W$ \\
\hline
Photometry & $22.03\pm0.05$ & $20.68\pm0.06$ & $20.00\pm0.08$ & $19.34\pm0.10$ & $18.98\pm0.10$ & $19.57\pm0.10$ & $19.79\pm0.10$ \\
Lens fraction &  0.944 & 0.904 & 0.927 & 0.727 & 0.784 & 0.786 & 0.630 \\
MW extinction & 0.573 & 0.410 & 0.279 & 0.148 & 0.095 & 0.021 & 0.016
\end{tabular} \\
\end{table*}

\subsubsection{\geu stellar mass}

To estimate the stellar mass of \geu's lens galaxy, we use reference observations in the $F390W$, $F475W$, $F625W$, $F814W$, $F110W$, and $F160W$ bands from \textit{HST}, taken after the SN faded away. We compute the photometry within an  aperture with radius $1''$ on the lens galaxy centre.
Measuring the photometry for \geu is complicated by the fact that the lens galaxy is blended with the lensed arcs from the host galaxy. To separate the light contributions from the lens and the host galaxy,  we simultaneously model the $F475W$, $F625W$, and $F814W$ filters from \textit{HST}, following the approach outlined in section~\ref{Sec:Macromodel}. In each filter, we compute the fraction of the flux coming from the lens galaxy inside the $1''$ aperture, and multiply that with the photometry measurements. 
The lens and host galaxy are difficult to disentangle in the \textit{HST} $F110W$ and $F160W$ bands, and therefore, we use the Keck $J$-band observation to estimate the lens galaxy flux fraction in the NIR. Details about the \geu Keck $J$-band light modelling are given in section~\ref{subsec:16geu_light}. The photometry and lens fraction measurements in all filters are given in Table~\ref{Table:16geu_zwicky_photometry}.
The resulting stellar mass estimate for \geu's lens galaxy is $\log(M_*/M_{\odot}) = 10.78^{+0.12}_{-0.18}$, for which the full stellar mass posterior and the best-fit SED from \texttt{prospector} are shown in Fig.~\ref{fig:SED_stellar_mass}.

\subsubsection{SN Zwicky stellar mass}

For SN Zwicky, the input for \texttt{prospector} are \textit{HST} observations in the $F475W$, $F625W$, and $F814W$ bands, and \textit{JWST} observations in the $F115W$, $F150W$, $F356W$, and $F444W$ bands, all taken after the SN faded away. In this case, the analysis of the lens galaxy is aided by the fact that the host galaxy is not strongly lensed and is located at a relatively large distance from the centre of the lens galaxy ($\sim 1.4''$). We compute the flux within an aperture of radius $1''$ (which excludes most of the host galaxy contribution) as input for \texttt{prospector}. To correct for small contributions of host galaxy flux, we model the lens and host galaxy to determine the lens galaxy flux fraction within the $1''$ aperture. The photometric measurements and lens galaxy flux fractions are listed in Table~\ref{Table:16geu_zwicky_photometry}.
The resulting stellar mass estimate for SN Zwicky's lens galaxy is $\log(M_*/M_{\odot}) = 10.23^{+0.10}_{-0.18}$, for which the full posterior and the best-fit SED from \texttt{prospector} are shown in Fig.~\ref{fig:SED_stellar_mass}.

\begin{figure*}
	\centering
{\includegraphics[width=\textwidth,clip=true]{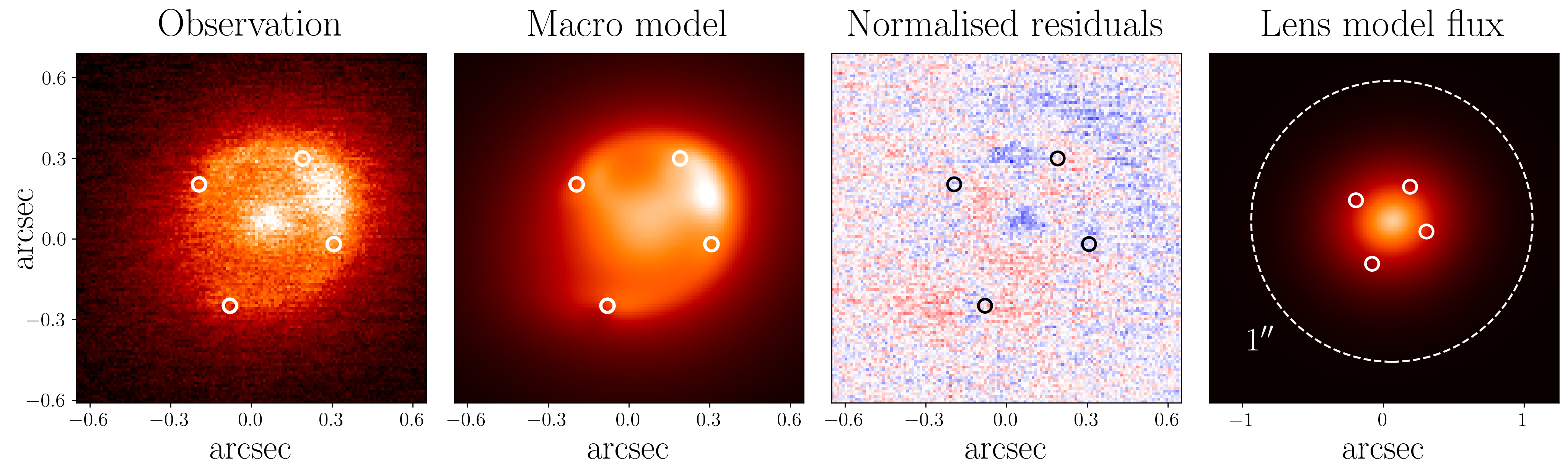}}
	\caption{\small Lens light modelling of \geu using Keck observations in the $J$-band, with the SN image positions projected as white/black open circles. From left to right: the reference observation of the lens and host galaxy, the best-fit macro model, normalised residuals, and the lens light model. To infer the stellar mass at the image positions, the stellar mass to light ratio is computed using the total flux in an aperture of $1''$, which is scaled with the flux density at the SN image positions. \\}
	\label{fig:16geu_lens_light}
\end{figure*}

\subsection{Lens galaxy light model for \geu}\label{subsec:16geu_light}

To model the lens galaxy light of \geu, we use Keck $J$-band observations of when the SN was active, as well as a reference observation after it had faded away. The Keck data provides the most detailed view of the lens galaxy’s light, with the NIR $J$-band selected for its ability to trace the older, redder stellar population that contributes the most to the stellar mass. While Keck observations in the $H$ and $K$ bands are also available, the SN images are not clearly visible in those bands, preventing the extraction of an accurate PSF to conduct the lens modelling.

\medskip
We model the lens system as described in section~\ref{subsect:fitting_meth}. Additionally, we include a uniform residual background surface brightness layer with the lens light model to prevent background over- or under-subtraction. We simultaneously fit the reference observation and the observation with the active SN. This fitting choice enforces that the lens galaxy mass distribution, the lens galaxy light profile, and the host galaxy light profile are each consistent across the two epochs. Because the host galaxy is faint in both observations, fitting the data jointly better constrains the model parameters. We carry out our analysis with a fixed lens galaxy mass slope of $\eta=1.8$, after ensuring that this value does not impact the final lens light predictions. The resulting model of the reference image is shown in Fig.~\ref{fig:16geu_lens_light}. We find that the Einstein radius is consistent with previous estimates of this parameter for \geu. To compute the flux densities at the SN image positions, we compare the lens light model flux at the position of each image to the total flux from the lens light model within an aperture of $1''$ radius from the lens galaxy centre. 

\subsection{Lens galaxy light model for SN Zwicky}\label{subsec:Zwicky_light}

To construct the light model for SN Zwicky's lens galaxy, we use \textit{JWST} observations made by NIRCam in the $F115W$ band. The observations were taken one year after the appearance of SN Zwicky, and hence the SN images are not visible anymore. In order to estimate the positions of the SN images on the lens galaxy, we use the \textit{HST} $F814W$ band observations from when SN Zwicky was active. We combine the observations by projecting the \textit{HST} data of SN Zwicky on top of the \textit{JWST} observations using the publicly available software \texttt{reproject}\footnote{\href{https://reproject.readthedocs.io/en/stable/}{https://reproject.readthedocs.io/en/stable/}} \citep{Reproject2020}. The projection is done through the World Coordinate Systems of both observations. We correct remaining shifts between the observations by fitting five stars in the field with a Moffat profile and correcting for their offsets. To ensure the clearest representation of SN Zwicky's image positions, we use the difference image between \textit{HST's} observations with the active SN and the reference observations taken a year later. Fig.~\ref{fig:jwst_zwicky} shows the \textit{JWST} data with the projected SN Zwicky image positions. 
Finally, we fit a Moffat profile to the projected SN images to accurately determine their positions within the \textit{JWST} observation.

\medskip
To estimate the lens galaxy flux at the image positions, we focus on the light profile in the centre of the lens galaxy, where the SN images are located. Therefore, we model the lens galaxy light within a square window of side length ($0.6'', 0.6''$) around the lens galaxy with a double Sérsic profile (Eq.~\ref{eq:sersic}), convolved with a Gaussian PSF with full width at half maximum $0.04''$, obtained from fitting a star in the field. We experimented with using a Moffat PSF and a cut-out of a star in the field, but the Gaussian PSF gave the best fit and smallest residuals for this smaller patch. After modelling, the image flux densities are computed by evaluating the lens light model at the projected SN image positions. The total lens galaxy flux corresponding to the stellar mass estimate is computed by summing the lens galaxy flux within a $1''$ radius from the lens centre. 

\medskip
For both \geu and SN Zwicky, we determine the stellar mass-to-light ratios by dividing their stellar mass estimates by the lens galaxy flux within the same aperture. We multiply the image position's flux densities with the stellar mass-to-light ratio to obtain the stellar surface mass density at the image positions. By dividing the stellar surface mass densities by the total surface mass densities ($\kappa$ obtained from the lens modelling), we obtain the stellar mass fractions at the image positions. In practice, the inferred stellar masses, fluxes, and flux densities are used as inputs in our joint Bayesian model to constrain the mass slopes and fraction of compact objects, as described in section~\ref{Sec:Model}. More details about the stellar mass fraction calculation are given in section~\ref{subseq:model_derived}.


\begin{figure*}[t]
\centering
{\includegraphics[width=1.04\textwidth,clip=true]{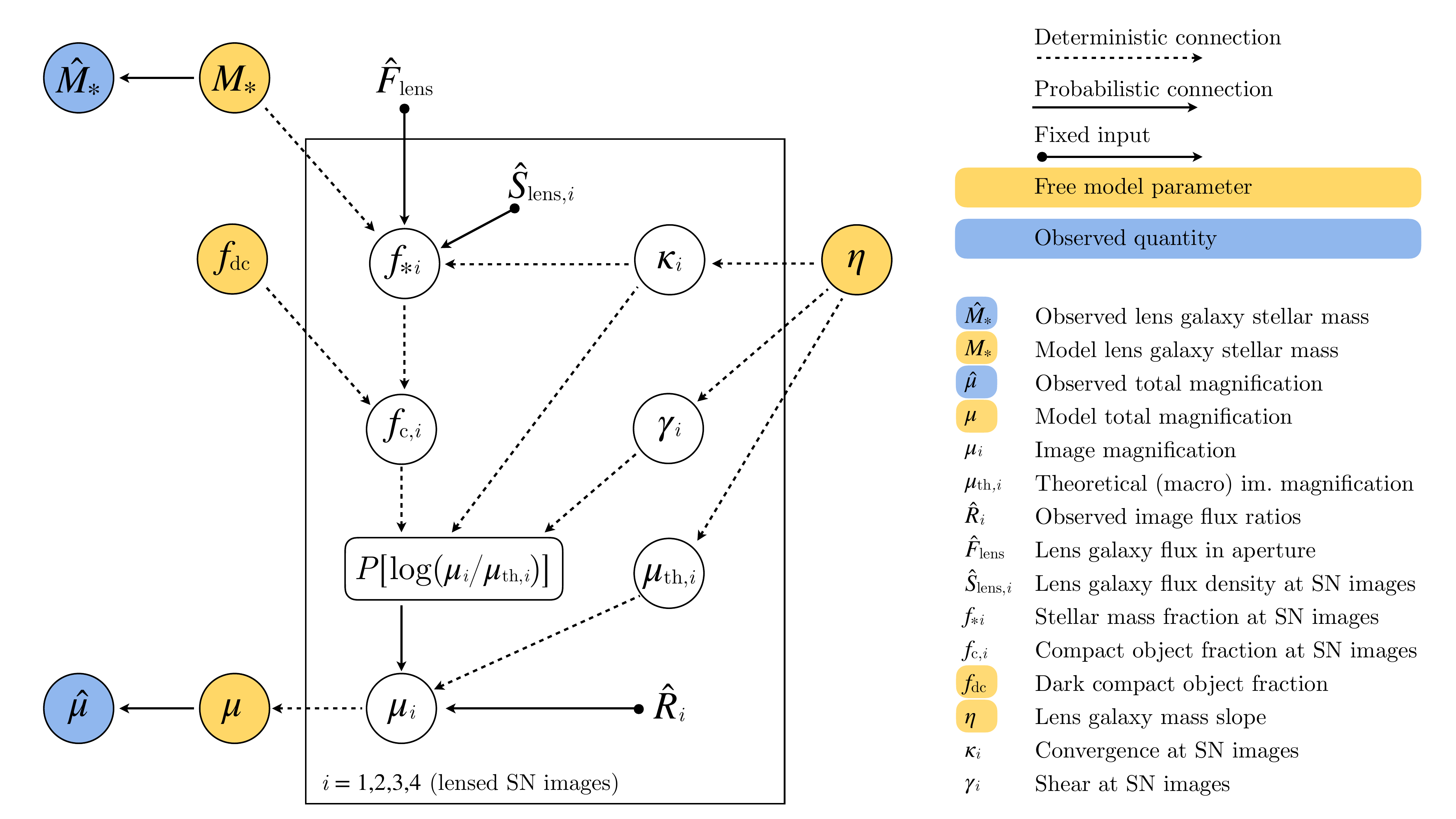}}
\caption{\small Probabilistic graphical model for the Bayesian analysis, depicting the connections between the free parameters (yellow shaded: $M_*$, $\mu$, $\eta$, \pbh), the observables with uncertainties (blue shaded: $\hat{M_*}$ and $\hat{\mu}$), and the observables treated as fixed inputs ($\hat{F}_{\textrm{lens}}, \hat{S}_{\textrm{lens},i}$ and $\hat{R}_{i}$). The arrows indicate relations of conditional probability, with solid arrows representing probabilistic connections and dashed ones deterministic connections. Parameters within the box, labelled $i = 1,2,3,4$, are repeated for every lensed SN image, whereas parameters outside the box describe the total lensed SN system.}
\label{fig:PGM}
\end{figure*}

\section{Bayesian model}\label{Sec:Model}

In this section, we introduce our Bayesian model to infer the lens galaxy mass slope, $\eta$, and dark compact object fraction, \pbh, based on quantities derived from the photometric observations, which have been described in the previous sections. The following subsections describe the observables, free parameters, derived parameters, the posterior, likelihood, priors, and inference method.
A probabilistic graphical model summarising the parameters of interest and their connections is displayed in Fig.~\ref{fig:PGM}.

\begin{figure*}[t]
\centering
{\includegraphics[width=1.04\textwidth,clip=true]{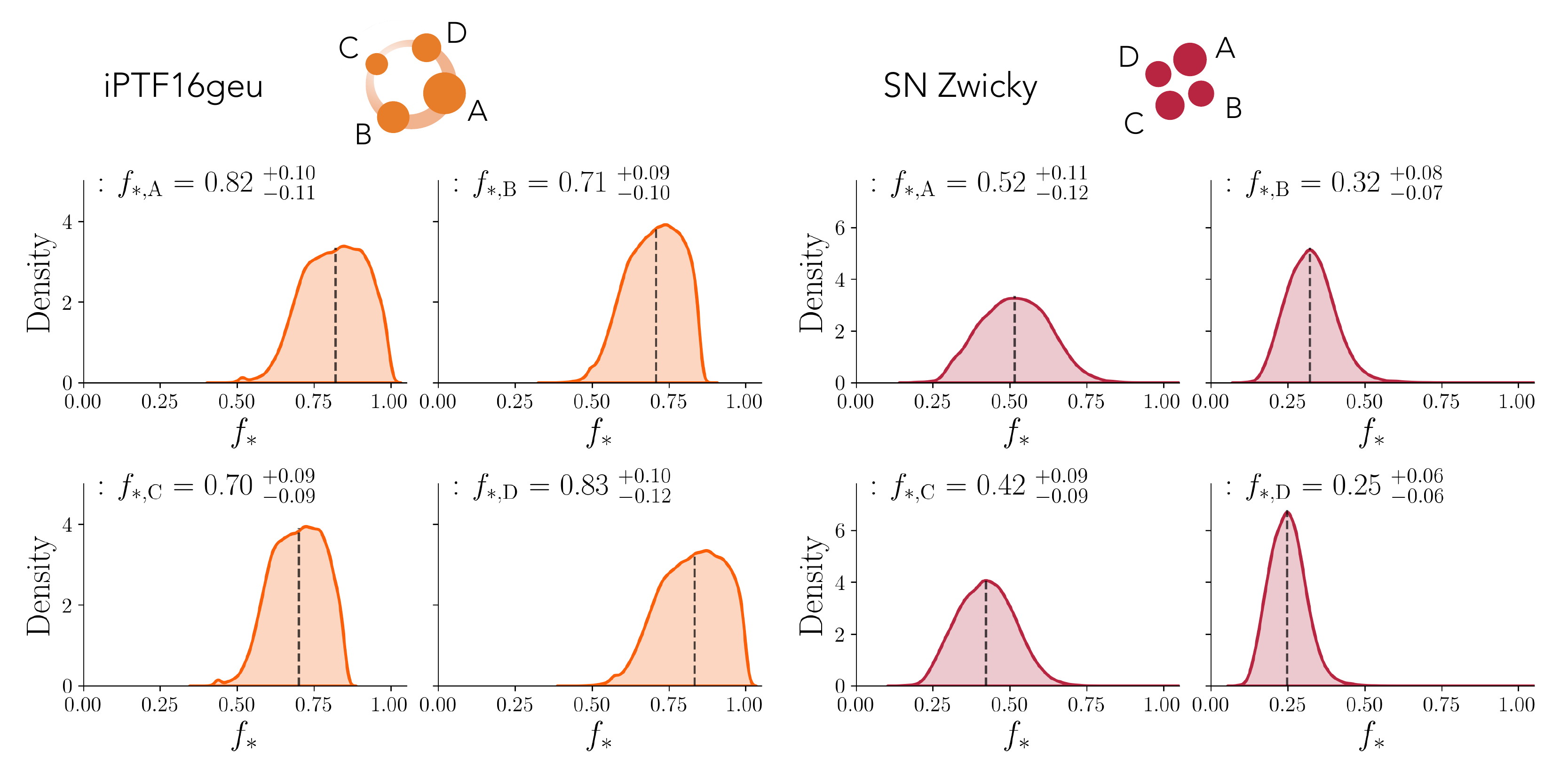}}
\caption{\small Posteriors for the stellar mass fraction $f_{*,i}$ at the image positions, for \geu (left) and SN Zwicky (right). The dashed line represents the median of the distribution, which is quoted above along with the 16th and 84th percentiles as uncertainties.}
\label{fig:f_star}
\end{figure*}

\begin{figure*}
\centering
{\includegraphics[width=0.95\textwidth,clip=true]{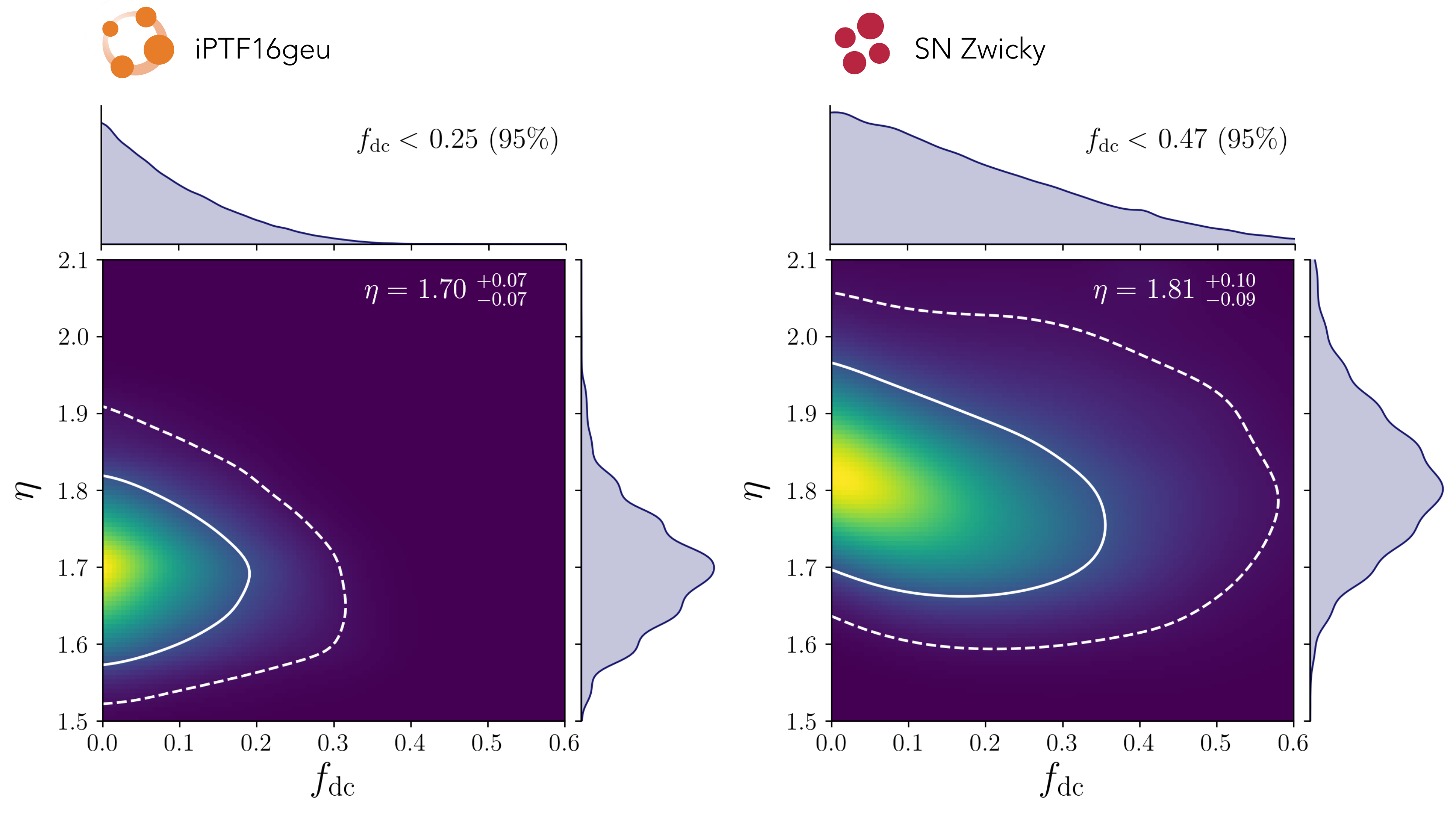}}
\caption{\small The joint probability of the galaxy mass slope, $\eta$, and the dark compact object fraction, \pbh, derived from the observations of \geu (left) and SN Zwicky (right). The solid and dashed contours correspond to the 68\% and 95\% confidence intervals, respectively.}
\label{fig:eta_fd}
\end{figure*}

\subsection{Observed quantities}\label{subseq:model_observed}

The photometric observations of SN Zwicky and \geu have been reduced to the following observed quantities. In all cases, a hat corresponds to an observed quantity, and a subscript $i$ refers to the quantity evaluated for SN image $i$. \\

\noindent \textit{Observed quantities with uncertainties}

\begin{itemize}[leftmargin=*,itemsep=1em]
    \item \textbf{Observed lens galaxy stellar mass $\boldsymbol{\hat{M_*}}$}: as described in section~\ref{subseq:stellar_mass}, we used photometric observations and SED modelling to infer the stellar masses of the lens galaxies within an aperture of $1''$ for \geu and SN Zwicky. The resulting estimates are $\log M_* / M_{\odot} = 10.78^{+ 0.12}_{- 0.18}$ for \geu and $\log M_* / M_{\odot} = 10.23^{+ 0.10}_{- 0.18}$ for SN Zwicky.
    \item \textbf{Observed total magnification $\boldsymbol{\hat{\mu}}$}: the observed magnification is determined by comparing the observed apparent magnitudes of the lensed SNe with the expected magnitude of a standard SNIa at that redshift. We take the results for the total magnification (sum of all images) from \citet{Dhawan2020_16geu} for \geu, $\hat{\mu} = 67.8 \pm 2.9$, and from \citet{Goobar2023_SNZwicky} for SN Zwicky, $\hat{\mu} = 23.7 \pm 3.2$.
\end{itemize}

\noindent \textit{Observed quantities as fixed inputs} \\

\noindent The following observed quantities are treated as fixed inputs in the Bayesian model, as their uncertainties are negligible relative to other sources of uncertainty. The uncertainties in the lens galaxy flux and flux density are negligible in comparison to the uncertainty of the stellar mass estimate. Similarly, the image flux ratio uncertainties are negligible compared to the uncertainty on the total magnification.

\begin{itemize}[leftmargin=*,itemsep=1em]
    \item \textbf{Lens galaxy flux $\boldsymbol{\hat{F}_{\textrm{lens}}}$}: the total lens light flux within an aperture of $1''$, as described in Sections~\ref{subsec:16geu_light} and \ref{subsec:Zwicky_light}.
    \item \textbf{Lens galaxy flux density at image positions $\boldsymbol{\hat{S}_{\textrm{lens},i}}$}: the flux density at the SN image positions, obtained from the lens light model. Described in Sections~\ref{subsec:16geu_light} and \ref{subsec:Zwicky_light}
    \item \textbf{SN image flux ratios $\boldsymbol{\hat{R}_{i}}$}: the observed flux ratios of the SN images, normalised to one. We take the results from \citet{Dhawan2020_16geu} for \geu, $\hat{R_{i}} = [0.52, 0.23, 0.11, 0.13]$, and from \citet{Goobar2023_SNZwicky} for SN Zwicky, $\hat{R_{i}} = [0.37, 0.15, 0.30, 0.18]$.
\end{itemize}

\subsection{Free parameters}\label{subseq:model_free}

The following parameters are sampled by the MCMC to obtain their posterior distributions. We expand on their priors in section~\ref{subseq:priors}.

\begin{itemize}[leftmargin=*,itemsep=1em]
    \item \textbf{Lens galaxy mass slope $\boldsymbol{\eta}$}: logarithmic slope of the lens galaxy's mass profile. 
    \item \textbf{Fraction of dark compact objects $\boldsymbol{f_{\rm dc}}$}: the fraction of the lens galaxy's total mass that is in the form of dark compact objects. In our analysis, we assume a constant \pbh for each of the SN image positions, a reasonable approximation because the images are formed at similar radii from the lens galaxy centre.
    \item \textbf{Lens galaxy stellar mass $\boldsymbol{M_*}$}: the true stellar mass of the lens galaxy, which the model aims to infer. It serves as the underlying quantity that generates the observed stellar mass, $\hat{M}_*$, through the likelihood function, accounting for observational uncertainties.
    \item \textbf{Total magnification $\boldsymbol{\mu}$}:  the true total magnification of the lensed SN system. It is the intrinsic quantity that generates the observed total magnification, $\hat{\mu}$, through the likelihood function, incorporating measurement uncertainties.
\end{itemize}

\subsection{Derived parameters}\label{subseq:model_derived}

In this section, we describe the derived parameters, which are quantities calculated from combinations of the observables and free parameters.

\begin{itemize}[leftmargin=*,itemsep=1em]
    \item \textbf{Stellar mass fractions $\boldsymbol{f_{*,i}}$}: the fraction of the lens galaxy's total mass that is in the form of stars, at the position of SN image $i$. Below, we describe how we calculate $f_{*,i}$ from $M_*, \eta, \hat{F}_{\textrm{lens}}$ and $\hat{S}_{\textrm{lens},i}$. \\
    From the lens galaxy's stellar mass, $M_*$, and flux measurement within the same aperture, $\hat{F}_{\textrm{lens}}$, the stellar mass-to-light ratio, $ML_*$, can be determined: 
    \begin{equation}
        ML_* = M_* / \hat{F}_{\textrm{lens}}.
    \end{equation}
    Combined with the lens galaxy flux density at the SN image positions, $\hat{S}_{\textrm{lens},i}$, the stellar mass surface density at the images can be calculated:
    \begin{equation}
        \Sigma_{*,i} = ML_* \cdot \hat{S}_{\textrm{lens},i},
    \end{equation}
    which can be converted into a stellar critical surface density, or stellar convergence, $\kappa_*$:
    \begin{equation}
        \kappa_* = \Sigma_* \frac{4 \pi G D_{\rm ds} D_{\rm d}}{c^2 D_{\rm s}},
    \end{equation}
    with $G$ the gravitational constant, $c$ the speed of light, and $D_{\rm l}$, $D_{\rm s}$ and $D_{\rm ls}$ the angular diameter distances between the observer and the lens, the observer and the source, and the lens and the source, respectively. Distances are calculated using a standard flat $\Lambda$CDM model with $H_0 = 67.4 \ \textrm{km} \, \textrm{s}^{-1} \textrm{Mpc}^{-1}$ and $\Omega_{\textrm{m}} = 0.315$ \citep{planck2018cosmo}.
    Finally, the stellar mass fraction is given by 
    \begin{equation}
        f_{*,i} = \kappa_{*,i} / \kappa_i
    \end{equation}
    where $\kappa_i$ is the total convergence for image $i$, calculated for different slope values as detailed in section~\ref{Sec:Macromodel} and in Table~\ref{Table:kappa_gamma_mu}. 
    \item \textbf{Fractions of compact objects $\boldsymbol{f_{\textrm{c},i}}$}: the lens galaxy's total mass fraction of compact objects at the SN image positions: $f_{\textrm{c},i} = f_{*,i} + f_{\rm dc}$.
    \item \textbf{Convergence $\boldsymbol{\kappa_i}$ and shear $\boldsymbol{\gamma_i}$}: the surface mass density (convergence) and shear at the SN image positions. As described in section~\ref{Sec:Macromodel}, we constrain the best-fit $\kappa_i$ and $\gamma_i$ values for a range of different mass slopes ($\eta$ ranging from 1.4 to 2.1 in steps of 0.05), which are displayed in Table~\ref{Table:kappa_gamma_mu}.
    \item \textbf{Predicted macro magnification per image $\boldsymbol{\mu_{\textrm{th},i}}$}: the magnification predicted by the best-fit macro lens model, at the SN image positions for a range of different mass slopes ($\eta$ ranging from 1.4 to 2.1 in steps of 0.05). Described in more detail in section~\ref{Sec:Macromodel}.
    \item \textbf{Magnification per image $\boldsymbol{\mu_{i}}$}: the combined macro + microlensing magnification for each SN image. $\mu_i$ is computed from the total magnification $\mu$ and the SN image flux ratios $\hat{R}_i$ as follows: $\mu_i = \mu \hat{R}_i$.
\end{itemize}

\subsection{The posterior}\label{subseq:model_posterior}

The posterior distribution provides the probabilistic constraints on the model parameters $M_*, \mu, \eta$ and \pbh, given the observed quantities $\hat{M_*}, \hat{\mu}, \hat{F}_{\textrm{lens}}, \hat{S}_{\textrm{lens},i}$ and $\hat{R}_{i}$. This distribution is defined by Bayes' theorem as the product of the likelihood and the prior terms, and can be expressed as follows:
\begin{align}
&P(M_*, \mu, \eta, f_{\rm dc},  \ | \ \hat{M_*}, \hat{\mu}, \hat{F}_{\textrm{lens}}, \hat{S}_{\textrm{lens},i}, \hat{R}_{i} ) \propto \notag \\
&P(\hat{\mu} \, | \, \mu) \ P(\hat{M_*} \, | \, M_*) \ P(\mu \, | \, M_*, \eta, f_{\rm dc}, \hat{F}_{\textrm{lens}}, \hat{S}_{\textrm{lens},i}, \hat{R}_{i} ) 
 \notag \\
&P(M_*, \eta, f_{\rm dc} \, | \, \hat{F}_{\textrm{lens}}, \hat{S}_{\textrm{lens},i})
\label{eq:posterior}
\end{align}
The likelihood and prior terms on the right-hand-side are described in more details below.

\subsection{Likelihoods on data}\label{subseq:llh_data}

The likelihood terms in the posterior distribution quantify how well the proposed model parameters explain the observed data. 

\begin{itemize}[leftmargin=*,itemsep=1em]
    \item $\boldsymbol{P(\hat{M_*} \, | \, M_*)}$.
    The observed stellar mass, $\hat{M_*}$, represents the stellar mass of the lens galaxy inferred within an aperture of $1''$. As detailed in section~\ref{subseq:stellar_mass}, the summaries of our stellar mass estimates are $\log M_* / M_{\odot} = 10.78^{+ 0.12}_{- 0.18}$ for \geu and $\log M_* / M_{\odot} = 10.23^{+ 0.10}_{- 0.18}$ for SN Zwicky. To incorporate the observed stellar mass into our inference model, we construct a kernel density estimate (KDE) of the stellar mass measurements. The likelihood $P(\hat{M_*} \, | \, M_*)$ is then evaluated by computing the probability density of the proposed stellar mass $M_*$ using the KDE.
    \item $\boldsymbol{P(\hat{\mu} \, | \, \mu)}$. 
    The observed total magnification, $\hat{\mu}$, is determined by comparing the apparent magnitudes of the lensed SNe with the expected magnitudes of standard SNe~Ia at the same redshifts. For \geu, we adopt the value $\hat{\mu} = 67.8 \pm 2.9$ from \citet{Dhawan2020_16geu}. For SN Zwicky, we use $\hat{\mu} = 23.7 \pm 3.2$ from \citet{Goobar2023_SNZwicky}. The likelihood $P(\hat{\mu} \, | \, \mu)$ is modelled as a Gaussian distribution with mean and standard deviation equal to the reported values in the respective studies.
\end{itemize}

\subsection{Priors}\label{subseq:priors}

The prior on the total magnification, $\mu$, is derived from the simulated microlensing magnification distributions. These distributions are evaluated for each SN image at its magnification, $\mu_i = \mu \hat{R}_i$, and combined multiplicatively across all images:
\begin{align}
&P(\mu \, | \, M_*, \eta, f_{\rm dc}, \hat{F}_{\textrm{lens}}, \hat{S}_{\textrm{lens},i}, \hat{R}_{i} ) = \notag \\
&\prod_{i=1}^{N} P(\mu \hat{R}_{i} \, | \, M_*, \eta, f_{\rm dc}, \hat{F}_{\textrm{lens}}, \hat{S}_{\textrm{lens},i}),
\end{align}
where $P(\mu \hat{R}_{i} \, | \, M_*, \eta, f_{\rm dc}, \hat{F}_{\textrm{lens}}, \hat{S}_{\textrm{lens},i})$ is the microlensing magnification distribution for image $i$ evaluated at $\mu_i = \mu \hat{R}_i$. In practice, the total magnification $\mu$ is sampled from a uniform proposal distribution, $\mathcal{U}(\mu \, ; \, 15, 50)$, and multiplied by the SN image flux ratios $\hat{R}_i$ to obtain $\mu_i$. For the remaining free parameters, we use the following priors:

\begin{align}
    &P(M_*, \eta, f_{\rm dc} \, | \, \hat{F}_{\textrm{lens}}, \hat{S}_{\textrm{lens},i} ) =  \notag \\
    &P(f_{\rm dc} \, | \, M_*, \eta, \hat{F}_{\textrm{lens}}, \hat{S}_{\textrm{lens},i} ) \ P(M_*) \ P(\eta),
\end{align}
with
\begin{align}
    &P(f_{\rm dc} \, | \, M_*, \eta, \hat{F}_{\textrm{lens}}, \hat{S}_{\textrm{lens},i}) = \notag \\
    &\prod_{i=1}^{N} \mathcal{U}(f_{\rm dc} + f_{*,i} \, ; \, 0,1).  \label{eq:prior_fdc}
\end{align}
The prior in eq.~\ref{eq:prior_fdc} ensures that for each SN image, the fraction of compact objects $f_{c,i} = f_{\rm dc} + f_{*,i}$ is between $0$ and $1$.
For $P(M_*)$ and $P(\eta)$, we initialise the walkers in a range of $\mathcal{U}(M_* \, ; \, 10^9, 10^{10.8})$ and $\mathcal{U}(\eta \, ; \, 1.4, 2.1)$, and we have uninformative priors over the entire parameter space. The values of $\eta$ are sampled from a continuous distribution but are rounded to the nearest $0.05$ to match the precision used in the microlensing simulations. Similarly, the compact object fraction per image, $f_{\textrm{c},i} = f_{\textrm{dc}} + f_{*,i}$ is rounded to the nearest $0.1$ to align with the simulation's precision.

\subsection{Inference} \label{subsect:model_inference}

We sample the posteriors of our parameters of interest using \texttt{emcee}. This approach allows us to incorporate observational uncertainties, apply prior constraints to the model parameters, and sample the parameters of interest jointly to assess their covariances. \texttt{emcee} is able to handle a combination of continuous and discrete parameters effectively, since it does not rely on gradient information to propose new states. This is particularly useful in our situation, where we have a discrete number of microlensing magnification distributions, corresponding to discrete values of $\eta$ and the compact object fraction per image.

\medskip
We run MCMC chains with 100 walkers and 3000 iterations to sample $\eta$, \pbh, ${M_*}$, and $\mu$. We verify the convergence of the chains by inspecting the trace plots. After discarding the first 500 samples for the burn-in phase, we combine the chains to derive the marginalized posterior distributions for each parameter, from which we report the median values and the corresponding 16th and 84th percentiles as uncertainties. Our results are summarised in the next Section.



\section{Results}\label{Sec:Results}

In this section, we present the main results obtained from the parameter inference model outlined in section~\ref{Sec:Model}.

\subsection{Stellar mass fractions}

The stellar mass fraction of the lens galaxy at the SN image positions, $f_{*,i}$, is a derived parameter in our model, calculated from the stellar mass, $M_*$, the lens galaxy flux, $\hat{F}_{\textrm{lens}}$, the lens galaxy flux density at the SN image positions, $\hat{S}_{\textrm{lens},i}$, and the convergence at the image positions, $\kappa_i$.
Fig.~\ref{fig:f_star} shows the inferred stellar mass fractions per lensed SN image, for \geu and SN Zwicky. 
Both \geu and SN Zwicky have high stellar mass fractions, which can be explained by the fact that the SN images are formed close to the lens centre, where the stellar density is high compared to the dark matter density.
The median $f_{*,i}$ values for \geu ($ \sim 0.7 - 0.8$) are higher than the ones inferred for \geu in \citet{Mortsell2020} ($\sim 0.2$), which is due to differences in the stellar mass determination between the papers.
In \citet{Mortsell2020}, the stellar mass was inferred from scaling relations in \citet{Kauffmann2003}, whereas in this work, we model the full lens galaxy SED with \texttt{prospector}. 

\subsection{Dark compact objects and galaxy mass slopes}

Fig.~\ref{fig:eta_fd} shows the joint $\eta$ and \pbh posterior distributions for \geu and SN Zwicky. For both lensed SNe, there is no preference for additional dark compact objects, and stars alone are sufficient to explain the observed microlensing effects. We put upper limits at the $95\%$ confidence level on the fraction of dark compact objects of \pbh $< 0.25$ for \geu and \pbh $< 0.47$ for SN Zwicky. The upper limits from \geu are tighter than those from SN Zwicky, due to the fact that \geu has high stellar mass fractions ($\sim 0.7-0.8$), leaving less room for the presence of dark compact objects.
These limits apply to compact objects with masses greater than $0.02 M_{\odot}$. Microlenses with lower masses have Einstein radii smaller than the size of the SN photosphere, causing their magnification contributions to average out.

\medskip
By utilising the standard candle nature of SNIa, while incorporating microlensing into our modelling, we were able to constrain the galaxy mass slopes for these systems, which are difficult to infer from lens modelling alone. For \geu, we find a slope of $\eta = 1.70 \pm 0.07$, which is slightly lower but consistent with the value derived in \citet{Mortsell2020} ($\eta \sim 1.8$). SN Zwicky's mass slope is $\eta = 1.81 \pm 0.10$. 
The step-like features in the posterior distribution of $\eta$ reflect the discreteness introduced by the microlensing simulations. 

\medskip
For the best-fit slope values, the microlensing magnification contributions for images A-D (in magnitudes) are the following: 
\begin{align*}
    &\textrm{\geu: } -0.72, \ -0.05, \  0.78, \ 0.97 \ \textrm{mag}. \\
&\textrm{SN Zwicky:  }-0.68, \ 0.79, \  0.10, \ 0.74 \ \textrm{mag}.
\end{align*}
For both \geu and SN Zwicky, the brightest observed images are $\sim 0.7$ magnitudes brighter due to microlensing, which is in line with microlensing expectations for lensed SNe \citep{dobler2006microlensing}, especially considering those images are both saddle points, which are more strongly affected by microlensing.

\begin{figure*}[tb]
\centering
{\includegraphics[width=0.95\textwidth,clip=true]{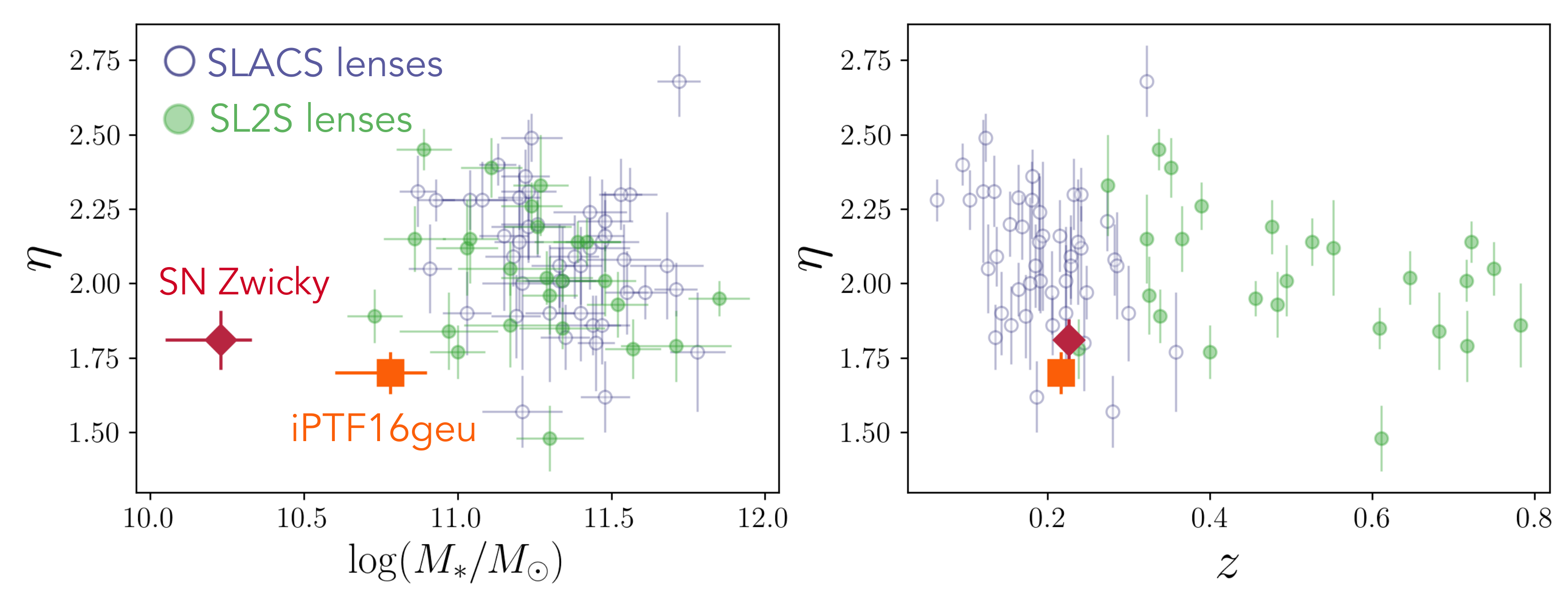}}
\caption{\small Lens galaxy mass slope, $\eta$, versus stellar mass, $M_*$, (left panel) and redshift (right panel), for \geu, SN Zwicky, and several lensed galaxies from the SLACS \citep{Auger2010} and SL2S \citep{Sonnenfeld2013} surveys. The lens galaxies of \geu and SN Zwicky have relatively low stellar masses (see section~\ref{subseq:stellar_mass}) and flat mass profiles (see section~\ref{Sec:Results}) compared to the population of observed lensed galaxies.}
\label{fig:slope_comparison}
\end{figure*}

\medskip
In Fig.~\ref{fig:slope_comparison}, we plot our inferred mass slopes for \geu and SN Zwicky together with those from lens galaxies in the Sloan Lens ACS Survey \citep[SLACS; ][]{Auger2010} and in the Strong Lensing Legacy Survey \citep[SL2S; ][]{Sonnenfeld2013}, as a function of redshift and stellar mass. \geu and SN Zwicky's mass slopes are relatively low compared to the lens galaxy population, while still being within the range observed for other galaxies. In terms of stellar mass, SN Zwicky is significantly lighter than the other lens galaxies.
As was pointed out in \citet{Goobar2023_SNZwicky}, \geu and SN Zwicky are the first discoveries of a new class of extremely low-mass and compact lens systems, identified primarily due to their high magnification. These discoveries were only possible because the lensed sources were SNIa with standardisable brightnesses. In this work, we have demonstrated that these lensing systems are highly suitable for microlensing studies and for constraining the matter in compact objects.

\subsection{Shared \pbh for \geu and SN Zwicky}

If we make the assumption that the fraction of dark compact objects is the same across different galaxies, we can combine the constraints from \geu and SN Zwicky. This could be the case if the dark compact objects were all created in the very early Universe, and are spread evenly throughout the Universe and across galaxies. We performed another inference run with a joint Bayesian model, where the free parameters are \pbh, $\eta$ Zwicky, $\log M_*$ Zwicky, $\mu$ Zwicky, $\eta$ \geu, $\log M_*$ \geu, and $\mu$ \geu. The data constraints we use are the same as in the individual fits. Fig.~\ref{fig:mcmc_joint_params} shows the resulting corner plots for all free parameters, where the results for $\eta$ are consistent with the individual fits. We obtain a tighter upper bound on the fraction of dark compact objects: \pbh $< 0.19$ at the $95\%$ confidence level.

\section{Discussion and conclusion}\label{Sec:Discussion}

In this work, we analysed the micro- and macro lens models of the two galaxy-scale lensed SNIa: \geu and SN Zwicky. Given that these systems both have standardizable candle sources that show strong hints of microlensing in their flux ratios -- deviating from macromodel predictions -- they are ideal to explore the potential of strongly lensed SNe for microlensing studies. 
We simulated realistic microlensing magnification distributions for a range of different galaxy mass slope values and compact object fractions.
By doing so, we evaluated which combination of slope and compact objects is the most likely, and whether there are signs of a population of dark compact objects in the lens galaxy.

\medskip
We developed a Bayesian model to simultaneously model our parameters of interest, taking into account constraints from observed quantities such as the stellar mass, magnification, lens galaxy light model, and SN image flux ratios. Our results show that there is no preference for additional dark compact objects in the lens galaxy, and we place $95\%$ upper limits of \pbh $< 0.25$ for \geu and \pbh $< 0.47$ for SN Zwicky. Our results are in agreement with \pbh measurements from strongly lensed quasars, which also find no evidence for any dark compact objects \citep[e.g. ][]{Pooley2009, Pooley2012, Schechter2004, Schechter2014, Mediavilla2009, Mediavilla2017, Esteban2020, Medavilla2024}. \revision{With a larger sample of lensed SNe Ia, we will eventually be able to test whether the initial mass function (assumed to be Chabrier in this work) is correct, as we become more sensitive to \pbh, or equivalently the stellar mass-to-light ratio \citep{2025MNRAS.539..393W}.}

\medskip
We also constrain the mass slope of the lens galaxies, which is made possible because they are standardizable candles for which we know their magnification. We measure $\eta = 1.70 \pm 0.07$ for \geu and $\eta = 1.81 \pm 0.10$ for SN Zwicky. Compared to other known lens galaxies from SLACS and SL2S, these slope measurements are on the lower side, but not outside of what has been observed before. For SN Zwicky, which does not have a lensed host galaxy, this measurement using the SNIa intrinsic brightness is the only way to place tight constraints on the mass slope. For \geu, we can cross-check this measurement with the one obtained by fitting a PEMD lens model with the slope as a free parameter, as is done in \citet{Mortsell2020}, which is in agreement.

\medskip
\revision{We also investigated an alternative explanation for the observed flux ratio anomalies in \geu and SN Zwicky: angular complexity in the lens model. We tested lens models with fourth-order multipole perturbations and found that they cannot account for the observed flux ratios, leaving microlensing as the most likely cause.}

\medskip
Future work on microlensing of lensed SNe can take into account time varying microlensing effects, as well as a spectrum of different microlens masses. Additionally, in our work we did not consider the effects of \textit{mililensing}, due to dark matter subhalos. The recent analysis of SN Zwicky by \citet{Larison2024}, which included \textit{HST} reference observations, investigated mililensing, and found its contribution to the total magnification to be negligible. Additionally, they investigated the flux ratio discrepancies of SN Zwicky in the $F475W$, $F625W$, $F814W$, and $F160W$ filters, confirming the evidence for microlensing and finding hints for a chromatic nature of the effects.

\medskip
This work demonstrates the power of strongly lensed SNe to obtain insights into compact objects and potential primordial black hole populations. With the advance of the next generation of telescopes such as the Vera C. Rubin Observatory \citep{ivezic2019lsst} and the Nancy Grace Roman Space Telescope \citep[e.g.,][]{pierel2020projected}, we will discover dozens more galaxy-scale lensed SNe, which will allow us to put tighter constraints on the population of dark compact objects. Moreover, these compact lensed SN systems offer one of the few avenues to probe the inner mass slope of lower-mass lens galaxies.

\section*{Acknowledgments}

We are grateful to Simon Birrer and Anowar Shajib for helpful suggestions regarding the lens modelling of iPTF16geu. 
We would like to thank Jorge Martin Camalich and Evencio Mediavilla for interesting and useful discussions, and Evencio Mediavilla for confirming our results with an independent re-analysis of the SN Zwicky data.
\revision{We thank the referee for suggesting the inclusion of the multipole analysis in the paper.}

\medskip
This work has been enabled by support from the research project grant ‘Understanding the Dynamic Universe’ funded by the Knut and Alice Wallenberg Foundation under Dnr KAW 2018.0067. EM acknowledges support from the
Swedish Research Council under Dnr VR 2020-03384. ST has been supported by funding from the European Research Council (ERC) under the European Union's Horizon 2020 research and innovation programmes (grant agreement no. 101018897 CosmicExplorer). SD acknowledges funding by UK Research and Innovation (UKRI) under the UK government’s Horizon Europe funding Guarantee EP/Z000475/1, a Kavli Fellowship and a JRF at Lucy Cavendish College. This project has received funding from the European Union’s Horizon Europe research and innovation programme under the Marie Sklodovska-Curie grant agreement No 101105725. AG acknowledges support from the Swedish Research Council (Dnr 2020-03444) and the Swedish National Space Agency (Dnr 2023-00226). EEH is supported by a Gates Cambridge Scholarship (\#OPP1144). This work has received funding from the European Research Council (ERC) under the European Union’s Horizon 2020 research and innovation programme (LensEra: grant agreement No. 945536). For the purpose of open access, the authors have applied a Creative Commons Attribution (CC BY) license to any Author Accepted Manuscript version arising.\\

\medskip
\textit{Software used:} AstroPy \citep{Astropy2013, Astropy2018, Astropy2022},
Jupyter \citep{JupyterNotebook},
Matplotlib \citep{Matplotlib2007, Matplotlib2020},
NumPy \citep{Numpy202},
Pandas \citep{Pandas2010, Pandas2023},
Pickle \citep{Pickle2020},
SciPy \citep{Scipy2020},
Seaborn \citep{Seaborn2020},
Emcee \citep{ForemanMackey2013, Goodman2010},
ChainConsumer \citep{Hinton2016},
Lenstronomy \citep{birrer2018lenstronomy, Birrer2021LenstronomyII},
Prospector \citep{Johnson2021},
FSPS \citep{Conroy2009, Conroy2010, Conroy2010ApJ},
Python-FSPS \citep{Johnson2024},
SEDPy \citep{SEDPy2021},
Dynesty \citep{Speagle2020, Koposov2024},
Reproject \citep{Reproject2020},
SAOImageDS9 \citep{DS9, Joye2003}.

\section*{Data Availability}

Code related to this article is available at \href{ https://github.com/Nikki1510/Microlensing}{\texttt{https://github.com/Nikki1510/Microlensing}}

\bibliographystyle{mnras}
\bibliography{references.bib}

@ARTICLE{bellm2019ZTF,
       author = {{Bellm}, Eric C. and {Kulkarni}, Shrinivas R. and {Graham}, Matthew J. and {Dekany}, Richard and {Smith}, Roger M. and {Riddle}, Reed and {Masci}, Frank J. and {Helou}, George and {Prince}, Thomas A. and {Adams}, Scott M. and {Barbarino}, C. and {Barlow}, Tom and {Bauer}, James and {Beck}, Ron and {Belicki}, Justin and {Biswas}, Rahul and {Blagorodnova}, Nadejda and {Bodewits}, Dennis and {Bolin}, Bryce and {Brinnel}, Valery and {Brooke}, Tim and {Bue}, Brian and {Bulla}, Mattia and {Burruss}, Rick and {Cenko}, S. Bradley and {Chang}, Chan-Kao and {Connolly}, Andrew and {Coughlin}, Michael and {Cromer}, John and {Cunningham}, Virginia and {De}, Kishalay and {Delacroix}, Alex and {Desai}, Vandana and {Duev}, Dmitry A. and {Eadie}, Gwendolyn and {Farnham}, Tony L. and {Feeney}, Michael and {Feindt}, Ulrich and {Flynn}, David and {Franckowiak}, Anna and {Frederick}, S. and {Fremling}, C. and {Gal-Yam}, Avishay and {Gezari}, Suvi and {Giomi}, Matteo and {Goldstein}, Daniel A. and {Golkhou}, V. Zach and {Goobar}, Ariel and {Groom}, Steven and {Hacopians}, Eugean and {Hale}, David and {Henning}, John and {Ho}, Anna Y.~Q. and {Hover}, David and {Howell}, Justin and {Hung}, Tiara and {Huppenkothen}, Daniela and {Imel}, David and {Ip}, Wing-Huen and {Ivezi{\'c}}, {\v{Z}}eljko and {Jackson}, Edward and {Jones}, Lynne and {Juric}, Mario and {Kasliwal}, Mansi M. and {Kaspi}, S. and {Kaye}, Stephen and {Kelley}, Michael S.~P. and {Kowalski}, Marek and {Kramer}, Emily and {Kupfer}, Thomas and {Landry}, Walter and {Laher}, Russ R. and {Lee}, Chien-De and {Lin}, Hsing Wen and {Lin}, Zhong-Yi and {Lunnan}, Ragnhild and {Giomi}, Matteo and {Mahabal}, Ashish and {Mao}, Peter and {Miller}, Adam A. and {Monkewitz}, Serge and {Murphy}, Patrick and {Ngeow}, Chow-Choong and {Nordin}, Jakob and {Nugent}, Peter and {Ofek}, Eran and {Patterson}, Maria T. and {Penprase}, Bryan and {Porter}, Michael and {Rauch}, Ludwig and {Rebbapragada}, Umaa and {Reiley}, Dan and {Rigault}, Mickael and {Rodriguez}, Hector and {van Roestel}, Jan and {Rusholme}, Ben and {van Santen}, Jakob and {Schulze}, S. and {Shupe}, David L. and {Singer}, Leo P. and {Soumagnac}, Maayane T. and {Stein}, Robert and {Surace}, Jason and {Sollerman}, Jesper and {Szkody}, Paula and {Taddia}, F. and {Terek}, Scott and {Van Sistine}, Angela and {van Velzen}, Sjoert and {Vestrand}, W. Thomas and {Walters}, Richard and {Ward}, Charlotte and {Ye}, Quan-Zhi and {Yu}, Po-Chieh and {Yan}, Lin and {Zolkower}, Jeffry},
        title = "{The Zwicky Transient Facility: System Overview, Performance, and First Results}",
      journal = {\pasp},
         year = 2019,
        month = jan,
       volume = {131},
       number = {995},
        pages = {018002},
          doi = {10.1088/1538-3873/aaecbe},
archivePrefix = {arXiv},
       eprint = {1902.01932},
 primaryClass = {astro-ph.IM},
       adsurl = {https://ui.adsabs.harvard.edu/abs/2019PASP..131a8002B}
}

@ARTICLE{Rigault2019,
       author = {{Rigault}, M. and {Neill}, J.~D. and {Blagorodnova}, N. and {Dugas}, A. and {Feeney}, M. and {Walters}, R. and {Brinnel}, V. and {Copin}, Y. and {Fremling}, C. and {Nordin}, J. and {Sollerman}, J.},
        title = "{Fully automated integral field spectrograph pipeline for the SEDMachine: pysedm}",
      journal = {\aap},
         year = 2019,
        month = jul,
       volume = {627},
          eid = {A115},
        pages = {A115},
          doi = {10.1051/0004-6361/201935344},
archivePrefix = {arXiv},
       eprint = {1902.08526},
 primaryClass = {astro-ph.IM},
       adsurl = {https://ui.adsabs.harvard.edu/abs/2019A&A...627A.115R}
}

@ARTICLE{Blago2018,
       author = {{Blagorodnova}, Nadejda and {Neill}, James D. and {Walters}, Richard and {Kulkarni}, Shrinivas R. and {Fremling}, Christoffer and {Ben-Ami}, Sagi and {Dekany}, Richard G. and {Fucik}, Jason R. and {Konidaris}, Nick and {Nash}, Reston and {Ngeow}, Chow-Choong and {Ofek}, Eran O. and {O' Sullivan}, Donal and {Quimby}, Robert and {Ritter}, Andreas and {Vyhmeister}, Karl E.},
        title = "{The SED Machine: A Robotic Spectrograph for Fast Transient Classification}",
      journal = {\pasp},
         year = 2018,
        month = mar,
       volume = {130},
       number = {985},
        pages = {035003},
          doi = {10.1088/1538-3873/aaa53f},
archivePrefix = {arXiv},
       eprint = {1710.02917},
 primaryClass = {astro-ph.IM},
       adsurl = {https://ui.adsabs.harvard.edu/abs/2018PASP..130c5003B}
}

@ARTICLE{foxley2018standardization,
       author = {{Foxley-Marrable}, Max and {Collett}, Thomas E. and {Vernardos}, Georgios and {Goldstein}, Daniel A. and {Bacon}, David},
        title = "{The impact of microlensing on the standardization of strongly lensed Type Ia supernovae}",
      journal = {\mnras},
         year = 2018,
        month = aug,
       volume = {478},
       number = {4},
        pages = {5081-5090},
          doi = {10.1093/mnras/sty1346},
archivePrefix = {arXiv},
       eprint = {1802.07738},
 primaryClass = {astro-ph.CO},
       adsurl = {https://ui.adsabs.harvard.edu/abs/2018MNRAS.478.5081F}
}

@ARTICLE{treu2016cosmography,
       author = {{Treu}, Tommaso and {Marshall}, Philip J.},
        title = "{Time delay cosmography}",
      journal = {\aapr},
         year = 2016,
        month = jul,
       volume = {24},
       number = {1},
          eid = {11},
        pages = {11},
          doi = {10.1007/s00159-016-0096-8},
archivePrefix = {arXiv},
       eprint = {1605.05333},
 primaryClass = {astro-ph.CO},
       adsurl = {https://ui.adsabs.harvard.edu/abs/2016A&ARv..24...11T}
}

@ARTICLE{Dhawan2020_16geu,
       author = {{Dhawan}, S. and {Johansson}, J. and {Goobar}, A. and {Amanullah}, R. and {M{\"o}rtsell}, E. and {Cenko}, S.~B. and {Cooray}, A. and {Fox}, O. and {Goldstein}, D. and {Kalender}, R. and {Kasliwal}, M. and {Kulkarni}, S.~R. and {Lee}, W.~H. and {Nayyeri}, H. and {Nugent}, P. and {Ofek}, E. and {Quimby}, R.},
        title = "{Magnification, dust, and time-delay constraints from the first resolved strongly lensed Type Ia supernova iPTF16geu}",
      journal = {\mnras},
         year = 2020,
        month = jan,
       volume = {491},
       number = {2},
        pages = {2639-2654},
          doi = {10.1093/mnras/stz2965},
archivePrefix = {arXiv},
       eprint = {1907.06756},
 primaryClass = {astro-ph.GA},
       adsurl = {https://ui.adsabs.harvard.edu/abs/2020MNRAS.491.2639D}
}

@ARTICLE{Calzetti2000,
       author = {{Calzetti}, Daniela and {Armus}, Lee and {Bohlin}, Ralph C. and {Kinney}, Anne L. and {Koornneef}, Jan and {Storchi-Bergmann}, Thaisa},
        title = "{The Dust Content and Opacity of Actively Star-forming Galaxies}",
      journal = {\apj},
         year = 2000,
        month = apr,
       volume = {533},
       number = {2},
        pages = {682-695},
          doi = {10.1086/308692},
archivePrefix = {arXiv},
       eprint = {astro-ph/9911459},
 primaryClass = {astro-ph},
       adsurl = {https://ui.adsabs.harvard.edu/abs/2000ApJ...533..682C}
}

@ARTICLE{Chabrier2003,
       author = {{Chabrier}, Gilles},
        title = "{Galactic Stellar and Substellar Initial Mass Function}",
      journal = {\pasp},
         year = 2003,
        month = jul,
       volume = {115},
       number = {809},
        pages = {763-795},
          doi = {10.1086/376392},
archivePrefix = {arXiv},
       eprint = {astro-ph/0304382},
 primaryClass = {astro-ph},
       adsurl = {https://ui.adsabs.harvard.edu/abs/2003PASP..115..763C}
}

@article{Johnson2021,
   title={Stellar Population Inference with Prospector},
   volume={254},
   ISSN={1538-4365},
   url={http://dx.doi.org/10.3847/1538-4365/abef67},
   DOI={10.3847/1538-4365/abef67},
   number={2},
   journal={\apjs},
   publisher={American Astronomical Society},
   author={Johnson, Benjamin D. and Leja, Joel and Conroy, Charlie and Speagle, Joshua S.},
   year={2021},
   month=may, pages={22},
adsurl = {https://ui.adsabs.harvard.edu/abs/2021ApJS..254...22J}
}

@ARTICLE{Goldstein2018microlensing,
       author = {{Goldstein}, Daniel A. and {Nugent}, Peter E. and {Kasen}, Daniel N. and {Collett}, Thomas E.},
        title = "{Precise Time Delays from Strongly Gravitationally Lensed Type Ia Supernovae with Chromatically Microlensed Images}",
      journal = {\apj},
         year = 2018,
        month = mar,
       volume = {855},
       number = {1},
          eid = {22},
        pages = {22},
          doi = {10.3847/1538-4357/aaa975},
archivePrefix = {arXiv},
       eprint = {1708.00003},
 primaryClass = {astro-ph.CO},
       adsurl = {https://ui.adsabs.harvard.edu/abs/2018ApJ...855...22G}
}

@ARTICLE{Huber2019LSSTcadence,
       author = {{Huber}, S. and {Suyu}, S.~H. and {Noebauer}, U.~M. and {Bonvin}, V. and {Rothchild}, D. and {Chan}, J.~H.~H. and {Awan}, H. and {Courbin}, F. and {Kromer}, M. and {Marshall}, P. and {Oguri}, M. and {Ribeiro}, T. and {LSST Dark Energy Science Collaboration}},
        title = "{Strongly lensed SNe Ia in the era of LSST: observing cadence for lens discoveries and time-delay measurements}",
      journal = {\aap},
         year = 2019,
        month = nov,
       volume = {631},
          eid = {A161},
        pages = {A161},
          doi = {10.1051/0004-6361/201935370},
archivePrefix = {arXiv},
       eprint = {1903.00510},
 primaryClass = {astro-ph.IM},
       adsurl = {https://ui.adsabs.harvard.edu/abs/2019A&A...631A.161H}
}

@ARTICLE{rodney2021gravitationally,
       author = {{Rodney}, Steven A. and {Brammer}, Gabriel B. and {Pierel}, Justin D.~R. and {Richard}, Johan and {Toft}, Sune and {O'Connor}, Kyle F. and {Akhshik}, Mohammad and {Whitaker}, Katherine E.},
        title = "{A gravitationally lensed supernova with an observable two-decade time delay}",
      journal = {Nature Astronomy},
         year = 2021,
        month = nov,
       volume = {5},
        pages = {1118-1125},
          doi = {10.1038/s41550-021-01450-9},
archivePrefix = {arXiv},
       eprint = {2106.08935},
 primaryClass = {astro-ph.CO},
       adsurl = {https://ui.adsabs.harvard.edu/abs/2021NatAs...5.1118R}
}

@ARTICLE{goldstein2019rates,
       author = {{Goldstein}, Daniel A. and {Nugent}, Peter E. and {Goobar}, Ariel},
        title = "{Rates and Properties of Supernovae Strongly Gravitationally Lensed by Elliptical Galaxies in Time-domain Imaging Surveys}",
      journal = {\apjs},
         year = 2019,
        month = jul,
       volume = {243},
       number = {1},
          eid = {6},
        pages = {6},
          doi = {10.3847/1538-4365/ab1fe0},
archivePrefix = {arXiv},
       eprint = {1809.10147},
 primaryClass = {astro-ph.GA},
       adsurl = {https://ui.adsabs.harvard.edu/abs/2019ApJS..243....6G}
}

@ARTICLE{refsdal1964hubble,
       author = {{Refsdal}, S.},
        title = "{On the possibility of determining Hubble's parameter and the masses of galaxies from the gravitational lens effect}",
      journal = {\mnras},
         year = 1964,
        month = jan,
       volume = {128},
        pages = {307},
          doi = {10.1093/mnras/128.4.307},
       adsurl = {https://ui.adsabs.harvard.edu/abs/1964MNRAS.128..307R}
}

@ARTICLE{dobler2006microlensing,
       author = {{Dobler}, Gregory and {Keeton}, Charles R.},
        title = "{Microlensing of Lensed Supernovae}",
      journal = {\apj},
         year = 2006,
        month = dec,
       volume = {653},
       number = {2},
        pages = {1391-1399},
          doi = {10.1086/508769},
archivePrefix = {arXiv},
       eprint = {astro-ph/0608391},
 primaryClass = {astro-ph},
       adsurl = {https://ui.adsabs.harvard.edu/abs/2006ApJ...653.1391D}
}

@ARTICLE{pierel2019turning,
       author = {{Pierel}, J.~D.~R. and {Rodney}, S.},
        title = "{Turning Gravitationally Lensed Supernovae into Cosmological Probes}",
      journal = {\apj},
         year = 2019,
        month = may,
       volume = {876},
       number = {2},
          eid = {107},
        pages = {107},
          doi = {10.3847/1538-4357/ab164a},
archivePrefix = {arXiv},
       eprint = {1902.01260},
 primaryClass = {astro-ph.CO},
       adsurl = {https://ui.adsabs.harvard.edu/abs/2019ApJ...876..107P}
}

@ARTICLE{oguri2010gravitationally,
       author = {{Oguri}, Masamune and {Marshall}, Philip J.},
        title = "{Gravitationally lensed quasars and supernovae in future wide-field optical imaging surveys}",
      journal = {\mnras},
         year = 2010,
        month = jul,
       volume = {405},
       number = {4},
        pages = {2579-2593},
          doi = {10.1111/j.1365-2966.2010.16639.x},
archivePrefix = {arXiv},
       eprint = {1001.2037},
 primaryClass = {astro-ph.CO},
       adsurl = {https://ui.adsabs.harvard.edu/abs/2010MNRAS.405.2579O}
}

@ARTICLE{more2017interpreting,
       author = {{More}, Anupreeta and {Suyu}, Sherry H. and {Oguri}, Masamune and {More}, Surhud and {Lee}, Chien-Hsiu},
        title = "{Interpreting the Strongly Lensed Supernova iPTF16geu: Time Delay Predictions, Microlensing, and Lensing Rates}",
      journal = {\apjl},
         year = 2017,
        month = feb,
       volume = {835},
       number = {2},
          eid = {L25},
        pages = {L25},
          doi = {10.3847/2041-8213/835/2/L25},
archivePrefix = {arXiv},
       eprint = {1611.04866},
 primaryClass = {astro-ph.CO},
       adsurl = {https://ui.adsabs.harvard.edu/abs/2017ApJ...835L..25M}
}

@ARTICLE{kelly2015multiple,
       author = {{Kelly}, Patrick L. and {Rodney}, Steven A. and {Treu}, Tommaso and {Foley}, Ryan J. and {Brammer}, Gabriel and {Schmidt}, Kasper B. and {Zitrin}, Adi and {Sonnenfeld}, Alessandro and {Strolger}, Louis-Gregory and {Graur}, Or and {Filippenko}, Alexei V. and {Jha}, Saurabh W. and {Riess}, Adam G. and {Bradac}, Marusa and {Weiner}, Benjamin J. and {Scolnic}, Daniel and {Malkan}, Matthew A. and {von der Linden}, Anja and {Trenti}, Michele and {Hjorth}, Jens and {Gavazzi}, Raphael and {Fontana}, Adriano and {Merten}, Julian C. and {McCully}, Curtis and {Jones}, Tucker and {Postman}, Marc and {Dressler}, Alan and {Patel}, Brandon and {Cenko}, S. Bradley and {Graham}, Melissa L. and {Tucker}, Bradley E.},
        title = "{Multiple images of a highly magnified supernova formed by an early-type cluster galaxy lens}",
      journal = {Science},
         year = 2015,
        month = mar,
       volume = {347},
       number = {6226},
        pages = {1123-1126},
          doi = {10.1126/science.aaa3350},
archivePrefix = {arXiv},
       eprint = {1411.6009},
 primaryClass = {astro-ph.CO},
       adsurl = {https://ui.adsabs.harvard.edu/abs/2015Sci...347.1123K}
}

@ARTICLE{goobar2017iptf16geu,
       author = {{Goobar}, A. and {Amanullah}, R. and {Kulkarni}, S.~R. and {Nugent}, P.~E. and {Johansson}, J. and {Steidel}, C. and {Law}, D. and {M{\"o}rtsell}, E. and {Quimby}, R. and {Blagorodnova}, N. and {Brandeker}, A. and {Cao}, Y. and {Cooray}, A. and {Ferretti}, R. and {Fremling}, C. and {Hangard}, L. and {Kasliwal}, M. and {Kupfer}, T. and {Lunnan}, R. and {Masci}, F. and {Miller}, A.~A. and {Nayyeri}, H. and {Neill}, J.~D. and {Ofek}, E.~O. and {Papadogiannakis}, S. and {Petrushevska}, T. and {Ravi}, V. and {Sollerman}, J. and {Sullivan}, M. and {Taddia}, F. and {Walters}, R. and {Wilson}, D. and {Yan}, L. and {Yaron}, O.},
        title = "{iPTF16geu: A multiply imaged, gravitationally lensed type Ia supernova}",
      journal = {Science},
         year = 2017,
        month = apr,
       volume = {356},
       number = {6335},
        pages = {291-295},
          doi = {10.1126/science.aal2729},
archivePrefix = {arXiv},
       eprint = {1611.00014},
 primaryClass = {astro-ph.CO},
       adsurl = {https://ui.adsabs.harvard.edu/abs/2017Sci...356..291G}
}

@ARTICLE{cano2018spectroscopic,
       author = {{Cano}, Zach and {Selsing}, Jonatan and {Hjorth}, Jens and {de Ugarte Postigo}, Antonio and {Christensen}, Lise and {Gall}, Christa and {Kann}, D.~A.},
        title = "{A spectroscopic look at the gravitationally lensed Type Ia supernova 2016geu at z = 0.409}",
      journal = {\mnras},
         year = 2018,
        month = jan,
       volume = {473},
       number = {3},
        pages = {4257-4267},
          doi = {10.1093/mnras/stx2624},
archivePrefix = {arXiv},
       eprint = {1708.05534},
 primaryClass = {astro-ph.HE},
       adsurl = {https://ui.adsabs.harvard.edu/abs/2018MNRAS.473.4257C}
}

@ARTICLE{pierel2020projected,
       author = {{Pierel}, J.~D.~R. and {Rodney}, S. and {Vernardos}, G. and {Oguri}, M. and {Kessler}, R. and {Anguita}, T.},
        title = "{Projected Cosmological Constraints from Strongly Lensed Supernovae with the Roman Space Telescope}",
      journal = {\apj},
         year = 2021,
        month = feb,
       volume = {908},
       number = {2},
          eid = {190},
        pages = {190},
          doi = {10.3847/1538-4357/abd8d3},
archivePrefix = {arXiv},
       eprint = {2010.12399},
 primaryClass = {astro-ph.CO},
       adsurl = {https://ui.adsabs.harvard.edu/abs/2021ApJ...908..190P}
}

@ARTICLE{birrer2018lenstronomy,
       author = {{Birrer}, Simon and {Amara}, Adam},
        title = "{lenstronomy: Multi-purpose gravitational lens modelling software package}",
      journal = {Phys.\ Dark Universe},
         year = 2018,
        month = dec,
       volume = {22},
        pages = {189-201},
          doi = {10.1016/j.dark.2018.11.002},
archivePrefix = {arXiv},
       eprint = {1803.09746},
 primaryClass = {astro-ph.CO},
       adsurl = {https://ui.adsabs.harvard.edu/abs/2018PDU....22..189B}
}

@ARTICLE{wojtak2019magnified,
       author = {{Wojtak}, Rados{\l}aw and {Hjorth}, Jens and {Gall}, Christa},
        title = "{Magnified or multiply imaged? - Search strategies for gravitationally lensed supernovae in wide-field surveys}",
      journal = {\mnras},
         year = 2019,
        month = aug,
       volume = {487},
       number = {3},
        pages = {3342-3355},
          doi = {10.1093/mnras/stz1516},
archivePrefix = {arXiv},
       eprint = {1903.07687},
 primaryClass = {astro-ph.CO},
       adsurl = {https://ui.adsabs.harvard.edu/abs/2019MNRAS.487.3342W}
}

@ARTICLE{ivezic2019lsst,
       author = {{Ivezi{\'c}}, {\v{Z}}eljko and {Kahn}, Steven M. and {Tyson}, J. Anthony and {Abel}, Bob and {Acosta}, Emily and {Allsman}, Robyn and {Alonso}, David and {AlSayyad}, Yusra and {Anderson}, Scott F. and {Andrew}, John and {Angel}, James Roger P. and {Angeli}, George Z. and {Ansari}, Reza and {Antilogus}, Pierre and {Araujo}, Constanza and {Armstrong}, Robert and {Arndt}, Kirk T. and {Astier}, Pierre and {Aubourg}, {\'E}ric and {Auza}, Nicole and {Axelrod}, Tim S. and {Bard}, Deborah J. and {Barr}, Jeff D. and {Barrau}, Aurelian and {Bartlett}, James G. and {Bauer}, Amanda E. and {Bauman}, Brian J. and {Baumont}, Sylvain and {Bechtol}, Ellen and {Bechtol}, Keith and {Becker}, Andrew C. and {Becla}, Jacek and {Beldica}, Cristina and {Bellavia}, Steve and {Bianco}, Federica B. and {Biswas}, Rahul and {Blanc}, Guillaume and {Blazek}, Jonathan and {Blandford}, Roger D. and {Bloom}, Josh S. and {Bogart}, Joanne and {Bond}, Tim W. and {Booth}, Michael T. and {Borgland}, Anders W. and {Borne}, Kirk and {Bosch}, James F. and {Boutigny}, Dominique and {Brackett}, Craig A. and {Bradshaw}, Andrew and {Brandt}, William Nielsen and {Brown}, Michael E. and {Bullock}, James S. and {Burchat}, Patricia and {Burke}, David L. and {Cagnoli}, Gianpietro and {Calabrese}, Daniel and {Callahan}, Shawn and {Callen}, Alice L. and {Carlin}, Jeffrey L. and {Carlson}, Erin L. and {Chandrasekharan}, Srinivasan and {Charles-Emerson}, Glenaver and {Chesley}, Steve and {Cheu}, Elliott C. and {Chiang}, Hsin-Fang and {Chiang}, James and {Chirino}, Carol and {Chow}, Derek and {Ciardi}, David R. and {Claver}, Charles F. and {Cohen-Tanugi}, Johann and {Cockrum}, Joseph J. and {Coles}, Rebecca and {Connolly}, Andrew J. and {Cook}, Kem H. and {Cooray}, Asantha and {Covey}, Kevin R. and {Cribbs}, Chris and {Cui}, Wei and {Cutri}, Roc and {Daly}, Philip N. and {Daniel}, Scott F. and {Daruich}, Felipe and {Daubard}, Guillaume and {Daues}, Greg and {Dawson}, William and {Delgado}, Francisco and {Dellapenna}, Alfred and {de Peyster}, Robert and {de Val-Borro}, Miguel and {Digel}, Seth W. and {Doherty}, Peter and {Dubois}, Richard and {Dubois-Felsmann}, Gregory P. and {Durech}, Josef and {Economou}, Frossie and {Eifler}, Tim and {Eracleous}, Michael and {Emmons}, Benjamin L. and {Fausti Neto}, Angelo and {Ferguson}, Henry and {Figueroa}, Enrique and {Fisher-Levine}, Merlin and {Focke}, Warren and {Foss}, Michael D. and {Frank}, James and {Freemon}, Michael D. and {Gangler}, Emmanuel and {Gawiser}, Eric and {Geary}, John C. and {Gee}, Perry and {Geha}, Marla and {Gessner}, Charles J.~B. and {Gibson}, Robert R. and {Gilmore}, D. Kirk and {Glanzman}, Thomas and {Glick}, William and {Goldina}, Tatiana and {Goldstein}, Daniel A. and {Goodenow}, Iain and {Graham}, Melissa L. and {Gressler}, William J. and {Gris}, Philippe and {Guy}, Leanne P. and {Guyonnet}, Augustin and {Haller}, Gunther and {Harris}, Ron and {Hascall}, Patrick A. and {Haupt}, Justine and {Hernandez}, Fabio and {Herrmann}, Sven and {Hileman}, Edward and {Hoblitt}, Joshua and {Hodgson}, John A. and {Hogan}, Craig and {Howard}, James D. and {Huang}, Dajun and {Huffer}, Michael E. and {Ingraham}, Patrick and {Innes}, Walter R. and {Jacoby}, Suzanne H. and {Jain}, Bhuvnesh and {Jammes}, Fabrice and {Jee}, M. James and {Jenness}, Tim and {Jernigan}, Garrett and {Jevremovi{\'c}}, Darko and {Johns}, Kenneth and {Johnson}, Anthony S. and {Johnson}, Margaret W.~G. and {Jones}, R. Lynne and {Juramy-Gilles}, Claire and {Juri{\'c}}, Mario and {Kalirai}, Jason S. and {Kallivayalil}, Nitya J. and {Kalmbach}, Bryce and {Kantor}, Jeffrey P. and {Karst}, Pierre and {Kasliwal}, Mansi M. and {Kelly}, Heather and {Kessler}, Richard and {Kinnison}, Veronica and {Kirkby}, David and {Knox}, Lloyd and {Kotov}, Ivan V. and {Krabbendam}, Victor L. and {Krughoff}, K. Simon and {Kub{\'a}nek}, Petr and {Kuczewski}, John and {Kulkarni}, Shri and {Ku}, John and {Kurita}, Nadine R. and {Lage}, Craig S. and {Lambert}, Ron and {Lange}, Travis and {Langton}, J. Brian and {Le Guillou}, Laurent and {Levine}, Deborah and {Liang}, Ming and {Lim}, Kian-Tat and {Lintott}, Chris J. and {Long}, Kevin E. and {Lopez}, Margaux and {Lotz}, Paul J. and {Lupton}, Robert H. and {Lust}, Nate B. and {MacArthur}, Lauren A. and {Mahabal}, Ashish and {Mandelbaum}, Rachel and {Markiewicz}, Thomas W. and {Marsh}, Darren S. and {Marshall}, Philip J. and {Marshall}, Stuart and {May}, Morgan and {McKercher}, Robert and {McQueen}, Michelle and {Meyers}, Joshua and {Migliore}, Myriam and {Miller}, Michelle and {Mills}, David J. and {Miraval}, Connor and {Moeyens}, Joachim and {Moolekamp}, Fred E. and {Monet}, David G. and {Moniez}, Marc and {Monkewitz}, Serge and {Montgomery}, Christopher and {Morrison}, Christopher B. and {Mueller}, Fritz and {Muller}, Gary P. and {Mu{\~n}oz Arancibia}, Freddy and {Neill}, Douglas R. and {Newbry}, Scott P. and {Nief}, Jean-Yves and {Nomerotski}, Andrei and {Nordby}, Martin and {O'Connor}, Paul and {Oliver}, John and {Olivier}, Scot S. and {Olsen}, Knut and {O'Mullane}, William and {Ortiz}, Sandra and {Osier}, Shawn and {Owen}, Russell E. and {Pain}, Reynald and {Palecek}, Paul E. and {Parejko}, John K. and {Parsons}, James B. and {Pease}, Nathan M. and {Peterson}, J. Matt and {Peterson}, John R. and {Petravick}, Donald L. and {Libby Petrick}, M.~E. and {Petry}, Cathy E. and {Pierfederici}, Francesco and {Pietrowicz}, Stephen and {Pike}, Rob and {Pinto}, Philip A. and {Plante}, Raymond and {Plate}, Stephen and {Plutchak}, Joel P. and {Price}, Paul A. and {Prouza}, Michael and {Radeka}, Veljko and {Rajagopal}, Jayadev and {Rasmussen}, Andrew P. and {Regnault}, Nicolas and {Reil}, Kevin A. and {Reiss}, David J. and {Reuter}, Michael A. and {Ridgway}, Stephen T. and {Riot}, Vincent J. and {Ritz}, Steve and {Robinson}, Sean and {Roby}, William and {Roodman}, Aaron and {Rosing}, Wayne and {Roucelle}, Cecille and {Rumore}, Matthew R. and {Russo}, Stefano and {Saha}, Abhijit and {Sassolas}, Benoit and {Schalk}, Terry L. and {Schellart}, Pim and {Schindler}, Rafe H. and {Schmidt}, Samuel and {Schneider}, Donald P. and {Schneider}, Michael D. and {Schoening}, William and {Schumacher}, German and {Schwamb}, Megan E. and {Sebag}, Jacques and {Selvy}, Brian and {Sembroski}, Glenn H. and {Seppala}, Lynn G. and {Serio}, Andrew and {Serrano}, Eduardo and {Shaw}, Richard A. and {Shipsey}, Ian and {Sick}, Jonathan and {Silvestri}, Nicole and {Slater}, Colin T. and {Smith}, J. Allyn and {Smith}, R. Chris and {Sobhani}, Shahram and {Soldahl}, Christine and {Storrie-Lombardi}, Lisa and {Stover}, Edward and {Strauss}, Michael A. and {Street}, Rachel A. and {Stubbs}, Christopher W. and {Sullivan}, Ian S. and {Sweeney}, Donald and {Swinbank}, John D. and {Szalay}, Alexander and {Takacs}, Peter and {Tether}, Stephen A. and {Thaler}, Jon J. and {Thayer}, John Gregg and {Thomas}, Sandrine and {Thornton}, Adam J. and {Thukral}, Vaikunth and {Tice}, Jeffrey and {Trilling}, David E. and {Turri}, Max and {Van Berg}, Richard and {Vanden Berk}, Daniel and {Vetter}, Kurt and {Virieux}, Francoise and {Vucina}, Tomislav and {Wahl}, William and {Walkowicz}, Lucianne and {Walsh}, Brian and {Walter}, Christopher W. and {Wang}, Daniel L. and {Wang}, Shin-Yawn and {Warner}, Michael and {Wiecha}, Oliver and {Willman}, Beth and {Winters}, Scott E. and {Wittman}, David and {Wolff}, Sidney C. and {Wood-Vasey}, W. Michael and {Wu}, Xiuqin and {Xin}, Bo and {Yoachim}, Peter and {Zhan}, Hu},
        title = "{LSST: From Science Drivers to Reference Design and Anticipated Data Products}",
      journal = {\apj},
         year = 2019,
        month = mar,
       volume = {873},
       number = {2},
          eid = {111},
        pages = {111},
          doi = {10.3847/1538-4357/ab042c},
archivePrefix = {arXiv},
       eprint = {0805.2366},
 primaryClass = {astro-ph},
       adsurl = {https://ui.adsabs.harvard.edu/abs/2019ApJ...873..111I}
}

@ARTICLE{planck2018cosmo,
       author = {{Planck Collaboration} and {Aghanim}, N. and {Akrami}, Y. and
         {Ashdown}, M. and {Aumont}, J. and {Baccigalupi}, C. and
         {Ballardini}, M. and {Banday}, A.~J. and {Barreiro}, R.~B. and
         {Bartolo}, N. and {Basak}, S. and {Battye}, R. and {Benabed}, K. and
         {Bernard}, J. -P. and {Bersanelli}, M. and {Bielewicz}, P. and
         {Bock}, J.~J. and {Bond}, J.~R. and {Borrill}, J. and {Bouchet}, F.~R. and
         {Boulanger}, F. and {Bucher}, M. and {Burigana}, C. and
         {Butler}, R.~C. and {Calabrese}, E. and {Cardoso}, J. -F. and
         {Carron}, J. and {Challinor}, A. and {Chiang}, H.~C. and {Chluba}, J. and
         {Colombo}, L.~P.~L. and {Combet}, C. and {Contreras}, D. and
         {Crill}, B.~P. and {Cuttaia}, F. and {de Bernardis}, P. and
         {de Zotti}, G. and {Delabrouille}, J. and {Delouis}, J. -M. and
         {Di Valentino}, E. and {Diego}, J.~M. and {Dor{\'e}}, O. and
         {Douspis}, M. and {Ducout}, A. and {Dupac}, X. and {Dusini}, S. and
         {Efstathiou}, G. and {Elsner}, F. and {En{\ss}lin}, T.~A. and
         {Eriksen}, H.~K. and {Fantaye}, Y. and {Farhang}, M. and
         {Fergusson}, J. and {Fernandez-Cobos}, R. and {Finelli}, F. and
         {Forastieri}, F. and {Frailis}, M. and {Franceschi}, E. and
         {Frolov}, A. and {Galeotta}, S. and {Galli}, S. and {Ganga}, K. and
         {G{\'e}nova-Santos}, R.~T. and {Gerbino}, M. and {Ghosh}, T. and
         {Gonz{\'a}lez-Nuevo}, J. and {G{\'o}rski}, K.~M. and {Gratton}, S. and
         {Gruppuso}, A. and {Gudmundsson}, J.~E. and {Hamann}, J. and {Hand
        ley}, W. and {Herranz}, D. and {Hivon}, E. and {Huang}, Z. and
         {Jaffe}, A.~H. and {Jones}, W.~C. and {Karakci}, A. and
         {Keih{\"a}nen}, E. and {Keskitalo}, R. and {Kiiveri}, K. and {Kim}, J. and
         {Kisner}, T.~S. and {Knox}, L. and {Krachmalnicoff}, N. and {Kunz}, M. and
         {Kurki-Suonio}, H. and {Lagache}, G. and {Lamarre}, J. -M. and
         {Lasenby}, A. and {Lattanzi}, M. and {Lawrence}, C.~R. and
         {Le Jeune}, M. and {Lemos}, P. and {Lesgourgues}, J. and {Levrier}, F. and
         {Lewis}, A. and {Liguori}, M. and {Lilje}, P.~B. and {Lilley}, M. and
         {Lindholm}, V. and {L{\'o}pez-Caniego}, M. and {Lubin}, P.~M. and
         {Ma}, Y. -Z. and {Mac{\'\i}as-P{\'e}rez}, J.~F. and {Maggio}, G. and
         {Maino}, D. and {Mandolesi}, N. and {Mangilli}, A. and
         {Marcos-Caballero}, A. and {Maris}, M. and {Martin}, P.~G. and
         {Martinelli}, M. and {Mart{\'\i}nez-Gonz{\'a}lez}, E. and
         {Matarrese}, S. and {Mauri}, N. and {McEwen}, J.~D. and
         {Meinhold}, P.~R. and {Melchiorri}, A. and {Mennella}, A. and
         {Migliaccio}, M. and {Millea}, M. and {Mitra}, S. and
         {Miville-Desch{\^e}nes}, M. -A. and {Molinari}, D. and {Montier}, L. and
         {Morgante}, G. and {Moss}, A. and {Natoli}, P. and
         {N{\o}rgaard-Nielsen}, H.~U. and {Pagano}, L. and {Paoletti}, D. and
         {Partridge}, B. and {Patanchon}, G. and {Peiris}, H.~V. and
         {Perrotta}, F. and {Pettorino}, V. and {Piacentini}, F. and
         {Polastri}, L. and {Polenta}, G. and {Puget}, J. -L. and
         {Rachen}, J.~P. and {Reinecke}, M. and {Remazeilles}, M. and
         {Renzi}, A. and {Rocha}, G. and {Rosset}, C. and {Roudier}, G. and
         {Rubi{\~n}o-Mart{\'\i}n}, J.~A. and {Ruiz-Granados}, B. and
         {Salvati}, L. and {Sandri}, M. and {Savelainen}, M. and {Scott}, D. and
         {Shellard}, E.~P.~S. and {Sirignano}, C. and {Sirri}, G. and
         {Spencer}, L.~D. and {Sunyaev}, R. and {Suur-Uski}, A. -S. and
         {Tauber}, J.~A. and {Tavagnacco}, D. and {Tenti}, M. and
         {Toffolatti}, L. and {Tomasi}, M. and {Trombetti}, T. and
         {Valenziano}, L. and {Valiviita}, J. and {Van Tent}, B. and
         {Vibert}, L. and {Vielva}, P. and {Villa}, F. and {Vittorio}, N. and {Wandelt}, B.~D. and {Wehus}, I.~K. and {White}, M. and {White}, S.~D.~M. and
         {Zacchei}, A. and {Zonca}, A.},
        title = "{Planck 2018 results. VI. Cosmological parameters}",
      journal = {\aap},
         year = 2020,
        month = sep,
       volume = {641},
          eid = {A6},
        pages = {A6},
          doi = {10.1051/0004-6361/201833910},
archivePrefix = {arXiv},
       eprint = {1807.06209},
 primaryClass = {astro-ph.CO},
       adsurl = {https://ui.adsabs.harvard.edu/abs/2020A&A...641A...6P}
}

@ARTICLE{Kelly2022riv,
       author = {{Kelly}, P. and {Zitrin}, A. and {Oguri}, M. and {Diego}, J. and {Williams}, H. and {Broadhurst}, T. and {Chen}, W. and {Koekemoer}, A. and {Pierel}, J. and {Strolger}, L. and {Treu}, T.},
        title = "{Strongly Lensed SN in MACS 2129 Galaxy-Cluster Field}",
      journal = {Transient Name Server AstroNote},
         year = 2022,
        month = aug,
       volume = {169},
        pages = {1},
       adsurl = {https://ui.adsabs.harvard.edu/abs/2022TNSAN.169....1K}
}

@ARTICLE{Chen2022_glSNAbell,
       author = {{Chen}, Wenlei and {Kelly}, Patrick L. and {Oguri}, Masamune and {Broadhurst}, Thomas J. and {Diego}, Jose M. and {Emami}, Najmeh and {Filippenko}, Alexei V. and {Treu}, Tommaso L. and {Zitrin}, Adi},
        title = "{Shock cooling of a red-supergiant supernova at redshift 3 in lensed images}",
      journal = {\nat},
         year = 2022,
        month = nov,
       volume = {611},
       number = {7935},
        pages = {256-259},
          doi = {10.1038/s41586-022-05252-5},
       adsurl = {https://ui.adsabs.harvard.edu/abs/2022Natur.611..256C}
}

@ARTICLE{Goobar2023_SNZwicky,
       author = {{Goobar}, Ariel and {Johansson}, Joel and {Schulze}, Steve and {Arendse}, Nikki and {Carracedo}, Ana Sagu{\'e}s and {Dhawan}, Suhail and {M{\"o}rtsell}, Edvard and {Fremling}, Christoffer and {Yan}, Lin and {Perley}, Daniel and {Sollerman}, Jesper and {Joseph}, R{\'e}my and {Hinds}, K. -Ryan and {Meynardie}, William and {Andreoni}, Igor and {Bellm}, Eric and {Bloom}, Josh and {Collett}, Thomas E. and {Drake}, Andrew and {Graham}, Matthew and {Kasliwal}, Mansi and {Kulkarni}, Shri R. and {Lemon}, Cameron and {Miller}, Adam A. and {Neill}, James D. and {Nordin}, Jakob and {Pierel}, Justin and {Richard}, Johan and {Riddle}, Reed and {Rigault}, Mickael and {Rusholme}, Ben and {Sharma}, Yashvi and {Stein}, Robert and {Stewart}, Gabrielle and {Townsend}, Alice and {Vinko}, Jozsef and {Wheeler}, J. Craig and {Wold}, Avery},
        title = "{Uncovering a population of gravitational lens galaxies with magnified standard candle SN Zwicky}",
      journal = {Nature Astronomy},
         year = 2023,
        month = jun,
       volume = {7},
        pages = {1098-1107},
          doi = {10.1038/s41550-023-01981-3},
archivePrefix = {arXiv},
       eprint = {2211.00656},
 primaryClass = {astro-ph.CO},
       adsurl = {https://ui.adsabs.harvard.edu/abs/2023NatAs...7.1098G}
}

@ARTICLE{Suyu2024,
       author = {{Suyu}, Sherry H. and {Goobar}, Ariel and {Collett}, Thomas and {More}, Anupreeta and {Vernardos}, Giorgos},
        title = "{Strong Gravitational Lensing and Microlensing of Supernovae}",
      journal = {\ssr},
         year = 2024,
        month = feb,
       volume = {220},
       number = {1},
          eid = {13},
        pages = {13},
          doi = {10.1007/s11214-024-01044-7},
archivePrefix = {arXiv},
       eprint = {2301.07729},
 primaryClass = {astro-ph.CO},
       adsurl = {https://ui.adsabs.harvard.edu/abs/2024SSRv..220...13S}
}

@ARTICLE{Birrer2021LenstronomyII,
       author = {{Birrer}, Simon and {Shajib}, Anowar and {Gilman}, Daniel and {Galan}, Aymeric and {Aalbers}, Jelle and {Millon}, Martin and {Morgan}, Robert and {Pagano}, Giulia and {Park}, Ji and {Teodori}, Luca and {Tessore}, Nicolas and {Ueland}, Madison and {Van de Vyvere}, Lyne and {Wagner-Carena}, Sebastian and {Wempe}, Ewoud and {Yang}, Lilan and {Ding}, Xuheng and {Schmidt}, Thomas and {Sluse}, Dominique and {Zhang}, Ming and {Amara}, Adam},
        title = "{lenstronomy II: A gravitational lensing software ecosystem}",
      journal = {J.\ Open Source Software},
         year = 2021,
        month = jun,
       volume = {6},
       number = {62},
          eid = {3283},
        pages = {3283},
          doi = {10.21105/joss.03283},
archivePrefix = {arXiv},
       eprint = {2106.05976},
 primaryClass = {astro-ph.CO},
       adsurl = {https://ui.adsabs.harvard.edu/abs/2021JOSS....6.3283B}
}

@ARTICLE{Astropy2013,
       author = {{Astropy Collaboration} and {Robitaille}, Thomas P. and {Tollerud}, Erik J. and {Greenfield}, Perry and {Droettboom}, Michael and {Bray}, Erik and {Aldcroft}, Tom and {Davis}, Matt and {Ginsburg}, Adam and {Price-Whelan}, Adrian M. and {Kerzendorf}, Wolfgang E. and {Conley}, Alexander and {Crighton}, Neil and {Barbary}, Kyle and {Muna}, Demitri and {Ferguson}, Henry and {Grollier}, Fr{\'e}d{\'e}ric and {Parikh}, Madhura M. and {Nair}, Prasanth H. and {Unther}, Hans M. and {Deil}, Christoph and {Woillez}, Julien and {Conseil}, Simon and {Kramer}, Roban and {Turner}, James E.~H. and {Singer}, Leo and {Fox}, Ryan and {Weaver}, Benjamin A. and {Zabalza}, Victor and {Edwards}, Zachary I. and {Azalee Bostroem}, K. and {Burke}, D.~J. and {Casey}, Andrew R. and {Crawford}, Steven M. and {Dencheva}, Nadia and {Ely}, Justin and {Jenness}, Tim and {Labrie}, Kathleen and {Lim}, Pey Lian and {Pierfederici}, Francesco and {Pontzen}, Andrew and {Ptak}, Andy and {Refsdal}, Brian and {Servillat}, Mathieu and {Streicher}, Ole},
        title = "{Astropy: A community Python package for astronomy}",
      journal = {\aap},
         year = 2013,
        month = oct,
       volume = {558},
          eid = {A33},
        pages = {A33},
          doi = {10.1051/0004-6361/201322068},
archivePrefix = {arXiv},
       eprint = {1307.6212},
 primaryClass = {astro-ph.IM},
       adsurl = {https://ui.adsabs.harvard.edu/abs/2013A&A...558A..33A}
}

@ARTICLE{Astropy2018,
       author = {{Astropy Collaboration} and {Price-Whelan}, A.~M. and {Sip{\H{o}}cz}, B.~M. and {G{\"u}nther}, H.~M. and {Lim}, P.~L. and {Crawford}, S.~M. and {Conseil}, S. and {Shupe}, D.~L. and {Craig}, M.~W. and {Dencheva}, N. and {Ginsburg}, A. and {VanderPlas}, J.~T. and {Bradley}, L.~D. and {P{\'e}rez-Su{\'a}rez}, D. and {de Val-Borro}, M. and {Aldcroft}, T.~L. and {Cruz}, K.~L. and {Robitaille}, T.~P. and {Tollerud}, E.~J. and {Ardelean}, C. and {Babej}, T. and {Bach}, Y.~P. and {Bachetti}, M. and {Bakanov}, A.~V. and {Bamford}, S.~P. and {Barentsen}, G. and {Barmby}, P. and {Baumbach}, A. and {Berry}, K.~L. and {Biscani}, F. and {Boquien}, M. and {Bostroem}, K.~A. and {Bouma}, L.~G. and {Brammer}, G.~B. and {Bray}, E.~M. and {Breytenbach}, H. and {Buddelmeijer}, H. and {Burke}, D.~J. and {Calderone}, G. and {Cano Rodr{\'\i}guez}, J.~L. and {Cara}, M. and {Cardoso}, J.~V.~M. and {Cheedella}, S. and {Copin}, Y. and {Corrales}, L. and {Crichton}, D. and {D'Avella}, D. and {Deil}, C. and {Depagne}, {\'E}. and {Dietrich}, J.~P. and {Donath}, A. and {Droettboom}, M. and {Earl}, N. and {Erben}, T. and {Fabbro}, S. and {Ferreira}, L.~A. and {Finethy}, T. and {Fox}, R.~T. and {Garrison}, L.~H. and {Gibbons}, S.~L.~J. and {Goldstein}, D.~A. and {Gommers}, R. and {Greco}, J.~P. and {Greenfield}, P. and {Groener}, A.~M. and {Grollier}, F. and {Hagen}, A. and {Hirst}, P. and {Homeier}, D. and {Horton}, A.~J. and {Hosseinzadeh}, G. and {Hu}, L. and {Hunkeler}, J.~S. and {Ivezi{\'c}}, {\v{Z}}. and {Jain}, A. and {Jenness}, T. and {Kanarek}, G. and {Kendrew}, S. and {Kern}, N.~S. and {Kerzendorf}, W.~E. and {Khvalko}, A. and {King}, J. and {Kirkby}, D. and {Kulkarni}, A.~M. and {Kumar}, A. and {Lee}, A. and {Lenz}, D. and {Littlefair}, S.~P. and {Ma}, Z. and {Macleod}, D.~M. and {Mastropietro}, M. and {McCully}, C. and {Montagnac}, S. and {Morris}, B.~M. and {Mueller}, M. and {Mumford}, S.~J. and {Muna}, D. and {Murphy}, N.~A. and {Nelson}, S. and {Nguyen}, G.~H. and {Ninan}, J.~P. and {N{\"o}the}, M. and {Ogaz}, S. and {Oh}, S. and {Parejko}, J.~K. and {Parley}, N. and {Pascual}, S. and {Patil}, R. and {Patil}, A.~A. and {Plunkett}, A.~L. and {Prochaska}, J.~X. and {Rastogi}, T. and {Reddy Janga}, V. and {Sabater}, J. and {Sakurikar}, P. and {Seifert}, M. and {Sherbert}, L.~E. and {Sherwood-Taylor}, H. and {Shih}, A.~Y. and {Sick}, J. and {Silbiger}, M.~T. and {Singanamalla}, S. and {Singer}, L.~P. and {Sladen}, P.~H. and {Sooley}, K.~A. and {Sornarajah}, S. and {Streicher}, O. and {Teuben}, P. and {Thomas}, S.~W. and {Tremblay}, G.~R. and {Turner}, J.~E.~H. and {Terr{\'o}n}, V. and {van Kerkwijk}, M.~H. and {de la Vega}, A. and {Watkins}, L.~L. and {Weaver}, B.~A. and {Whitmore}, J.~B. and {Woillez}, J. and {Zabalza}, V. and {Astropy Contributors}},
        title = "{The Astropy Project: Building an Open-science Project and Status of the v2.0 Core Package}",
      journal = {\aj},
         year = 2018,
        month = sep,
       volume = {156},
       number = {3},
          eid = {123},
        pages = {123},
          doi = {10.3847/1538-3881/aabc4f},
archivePrefix = {arXiv},
       eprint = {1801.02634},
 primaryClass = {astro-ph.IM},
       adsurl = {https://ui.adsabs.harvard.edu/abs/2018AJ....156..123A}
}

@ARTICLE{Astropy2022,
       author = {{Astropy Collaboration} and {Price-Whelan}, Adrian M. and {Lim}, Pey Lian and {Earl}, Nicholas and {Starkman}, Nathaniel and {Bradley}, Larry and {Shupe}, David L. and {Patil}, Aarya A. and {Corrales}, Lia and {Brasseur}, C.~E. and {N{\"o}the}, Maximilian and {Donath}, Axel and {Tollerud}, Erik and {Morris}, Brett M. and {Ginsburg}, Adam and {Vaher}, Eero and {Weaver}, Benjamin A. and {Tocknell}, James and {Jamieson}, William and {van Kerkwijk}, Marten H. and {Robitaille}, Thomas P. and {Merry}, Bruce and {Bachetti}, Matteo and {G{\"u}nther}, H. Moritz and {Aldcroft}, Thomas L. and {Alvarado-Montes}, Jaime A. and {Archibald}, Anne M. and {B{\'o}di}, Attila and {Bapat}, Shreyas and {Barentsen}, Geert and {Baz{\'a}n}, Juanjo and {Biswas}, Manish and {Boquien}, M{\'e}d{\'e}ric and {Burke}, D.~J. and {Cara}, Daria and {Cara}, Mihai and {Conroy}, Kyle E. and {Conseil}, Simon and {Craig}, Matthew W. and {Cross}, Robert M. and {Cruz}, Kelle L. and {D'Eugenio}, Francesco and {Dencheva}, Nadia and {Devillepoix}, Hadrien A.~R. and {Dietrich}, J{\"o}rg P. and {Eigenbrot}, Arthur Davis and {Erben}, Thomas and {Ferreira}, Leonardo and {Foreman-Mackey}, Daniel and {Fox}, Ryan and {Freij}, Nabil and {Garg}, Suyog and {Geda}, Robel and {Glattly}, Lauren and {Gondhalekar}, Yash and {Gordon}, Karl D. and {Grant}, David and {Greenfield}, Perry and {Groener}, Austen M. and {Guest}, Steve and {Gurovich}, Sebastian and {Handberg}, Rasmus and {Hart}, Akeem and {Hatfield-Dodds}, Zac and {Homeier}, Derek and {Hosseinzadeh}, Griffin and {Jenness}, Tim and {Jones}, Craig K. and {Joseph}, Prajwel and {Kalmbach}, J. Bryce and {Karamehmetoglu}, Emir and {Ka{\l}uszy{\'n}ski}, Miko{\l}aj and {Kelley}, Michael S.~P. and {Kern}, Nicholas and {Kerzendorf}, Wolfgang E. and {Koch}, Eric W. and {Kulumani}, Shankar and {Lee}, Antony and {Ly}, Chun and {Ma}, Zhiyuan and {MacBride}, Conor and {Maljaars}, Jakob M. and {Muna}, Demitri and {Murphy}, N.~A. and {Norman}, Henrik and {O'Steen}, Richard and {Oman}, Kyle A. and {Pacifici}, Camilla and {Pascual}, Sergio and {Pascual-Granado}, J. and {Patil}, Rohit R. and {Perren}, Gabriel I. and {Pickering}, Timothy E. and {Rastogi}, Tanuj and {Roulston}, Benjamin R. and {Ryan}, Daniel F. and {Rykoff}, Eli S. and {Sabater}, Jose and {Sakurikar}, Parikshit and {Salgado}, Jes{\'u}s and {Sanghi}, Aniket and {Saunders}, Nicholas and {Savchenko}, Volodymyr and {Schwardt}, Ludwig and {Seifert-Eckert}, Michael and {Shih}, Albert Y. and {Jain}, Anany Shrey and {Shukla}, Gyanendra and {Sick}, Jonathan and {Simpson}, Chris and {Singanamalla}, Sudheesh and {Singer}, Leo P. and {Singhal}, Jaladh and {Sinha}, Manodeep and {Sip{\H{o}}cz}, Brigitta M. and {Spitler}, Lee R. and {Stansby}, David and {Streicher}, Ole and {{\v{S}}umak}, Jani and {Swinbank}, John D. and {Taranu}, Dan S. and {Tewary}, Nikita and {Tremblay}, Grant R. and {de Val-Borro}, Miguel and {Van Kooten}, Samuel J. and {Vasovi{\'c}}, Zlatan and {Verma}, Shresth and {de Miranda Cardoso}, Jos{\'e} Vin{\'\i}cius and {Williams}, Peter K.~G. and {Wilson}, Tom J. and {Winkel}, Benjamin and {Wood-Vasey}, W.~M. and {Xue}, Rui and {Yoachim}, Peter and {Zhang}, Chen and {Zonca}, Andrea and {Astropy Project Contributors}},
        title = "{The Astropy Project: Sustaining and Growing a Community-oriented Open-source Project and the Latest Major Release (v5.0) of the Core Package}",
      journal = {\apj},
         year = 2022,
        month = aug,
       volume = {935},
       number = {2},
          eid = {167},
        pages = {167},
          doi = {10.3847/1538-4357/ac7c74},
archivePrefix = {arXiv},
       eprint = {2206.14220},
 primaryClass = {astro-ph.IM},
       adsurl = {https://ui.adsabs.harvard.edu/abs/2022ApJ...935..167A}
}

@INPROCEEDINGS{JupyterNotebook,
       author = {{Kluyver}, Thomas and {Ragan-Kelley}, Benjain and {P{\'e}rez}, Fernando and {Granger}, Brian and {Bussonnier}, Matthias and {Frederic}, Jonathan and {Kelley}, Kyle and {Hamrick}, Jessica and {Grout}, Jason and {Corlay}, Sylvain and {Ivanov}, Paul and {Avila}, Dami{\'a}n and {Abdalla}, Safia and {Willing}, Carol and {Jupyter Development Team}},
        title = "{Jupyter Notebooks{\textemdash}a publishing format for reproducible computational workflows}",
        editor = {{Loizides}, F. and {Schmidt}, B. },
    booktitle = {Positioning and Power in Academic Publishing: Players, Agents and Agendas},
    publisher = {IOS Press},
         year = 2016,
        pages = {87-90},
          doi = {10.3233/978-1-61499-649-1-87},
       adsurl = {https://ui.adsabs.harvard.edu/abs/2016ppap.book...87K}
}

@ARTICLE{Matplotlib2007,
  author={Hunter, John D.},
  journal={Comput.\ Sci.\ Eng.}, 
  title={Matplotlib: A 2D Graphics Environment}, 
  year={2007},
  volume={9},
  number={3},
  pages={90-95},
  doi={10.1109/MCSE.2007.55}}

@software{Matplotlib2020,
  author       = {Thomas A Caswell and
                  Michael Droettboom and
                  Antony Lee and
                  John Hunter and
                  Elliott Sales de Andrade and
                  Eric Firing and
                  Tim Hoffmann and
                  Jody Klymak and
                  David Stansby and
                  Nelle Varoquaux and
                  Jens Hedegaard Nielsen and
                  Benjamin Root and
                  Ryan May and
                  Phil Elson and
                  Jouni K. Seppänen and
                  Darren Dale and
                  Jae-Joon Lee and
                  Damon McDougall and
                  Andrew Straw and
                  Paul Hobson and
                  Christoph Gohlke and
                  Tony S Yu and
                  Eric Ma and
                  Adrien F. Vincent and
                  Steven Silvester and
                  Charlie Moad and
                  Nikita Kniazev and
                  hannah and
                  Elan Ernest and
                  Paul Ivanov},
  title        = {matplotlib/matplotlib: v3.3.2},
  month        = sep,
  year         = 2020,
  publisher    = {Zenodo},
  version      = {v3.3.2},
  doi = {10.5281/zenodo.4030140}
}

@ARTICLE{Numpy202,
       author = {{Harris}, Charles R. and {Millman}, K. Jarrod and {van der Walt}, St{\'e}fan J. and {Gommers}, Ralf and {Virtanen}, Pauli and {Cournapeau}, David and {Wieser}, Eric and {Taylor}, Julian and {Berg}, Sebastian and {Smith}, Nathaniel J. and {Kern}, Robert and {Picus}, Matti and {Hoyer}, Stephan and {van Kerkwijk}, Marten H. and {Brett}, Matthew and {Haldane}, Allan and {del R{\'\i}o}, Jaime Fern{\'a}ndez and {Wiebe}, Mark and {Peterson}, Pearu and {G{\'e}rard-Marchant}, Pierre and {Sheppard}, Kevin and {Reddy}, Tyler and {Weckesser}, Warren and {Abbasi}, Hameer and {Gohlke}, Christoph and {Oliphant}, Travis E.},
        title = "{Array programming with NumPy}",
      journal = {\nat},
         year = 2020,
        month = sep,
       volume = {585},
       number = {7825},
        pages = {357-362},
          doi = {10.1038/s41586-020-2649-2},
archivePrefix = {arXiv},
       eprint = {2006.10256},
 primaryClass = {cs.MS},
       adsurl = {https://ui.adsabs.harvard.edu/abs/2020Natur.585..357H}
}

@InProceedings{Pandas2010,
  author    = {{McKinney}, W},
  title     = {{D}ata {S}tructures for {S}tatistical {C}omputing in {P}ython},
  booktitle = {{P}roceedings of the 9th {P}ython in {S}cience {C}onference},
  pages     = {56-61},
  year      = {2010},
  editor    = { {van der Walt}, S. and {Millman}, J. },
  doi       = {10.25080/Majora-92bf1922-00a}
}

@MISC{Pandas2023,
       author = {{Pandas Development Team}},
        title = "{pandas-dev/pandas: v2.3.3}",
 howpublished = {Zenodo},
         year = 2023,
        month = sep,
          eid = {10.5281/zenodo.3509134},
          doi = {10.5281/zenodo.3509134},
      version = {v2.1.1},
    publisher = {Zenodo},
       adsurl = {https://ui.adsabs.harvard.edu/abs/2022zndo...3509134T}
}

@book{Pickle2020, 
  title={The Python Library Reference, Release 3.8.2},
  author={Van Rossum, Guido}, 
  year={2020}, 
  publisher={Python Software Foundation} 
}

@ARTICLE{Scipy2020,
       author = {{Virtanen}, Pauli and {Gommers}, Ralf and {Oliphant}, Travis E. and {Haberland}, Matt and {Reddy}, Tyler and {Cournapeau}, David and {Burovski}, Evgeni and {Peterson}, Pearu and {Weckesser}, Warren and {Bright}, Jonathan and {van der Walt}, St{\'e}fan J. and {Brett}, Matthew and {Wilson}, Joshua and {Millman}, K. Jarrod and {Mayorov}, Nikolay and {Nelson}, Andrew R.~J. and {Jones}, Eric and {Kern}, Robert and {Larson}, Eric and {Carey}, C.~J. and {Polat}, {\.I}lhan and {Feng}, Yu and {Moore}, Eric W. and {VanderPlas}, Jake and {Laxalde}, Denis and {Perktold}, Josef and {Cimrman}, Robert and {Henriksen}, Ian and {Quintero}, E.~A. and {Harris}, Charles R. and {Archibald}, Anne M. and {Ribeiro}, Ant{\^o}nio H. and {Pedregosa}, Fabian and {van Mulbregt}, Paul and {SciPy 1. 0 Contributors}},
        title = "{SciPy 1.0: fundamental algorithms for scientific computing in Python}",
      journal = {Nature Methods},
         year = 2020,
        month = feb,
       volume = {17},
        pages = {261-272},
          doi = {10.1038/s41592-019-0686-2},
archivePrefix = {arXiv},
       eprint = {1907.10121},
 primaryClass = {cs.MS},
       adsurl = {https://ui.adsabs.harvard.edu/abs/2020NatMe..17..261V}
}

@software{Seaborn2020,
  author       = {Michael Waskom and
                  Olga Botvinnik and
                  Maoz Gelbart and
                  Joel Ostblom and
                  Paul Hobson and
                  Saulius Lukauskas and
                  David C Gemperline and
                  Tom Augspurger and
                  Yaroslav Halchenko and
                  Jordi Warmenhoven and
                  John B. Cole and
                  Julian de Ruiter and
                  Jake Vanderplas and
                  Stephan Hoyer and
                  Cameron Pye and
                  Alistair Miles and
                  Corban Swain and
                  Kyle Meyer and
                  Marcel Martin and
                  Pete Bachant and
                  Eric Quintero and
                  Gero Kunter and
                  Santi Villalba and
                  Brian and
                  Clark Fitzgerald and
                  C.G. Evans and
                  Mike Lee Williams and
                  Drew O'Kane and
                  Tal Yarkoni and
                  Thomas Brunner},
  title        = {mwaskom/seaborn: v0.11.0},
  month        = sep,
  year         = 2020,
  publisher    = {Zenodo},
  version      = {v0.11.0},
  doi = {10.5281/zenodo.4019146}
}

@ARTICLE{Hinton2016,
  author = {{Hinton}, S.~R.},
   title = "{ChainConsumer}",
 journal = {J.\ Open Source Software},
    year = 2016,
   month = aug,
  volume = 1,
     eid = {00045},
   pages = {00045},
     doi = {10.21105/joss.00045},
  adsurl = {http://adsabs.harvard.edu/abs/2016JOSS....1...45H},
}

@ARTICLE{SainzDeMurieta2023,
       author = {{Sainz de Murieta}, Ana and {Collett}, Thomas E. and {Magee}, Mark R. and {Weisenbach}, Luke and {Krawczyk}, Coleman M. and {Enzi}, Wolfgang},
        title = "{Lensed Type Ia supernovae in light of SN Zwicky and iPTF16geu}",
      journal = {\mnras},
         year = 2023,
        month = dec,
       volume = {526},
       number = {3},
        pages = {4296-4307},
          doi = {10.1093/mnras/stad3031},
archivePrefix = {arXiv},
       eprint = {2307.12881},
 primaryClass = {astro-ph.CO},
       adsurl = {https://ui.adsabs.harvard.edu/abs/2023MNRAS.526.4296S}
}

@ARTICLE{2022Fremling,
       author = {{Fremling}, C. and {Meynardie}, W. and {Yan}, L. and {Salama}, M. and {Jensen-Clem}},
        title = "{Keck NIRC2+LGS Imaging Observations of SN 2022qmx (``SN Zwicky'')}",
      journal = {Transient Name Server AstroNote},
         year = 2022,
        month = sep,
       volume = {194},
        pages = {1},
       adsurl = {https://ui.adsabs.harvard.edu/abs/2022TNSAN.194....1F}
}

@ARTICLE{2022Pierel,
       author = {{Pierel}, J. and {Strolger}, L. and {Hjorth}, J. and {Perez-Fournon}, I. and {Suyu}, S. and {Agnello}, A. and {Angel}, C.~J. and {Bolton}, A. and {Canameras}, R. and {Chakrabarti}, S. and {Christensen}, L. and {Courbin}, F. and {Dhawan}, S. and {Engesser}, M. and {Foley}, R.~J. and {Gall}, C. and {Geier}, S. and {Goobar}, A. and {Lee}, C. and {Huang}, X. and {Jha}, S. and {Johansson}, J. and {Larison}, C. and {Marques-Chaves}, R. and {Mazzali}, P. and {McCully}, C. and {Moustakas}, L. and {Nonino}, M. and {Poidevin}, F. and {Riess}, A. and {Shirley}, R. and {Shu}, Y. and {Soraisam}, M. and {Storfer}, C. and {Taubenberger}, S. and {Treu}, T. and {Vegetti}, S. and {Wojtak}, R.},
        title = "{Multiple Images of SN 2022qmx (``SN Zwicky'') Resolved in HST Observations}",
      journal = {Transient Name Server AstroNote},
         year = 2022,
        month = sep,
       volume = {196},
        pages = {1},
       adsurl = {https://ui.adsabs.harvard.edu/abs/2022TNSAN.196....1P}
}

@ARTICLE{Pierel2023,
       author = {{Pierel}, J.~D.~R. and {Arendse}, N. and {Ertl}, S. and {Huang}, X. and {Moustakas}, L.~A. and {Schuldt}, S. and {Shajib}, A.~J. and {Shu}, Y. and {Birrer}, S. and {Bronikowski}, M. and {Hjorth}, J. and {Suyu}, S.~H. and {Agarwal}, S. and {Agnello}, A. and {Bolton}, A.~S. and {Chakrabarti}, S. and {Cold}, C. and {Courbin}, F. and {Della Costa}, J.~M. and {Dhawan}, S. and {Engesser}, M. and {Fox}, Ori D. and {Gall}, C. and {Gomez}, S. and {Goobar}, A. and {Jha}, S.~W. and {Jimenez}, C. and {Johansson}, J. and {Larison}, C. and {Li}, G. and {Marques-Chaves}, R. and {Mao}, S. and {Mazzali}, P.~A. and {Perez-Fournon}, I. and {Petrushevska}, T. and {Poidevin}, F. and {Rest}, A. and {Sheu}, W. and {Shirley}, R. and {Silver}, E. and {Storfer}, C. and {Strolger}, L.~G. and {Treu}, T. and {Wojtak}, R. and {Zenati}, Y.},
        title = "{LensWatch. I. Resolved HST Observations and Constraints on the Strongly Lensed Type Ia Supernova 2022qmx (``SN Zwicky'')}",
      journal = {\apj},
         year = 2023,
        month = may,
       volume = {948},
       number = {2},
          eid = {115},
        pages = {115},
          doi = {10.3847/1538-4357/acc7a6},
archivePrefix = {arXiv},
       eprint = {2211.03772},
 primaryClass = {astro-ph.CO},
       adsurl = {https://ui.adsabs.harvard.edu/abs/2023ApJ...948..115P}
}

@ARTICLE{Pierel2023_Encore,
       author = {{Pierel}, J.~D.~R. and {Newman}, A.~B. and {Dhawan}, S. and {Gu}, M. and {Joshi}, B.~A. and {Li}, T. and {Schuldt}, S. and {Strolger}, L.~G. and {Suyu}, S.~H. and {Caminha}, G.~B. and {Cohen}, S.~H. and {Diego}, J.~M. and {D{\'S}ilva}, J.~C.~J. and {Ertl}, S. and {Frye}, B.~L. and {Granata}, G. and {Grillo}, C. and {Koekemoer}, A.~M. and {Li}, J. and {Robotham}, A. and {Summers}, J. and {Treu}, T. and {Windhorst}, R.~A. and {Zitrin}, A. and {Agarwal}, S. and {Agrawal}, A. and {Arendse}, N. and {Belli}, S. and {Burns}, C. and {Ca{\~n}ameras}, R. and {Chakrabarti}, S. and {Chen}, W. and {Collett}, T.~E. and {Coulter}, D.~A. and {Ellis}, R.~S. and {Engesser}, M. and {Foo}, N. and {Fox}, O.~D. and {Gall}, C. and {Garuda}, N. and {Gezari}, S. and {Gomez}, S. and {Glazebrook}, K. and {Hjorth}, J. and {Huang}, X. and {Jha}, S.~W. and {Kamieneski}, P.~S. and {Kelly}, P. and {Larison}, C. and {Moustakas}, L.~A. and {Pascale}, M. and {P{\'e}rez-Fournon}, I. and {Petrushevska}, T. and {Poidevin}, F. and {Rest}, A. and {Shahbandeh}, M. and {Shajib}, A.~J. and {Siebert}, M. and {Storfer}, C. and {Talbot}, M. and {Wang}, Q. and {Wevers}, T. and {Zenati}, Y.},
        title = "{Lensed Type Ia Supernova ``Encore'' at z = 2: The First Instance of Two Multiply Imaged Supernovae in the Same Host Galaxy}",
      journal = {\apjl},
         year = 2024,
        month = jun,
       volume = {967},
       number = {2},
          eid = {L37},
        pages = {L37},
          doi = {10.3847/2041-8213/ad4648},
archivePrefix = {arXiv},
       eprint = {2404.02139},
 primaryClass = {astro-ph.CO},
       adsurl = {https://ui.adsabs.harvard.edu/abs/2024ApJ...967L..37P}
}

@ARTICLE{Frye2024,
       author = {{Frye}, Brenda L. and {Pascale}, Massimo and {Pierel}, Justin and {Chen}, Wenlei and {Foo}, Nicholas and {Leimbach}, Reagen and {Garuda}, Nikhil and {Cohen}, Seth H. and {Kamieneski}, Patrick S. and {Windhorst}, Rogier A. and {Koekemoer}, Anton M. and {Kelly}, Pat and {Summers}, Jake and {Engesser}, Michael and {Liu}, Daizhong and {Furtak}, Lukas J. and {del Carmen Polletta}, Maria and {Harrington}, Kevin C. and {Willner}, S.~P. and {Diego}, Jose M. and {Jansen}, Rolf A. and {Coe}, Dan and {Conselice}, Christopher J. and {Dai}, Liang and {Dole}, Herv{\'e} and {D'Silva}, Jordan C.~J. and {Driver}, Simon P. and {Grogin}, Norman A. and {Marshall}, Madeline A. and {Meena}, Ashish K. and {Nonino}, Mario and {Ortiz}, Rafael and {Pirzkal}, Nor and {Robotham}, Aaron and {Ryan}, Russell E. and {Strolger}, Lou and {Tompkins}, Scott and {Willmer}, Christopher N.~A. and {Yan}, Haojing and {Yun}, Min S. and {Zitrin}, Adi},
        title = "{The JWST Discovery of the Triply Imaged Type Ia ``Supernova H0pe'' and Observations of the Galaxy Cluster PLCK G165.7+67.0}",
      journal = {\apj},
         year = 2024,
        month = feb,
       volume = {961},
       number = {2},
          eid = {171},
        pages = {171},
          doi = {10.3847/1538-4357/ad1034},
archivePrefix = {arXiv},
       eprint = {2309.07326},
 primaryClass = {astro-ph.GA},
       adsurl = {https://ui.adsabs.harvard.edu/abs/2024ApJ...961..171F}
}

@ARTICLE{Kulkarni2013_iPTF,
       author = {{Kulkarni}, S.~R.},
        title = "{The intermediate Palomar Transient Factory (iPTF) begins}",
      journal = {The Astronomer's Telegram},
         year = 2013,
        month = feb,
       volume = {4807},
        pages = {1},
       adsurl = {https://ui.adsabs.harvard.edu/abs/2013ATel.4807....1K}
}

@ARTICLE{Graham2019_ZTF,
       author = {{Graham}, Matthew J. and {Kulkarni}, S.~R. and {Bellm}, Eric C. and {Adams}, Scott M. and {Barbarino}, Cristina and {Blagorodnova}, Nadejda and {Bodewits}, Dennis and {Bolin}, Bryce and {Brady}, Patrick R. and {Cenko}, S. Bradley and {Chang}, Chan-Kao and {Coughlin}, Michael W. and {De}, Kishalay and {Eadie}, Gwendolyn and {Farnham}, Tony L. and {Feindt}, Ulrich and {Franckowiak}, Anna and {Fremling}, Christoffer and {Gezari}, Suvi and {Ghosh}, Shaon and {Goldstein}, Daniel A. and {Golkhou}, V. Zach and {Goobar}, Ariel and {Ho}, Anna Y.~Q. and {Huppenkothen}, Daniela and {Ivezi{\'c}}, {\v{Z}}eljko and {Jones}, R. Lynne and {Juric}, Mario and {Kaplan}, David L. and {Kasliwal}, Mansi M. and {Kelley}, Michael S.~P. and {Kupfer}, Thomas and {Lee}, Chien-De and {Lin}, Hsing Wen and {Lunnan}, Ragnhild and {Mahabal}, Ashish A. and {Miller}, Adam A. and {Ngeow}, Chow-Choong and {Nugent}, Peter and {Ofek}, Eran O. and {Prince}, Thomas A. and {Rauch}, Ludwig and {van Roestel}, Jan and {Schulze}, Steve and {Singer}, Leo P. and {Sollerman}, Jesper and {Taddia}, Francesco and {Yan}, Lin and {Ye}, Quan-Zhi and {Yu}, Po-Chieh and {Barlow}, Tom and {Bauer}, James and {Beck}, Ron and {Belicki}, Justin and {Biswas}, Rahul and {Brinnel}, Valery and {Brooke}, Tim and {Bue}, Brian and {Bulla}, Mattia and {Burruss}, Rick and {Connolly}, Andrew and {Cromer}, John and {Cunningham}, Virginia and {Dekany}, Richard and {Delacroix}, Alex and {Desai}, Vandana and {Duev}, Dmitry A. and {Feeney}, Michael and {Flynn}, David and {Frederick}, Sara and {Gal-Yam}, Avishay and {Giomi}, Matteo and {Groom}, Steven and {Hacopians}, Eugean and {Hale}, David and {Helou}, George and {Henning}, John and {Hover}, David and {Hillenbrand}, Lynne A. and {Howell}, Justin and {Hung}, Tiara and {Imel}, David and {Ip}, Wing-Huen and {Jackson}, Edward and {Kaspi}, Shai and {Kaye}, Stephen and {Kowalski}, Marek and {Kramer}, Emily and {Kuhn}, Michael and {Landry}, Walter and {Laher}, Russ R. and {Mao}, Peter and {Masci}, Frank J. and {Monkewitz}, Serge and {Murphy}, Patrick and {Nordin}, Jakob and {Patterson}, Maria T. and {Penprase}, Bryan and {Porter}, Michael and {Rebbapragada}, Umaa and {Reiley}, Dan and {Riddle}, Reed and {Rigault}, Mickael and {Rodriguez}, Hector and {Rusholme}, Ben and {van Santen}, Jakob and {Shupe}, David L. and {Smith}, Roger M. and {Soumagnac}, Maayane T. and {Stein}, Robert and {Surace}, Jason and {Szkody}, Paula and {Terek}, Scott and {Van Sistine}, Angela and {van Velzen}, Sjoert and {Vestrand}, W. Thomas and {Walters}, Richard and {Ward}, Charlotte and {Zhang}, Chaoran and {Zolkower}, Jeffry},
        title = "{The Zwicky Transient Facility: Science Objectives}",
      journal = {\pasp},
         year = 2019,
        month = jul,
       volume = {131},
       number = {1001},
        pages = {078001},
          doi = {10.1088/1538-3873/ab006c},
archivePrefix = {arXiv},
       eprint = {1902.01945},
 primaryClass = {astro-ph.IM},
       adsurl = {https://ui.adsabs.harvard.edu/abs/2019PASP..131g8001G}
}

@ARTICLE{Magee2023,
       author = {{Magee}, M.~R. and {Sainz de Murieta}, A. and {Collett}, T.~E. and {Enzi}, W.},
        title = "{A search for gravitationally lensed supernovae within the Zwicky transient facility public survey}",
      journal = {\mnras},
         year = 2023,
        month = oct,
       volume = {525},
       number = {1},
        pages = {542-560},
          doi = {10.1093/mnras/stad2263},
archivePrefix = {arXiv},
       eprint = {2303.15439},
 primaryClass = {astro-ph.HE},
       adsurl = {https://ui.adsabs.harvard.edu/abs/2023MNRAS.525..542M}
}

@ARTICLE{Townsend2024,
       author = {{Townsend}, A. and {Nordin}, J. and {Sagu{\'e}s Carracedo}, A. and {Kowalski}, M. and {Arendse}, N. and {Dhawan}, S. and {Goobar}, A. and {Johansson}, J. and {M{\"o}rtsell}, E. and {Schulze}, S. and {Andreoni}, I. and {Fern{\'a}ndez}, E. and {Kim}, A.~G. and {Nugent}, P.~E. and {Prada}, F. and {Rigault}, M. and {Sarin}, N. and {Sharma}, D. and {Bellm}, E.~C. and {Coughlin}, M.~W. and {Dekany}, R. and {Groom}, S.~L. and {Lacroix}, L. and {Laher}, R.~R. and {Riddle}, R. and {Aguilar}, J. and {Ahlen}, S. and {Bailey}, S. and {Brooks}, D. and {Claybaugh}, T. and {de la Macorra}, A. and {Dey}, A. and {Dey}, B. and {Doel}, P. and {Fanning}, K. and {Forero-Romero}, J.~E. and {Gazta{\~n}aga}, E. and {Gontcho A Gontcho}, S. and {Honscheid}, K. and {Howlett}, C. and {Kisner}, T. and {Kremin}, A. and {Lambert}, A. and {Landriau}, M. and {Le Guillou}, L. and {Levi}, M.~E. and {Manera}, M. and {Meisner}, A. and {Miquel}, R. and {Moustakas}, J. and {Mueller}, E. and {Myers}, A.~D. and {Nie}, J. and {Palanque-Delabrouille}, N. and {Poppett}, C. and {Rezaie}, M. and {Rossi}, G. and {Sanchez}, E. and {Schlegel}, D. and {Schubnell}, M. and {Seo}, H. and {Sprayberry}, D. and {Tarl{\'e}}, G. and {Zou}, H.},
        title = "{Candidate strongly lensed type Ia supernovae in the Zwicky Transient Facility archive}",
      journal = {\aap},
         year = 2025,
        month = feb,
       volume = {694},
          eid = {A146},
        pages = {A146},
          doi = {10.1051/0004-6361/202451082},
archivePrefix = {arXiv},
       eprint = {2405.18589},
 primaryClass = {astro-ph.HE},
       adsurl = {https://ui.adsabs.harvard.edu/abs/2025A&A...694A.146T}
}

@ARTICLE{Sagues2024,
       author = {{Sagu{\'e}s Carracedo}, A. and {Goobar}, A. and {M{\"o}rtsell}, E. and {Arendse}, N. and {Johansson}, J. and {Townsend}, A. and {Dhawan}, S. and {Nordin}, J. and {Sollerman}, J. and {Schulze}, S.},
        title = "{Detectability and Characterisation of Strongly Lensed Supernova Lightcurves in the Zwicky Transient Facility}",
      journal = {ArXiv e-prints},
         year = 2024,
        month = may,
          eid = {arXiv:2406.00052},
          doi = {10.48550/arXiv.2406.00052},
archivePrefix = {arXiv},
       eprint = {2406.00052},
 primaryClass = {astro-ph.HE},
       adsurl = {https://ui.adsabs.harvard.edu/abs/2024arXiv240600052S}
}

@ARTICLE{Young1981,
       author = {{Young}, P.},
        title = "{Q0957+561 : effects of random stars on the gravitational lens.}",
      journal = {\apj},
         year = 1981,
        month = mar,
       volume = {244},
        pages = {756-767},
          doi = {10.1086/158752},
       adsurl = {https://ui.adsabs.harvard.edu/abs/1981ApJ...244..756Y}
}

@ARTICLE{Paczynski1986,
       author = {{Paczynski}, B.},
        title = "{Gravitational Microlensing at Large Optical Depth}",
      journal = {\apj},
         year = 1986,
        month = feb,
       volume = {301},
        pages = {503},
          doi = {10.1086/163919},
       adsurl = {https://ui.adsabs.harvard.edu/abs/1986ApJ...301..503P}
}

@ARTICLE{Vernardos2024,
       author = {{Vernardos}, G. and {Sluse}, D. and {Pooley}, D. and {Schmidt}, R.~W. and {Millon}, M. and {Weisenbach}, L. and {Motta}, V. and {Anguita}, T. and {Saha}, P. and {O'Dowd}, M. and {Peel}, A. and {Schechter}, P.~L.},
        title = "{Microlensing of Strongly Lensed Quasars}",
      journal = {\ssr},
         year = 2024,
        month = feb,
       volume = {220},
       number = {1},
          eid = {14},
        pages = {14},
          doi = {10.1007/s11214-024-01043-8},
archivePrefix = {arXiv},
       eprint = {2312.00931},
 primaryClass = {astro-ph.GA},
       adsurl = {https://ui.adsabs.harvard.edu/abs/2024SSRv..220...14V}
}

@ARTICLE{Yahalomi2017,
       author = {{Yahalomi}, Daniel A. and {Schechter}, Paul L. and {Wambsganss}, Joachim},
        title = "{A Quadruply Lensed SN Ia: Gaining a Time-Delay...Losing a Standard Candle}",
      journal = {ArXiv e-prints},
         year = 2017,
        month = nov,
          eid = {arXiv:1711.07919},
          doi = {10.48550/arXiv.1711.07919},
archivePrefix = {arXiv},
       eprint = {1711.07919},
 primaryClass = {astro-ph.CO},
       adsurl = {https://ui.adsabs.harvard.edu/abs/2017arXiv171107919Y}
}

@ARTICLE{Weisenbach_2024,
       author = {{Weisenbach}, Luke and {Collett}, Thomas and {de Murieta}, Ana Sainz and {Krawczyk}, Coleman and {Vernardos}, Georgios and {Enzi}, Wolfgang and {Lundgren}, Andrew},
        title = "{How to break the mass sheet degeneracy with the light curves of microlensed Type Ia supernovae}",
      journal = {\mnras},
         year = 2024,
        month = jul,
       volume = {531},
       number = {4},
        pages = {4349-4362},
          doi = {10.1093/mnras/stae1396},
archivePrefix = {arXiv},
       eprint = {2403.03264},
 primaryClass = {astro-ph.GA},
       adsurl = {https://ui.adsabs.harvard.edu/abs/2024MNRAS.531.4349W}
}

@ARTICLE{Weisenbach2021,
       author = {{Weisenbach}, Luke and {Schechter}, Paul L. and {Pontula}, Sahil},
        title = "{``Worst-case'' Microlensing in the Identification and Modeling of Lensed Quasars}",
      journal = {\apj},
         year = 2021,
        month = nov,
       volume = {922},
       number = {1},
          eid = {70},
        pages = {70},
          doi = {10.3847/1538-4357/ac2228},
archivePrefix = {arXiv},
       eprint = {2105.08690},
 primaryClass = {astro-ph.GA},
       adsurl = {https://ui.adsabs.harvard.edu/abs/2021ApJ...922...70W}
}

@ARTICLE{Chang1979,
       author = {{Chang}, K. and {Refsdal}, S.},
        title = "{Flux variations of QSO 0957 + 561 A, B and image splitting by stars near the light path}",
      journal = {\nat},
         year = 1979,
        month = dec,
       volume = {282},
       number = {5739},
        pages = {561-564},
          doi = {10.1038/282561a0},
       adsurl = {https://ui.adsabs.harvard.edu/abs/1979Natur.282..561C}
}

@inproceedings{Wambsganss2006,
       author = {{Wambsganss}, Joachim},
        title = "{Gravitational Microlensing}",
        booktitle = {Gravitational Lensing: Strong, Weak \& Micro},
      series = {Saas-Fee Advanced Course},
    volume = {33},
    editor = {G. Meylan and P. Jetzer and P. North},
pages = {453--540},
 publisher = {Springer, Berlin, Heidelberg},
         year = 2006,
        month = apr,
          eid = {astro-ph/0604278},
archivePrefix = {arXiv},
       eprint = {astro-ph/0604278},
 primaryClass = {astro-ph},
       adsurl = {https://ui.adsabs.harvard.edu/abs/2006astro.ph..4278W}
}

@ARTICLE{Schmidt2010,
       author = {{Schmidt}, R.~W. and {Wambsganss}, J.},
        title = "{Quasar microlensing}",
      journal = {Gen.\ Relativ.\ Gravitation},
         year = 2010,
        month = sep,
       volume = {42},
       number = {9},
        pages = {2127-2150},
          doi = {10.1007/s10714-010-0956-x},
       adsurl = {https://ui.adsabs.harvard.edu/abs/2010GReGr..42.2127S}
}

@ARTICLE{Vernardos2019,
       author = {{Vernardos}, Georgios and {Tsagkatakis}, Grigorios},
        title = "{Quasar microlensing light-curve analysis using deep machine learning}",
      journal = {\mnras},
         year = 2019,
        month = jun,
       volume = {486},
       number = {2},
        pages = {1944-1952},
          doi = {10.1093/mnras/stz868},
archivePrefix = {arXiv},
       eprint = {1903.09170},
 primaryClass = {astro-ph.IM},
       adsurl = {https://ui.adsabs.harvard.edu/abs/2019MNRAS.486.1944V}
}

@ARTICLE{Pooley2012,
       author = {{Pooley}, David and {Rappaport}, Saul and {Blackburne}, Jeffrey A. and {Schechter}, Paul L. and {Wambsganss}, Joachim},
        title = "{X-Ray and Optical Flux Ratio Anomalies in Quadruply Lensed Quasars. II. Mapping the Dark Matter Content in Elliptical Galaxies}",
      journal = {\apj},
         year = 2012,
        month = jan,
       volume = {744},
       number = {2},
          eid = {111},
        pages = {111},
          doi = {10.1088/0004-637X/744/2/111},
archivePrefix = {arXiv},
       eprint = {1108.2725},
 primaryClass = {astro-ph.HE},
       adsurl = {https://ui.adsabs.harvard.edu/abs/2012ApJ...744..111P}
}

@ARTICLE{Schechter2014,
       author = {{Schechter}, Paul L. and {Pooley}, David and {Blackburne}, Jeffrey A. and {Wambsganss}, Joachim},
        title = "{A Calibration of the Stellar Mass Fundamental Plane at z \raisebox{-0.5ex}\textasciitilde 0.5 Using the Micro-lensing-induced Flux Ratio Anomalies of Macro-lensed Quasars}",
      journal = {\apj},
         year = 2014,
        month = oct,
       volume = {793},
       number = {2},
          eid = {96},
        pages = {96},
          doi = {10.1088/0004-637X/793/2/96},
archivePrefix = {arXiv},
       eprint = {1405.0038},
 primaryClass = {astro-ph.GA},
       adsurl = {https://ui.adsabs.harvard.edu/abs/2014ApJ...793...96S}
}

@ARTICLE{Pooley2009,
       author = {{Pooley}, D. and {Rappaport}, S. and {Blackburne}, J. and {Schechter}, P.~L. and {Schwab}, J. and {Wambsganss}, J.},
        title = "{The Dark-matter Fraction in the Elliptical Galaxy Lensing the Quasar PG 1115+080}",
      journal = {\apj},
         year = 2009,
        month = jun,
       volume = {697},
       number = {2},
        pages = {1892-1900},
          doi = {10.1088/0004-637X/697/2/1892},
archivePrefix = {arXiv},
       eprint = {0808.3299},
 primaryClass = {astro-ph},
       adsurl = {https://ui.adsabs.harvard.edu/abs/2009ApJ...697.1892P}
}

@INPROCEEDINGS{Schechter2004,
       author = {{Schechter}, P.~L. and {Wambsganss}, J.},
        title = "{The dark matter content of lensing galaxies at 1.5 R\_e}",
    booktitle = {Dark Matter in Galaxies},
         year = 2004,
       editor = {{Ryder}, S. and {Pisano}, D. and {Walker}, M. and {Freeman}, K.},
        series = {IAU Symp.},
       volume = {220},
        month = jul,
        pages = {103},
archivePrefix = {arXiv},
       eprint = {astro-ph/0309163},
 primaryClass = {astro-ph},
       adsurl = {https://ui.adsabs.harvard.edu/abs/2004IAUS..220..103S}
}

@ARTICLE{Mediavilla2017,
       author = {{Mediavilla}, E. and {Jim{\'e}nez-Vicente}, J. and {Mu{\~n}oz}, J.~A. and {Vives-Arias}, H. and {Calder{\'o}n-Infante}, J.},
        title = "{Limits on the Mass and Abundance of Primordial Black Holes from Quasar Gravitational Microlensing}",
      journal = {\apjl},
         year = 2017,
        month = feb,
       volume = {836},
       number = {2},
          eid = {L18},
        pages = {L18},
          doi = {10.3847/2041-8213/aa5dab},
archivePrefix = {arXiv},
       eprint = {1702.00947},
 primaryClass = {astro-ph.GA},
       adsurl = {https://ui.adsabs.harvard.edu/abs/2017ApJ...836L..18M}
}

@ARTICLE{Mediavilla2009,
       author = {{Mediavilla}, E. and {Mu{\~n}oz}, J.~A. and {Falco}, E. and {Motta}, V. and {Guerras}, E. and {Canovas}, H. and {Jean}, C. and {Oscoz}, A. and {Mosquera}, A.~M.},
        title = "{Microlensing-based Estimate of the Mass Fraction in Compact Objects in Lens Galaxies}",
      journal = {\apj},
         year = 2009,
        month = dec,
       volume = {706},
       number = {2},
        pages = {1451-1462},
          doi = {10.1088/0004-637X/706/2/1451},
archivePrefix = {arXiv},
       eprint = {0910.3645},
 primaryClass = {astro-ph.CO},
       adsurl = {https://ui.adsabs.harvard.edu/abs/2009ApJ...706.1451M}
}

@ARTICLE{Esteban2020,
       author = {{Esteban-Guti{\'e}rrez}, A. and {Ag{\"u}es-Paszkowsky}, N. and {Mediavilla}, E. and {Jim{\'e}nez-Vicente}, J. and {Mu{\~n}oz}, J.~A. and {Heydenreich}, S.},
        title = "{The Impact of the Mass Spectrum of Lenses in Quasar Microlensing Studies. Constraints on a Mixed Population of Primordial Black Holes and Stars}",
      journal = {\apj},
         year = 2020,
        month = dec,
       volume = {904},
       number = {2},
          eid = {176},
        pages = {176},
          doi = {10.3847/1538-4357/abbdf7},
archivePrefix = {arXiv},
       eprint = {2011.05751},
 primaryClass = {astro-ph.CO},
       adsurl = {https://ui.adsabs.harvard.edu/abs/2020ApJ...904..176E}
}

@ARTICLE{Clesse2015,
       author = {{Clesse}, S{\'e}bastien and {Garc{\'\i}a-Bellido}, Juan},
        title = "{Massive primordial black holes from hybrid inflation as dark matter and the seeds of galaxies}",
      journal = {\prd},
         year = 2015,
        month = jul,
       volume = {92},
       number = {2},
          eid = {023524},
        pages = {023524},
          doi = {10.1103/PhysRevD.92.023524},
archivePrefix = {arXiv},
       eprint = {1501.07565},
 primaryClass = {astro-ph.CO},
       adsurl = {https://ui.adsabs.harvard.edu/abs/2015PhRvD..92b3524C}
}

@ARTICLE{Hawkins2020,
       author = {{Hawkins}, M.~R.~S.},
        title = "{The signature of primordial black holes in the dark matter halos of galaxies}",
      journal = {\aap},
         year = 2020,
        month = jan,
       volume = {633},
          eid = {A107},
        pages = {A107},
          doi = {10.1051/0004-6361/201936462},
archivePrefix = {arXiv},
       eprint = {2001.07633},
 primaryClass = {astro-ph.GA},
       adsurl = {https://ui.adsabs.harvard.edu/abs/2020A&A...633A.107H}
}

@ARTICLE{Kashlinsky2016,
       author = {{Kashlinsky}, A.},
        title = "{LIGO Gravitational Wave Detection, Primordial Black Holes, and the Near-IR Cosmic Infrared Background Anisotropies}",
      journal = {\apjl},
         year = 2016,
        month = jun,
       volume = {823},
       number = {2},
          eid = {L25},
        pages = {L25},
          doi = {10.3847/2041-8205/823/2/L25},
archivePrefix = {arXiv},
       eprint = {1605.04023},
 primaryClass = {astro-ph.CO},
       adsurl = {https://ui.adsabs.harvard.edu/abs/2016ApJ...823L..25K}
}

@inproceedings{Escriva2022,
       author = {{Escriv{\`a}}, Albert and {K{\"u}hnel}, Florian and {Tada}, Yuichiro},
        title = "{Primordial Black Holes}",
    booktitle = {Black Holes in the Era of Gravitational-Wave Astronomy},
        editor = {Manuel Arca Sedda and Elisa Bortolas and Mario Spera},
        publisher = {Elsevier},
        pages = {261-377},
         year = 2024,
        month = nov,
          eid = {arXiv:2211.05767},
archivePrefix = {arXiv},
       eprint = {2211.05767},
 primaryClass = {astro-ph.CO},
       adsurl = {https://ui.adsabs.harvard.edu/abs/2022arXiv221105767E}
}

@ARTICLE{Khlopov2010,
       author = {{Khlopov}, Maxim Yu.},
        title = "{Primordial black holes}",
      journal = {Res.\ Astron.\ Astrophys.},
         year = 2010,
        month = jun,
       volume = {10},
       number = {6},
        pages = {495-528},
          doi = {10.1088/1674-4527/10/6/001},
archivePrefix = {arXiv},
       eprint = {0801.0116},
 primaryClass = {astro-ph},
       adsurl = {https://ui.adsabs.harvard.edu/abs/2010RAA....10..495K}
}

@ARTICLE{Carr2016,
       author = {{Carr}, Bernard and {K{\"u}hnel}, Florian and {Sandstad}, Marit},
        title = "{Primordial black holes as dark matter}",
      journal = {\prd},
         year = 2016,
        month = oct,
       volume = {94},
       number = {8},
          eid = {083504},
        pages = {083504},
          doi = {10.1103/PhysRevD.94.083504},
archivePrefix = {arXiv},
       eprint = {1607.06077},
 primaryClass = {astro-ph.CO},
       adsurl = {https://ui.adsabs.harvard.edu/abs/2016PhRvD..94h3504C}
}

@ARTICLE{Mortsell2020,
       author = {{M{\"o}rtsell}, E. and {Johansson}, J. and {Dhawan}, S. and {Goobar}, A. and {Amanullah}, R. and {Goldstein}, D.~A.},
        title = "{Lens modelling of the strongly lensed Type Ia supernova iPTF16geu}",
      journal = {\mnras},
         year = 2020,
        month = aug,
       volume = {496},
       number = {3},
        pages = {3270-3280},
          doi = {10.1093/mnras/staa1600},
archivePrefix = {arXiv},
       eprint = {1907.06609},
 primaryClass = {astro-ph.CO},
       adsurl = {https://ui.adsabs.harvard.edu/abs/2020MNRAS.496.3270M}
}

@ARTICLE{Diego2022,
       author = {{Diego}, J.~M. and {Bernstein}, G. and {Chen}, W. and {Goobar}, A. and {Johansson}, J.~P. and {Kelly}, P.~L. and {M{\"o}rtsell}, E. and {Nightingale}, J.~W.},
        title = "{Microlensing and the type Ia supernova iPTF16geu}",
      journal = {\aap},
         year = 2022,
        month = jun,
       volume = {662},
          eid = {A34},
        pages = {A34},
          doi = {10.1051/0004-6361/202143009},
archivePrefix = {arXiv},
       eprint = {2112.04524},
 primaryClass = {astro-ph.CO},
       adsurl = {https://ui.adsabs.harvard.edu/abs/2022A&A...662A..34D}
}

@ARTICLE{Conroy2010,
       author = {{Conroy}, Charlie and {White}, Martin and {Gunn}, James E.},
        title = "{The Propagation of Uncertainties in Stellar Population Synthesis Modeling. II. The Challenge of Comparing Galaxy Evolution Models to Observations}",
      journal = {\apj},
         year = 2010,
        month = jan,
       volume = {708},
       number = {1},
        pages = {58-70},
          doi = {10.1088/0004-637X/708/1/58},
archivePrefix = {arXiv},
       eprint = {0904.0002},
 primaryClass = {astro-ph.CO},
       adsurl = {https://ui.adsabs.harvard.edu/abs/2010ApJ...708...58C}
}

@ARTICLE{Goodman2010,
       author = {{Goodman}, Jonathan and {Weare}, Jonathan},
        title = "{Ensemble samplers with affine invariance}",
      journal = {Communications Applied Math.\ Comput.\ Sci.},
         year = 2010,
        month = jan,
       volume = {5},
       number = {1},
        pages = {65-80},
          doi = {10.2140/camcos.2010.5.65},
       adsurl = {https://ui.adsabs.harvard.edu/abs/2010CAMCS...5...65G}
}

@ARTICLE{ForemanMackey2013,
       author = {{Foreman-Mackey}, Daniel and {Hogg}, David W. and {Lang}, Dustin and {Goodman}, Jonathan},
        title = "{emcee: The MCMC Hammer}",
      journal = {\pasp},
         year = 2013,
        month = mar,
       volume = {125},
       number = {925},
        pages = {306},
          doi = {10.1086/670067},
archivePrefix = {arXiv},
       eprint = {1202.3665},
 primaryClass = {astro-ph.IM},
       adsurl = {https://ui.adsabs.harvard.edu/abs/2013PASP..125..306F}
}

@ARTICLE{Kauffmann2003,
       author = {{Kauffmann}, Guinevere and {Heckman}, Timothy M. and {White}, Simon D.~M. and {Charlot}, St{\'e}phane and {Tremonti}, Christy and {Brinchmann}, Jarle and {Bruzual}, Gustavo and {Peng}, Eric W. and {Seibert}, Mark and {Bernardi}, Mariangela and {Blanton}, Michael and {Brinkmann}, Jon and {Castander}, Francisco and {Cs{\'a}bai}, Istvan and {Fukugita}, Masataka and {Ivezic}, Zeljko and {Munn}, Jeffrey A. and {Nichol}, Robert C. and {Padmanabhan}, Nikhil and {Thakar}, Aniruddha R. and {Weinberg}, David H. and {York}, Donald},
        title = "{Stellar masses and star formation histories for {}10$^{5}$ galaxies from the Sloan Digital Sky Survey}",
      journal = {\mnras},
         year = 2003,
        month = may,
       volume = {341},
       number = {1},
        pages = {33-53},
          doi = {10.1046/j.1365-8711.2003.06291.x},
archivePrefix = {arXiv},
       eprint = {astro-ph/0204055},
 primaryClass = {astro-ph},
       adsurl = {https://ui.adsabs.harvard.edu/abs/2003MNRAS.341...33K}
}

@ARTICLE{Auger2010,
       author = {{Auger}, M.~W. and {Treu}, T. and {Bolton}, A.~S. and {Gavazzi}, R. and {Koopmans}, L.~V.~E. and {Marshall}, P.~J. and {Moustakas}, L.~A. and {Burles}, S.},
        title = "{The Sloan Lens ACS Survey. X. Stellar, Dynamical, and Total Mass Correlations of Massive Early-type Galaxies}",
      journal = {\apj},
         year = 2010,
        month = nov,
       volume = {724},
       number = {1},
        pages = {511-525},
          doi = {10.1088/0004-637X/724/1/511},
archivePrefix = {arXiv},
       eprint = {1007.2880},
 primaryClass = {astro-ph.CO},
       adsurl = {https://ui.adsabs.harvard.edu/abs/2010ApJ...724..511A}
}

@ARTICLE{Sonnenfeld2013,
       author = {{Sonnenfeld}, Alessandro and {Treu}, Tommaso and {Gavazzi}, Rapha{\"e}l and {Suyu}, Sherry H. and {Marshall}, Philip J. and {Auger}, Matthew W. and {Nipoti}, Carlo},
        title = "{The SL2S Galaxy-scale Lens Sample. IV. The Dependence of the Total Mass Density Profile of Early-type Galaxies on Redshift, Stellar Mass, and Size}",
      journal = {\apj},
         year = 2013,
        month = nov,
       volume = {777},
       number = {2},
          eid = {98},
        pages = {98},
          doi = {10.1088/0004-637X/777/2/98},
archivePrefix = {arXiv},
       eprint = {1307.4759},
 primaryClass = {astro-ph.CO},
       adsurl = {https://ui.adsabs.harvard.edu/abs/2013ApJ...777...98S}
}

@Inproceedings{Medavilla2024,
author="Mediavilla, Evencio
and Jim{\'e}nez-Vicente, Jorge",
editor="Byrnes, Christian
and Franciolini, Gabriele
and Harada, Tomohiro
and Pani, Paolo
and Sasaki, Misao",
title="Lensing Constraints: Substellar to Intermediate Masses",
booktitle="Primordial Black Holes",
year="2025",
publisher="Springer Nature",
address="Singapore",
pages="577--593",
doi="10.1007/978-981-97-8887-3_23",
archivePrefix = {arXiv},
       eprint = {2405.14984},
 primaryClass = {astro-ph.CO},
       adsurl = {https://ui.adsabs.harvard.edu/abs/2024arXiv240514984M}
}

@software{SEDPy2021,
       author = {{Johnson}, Benjamin D.},
        title = "{bd-j/sedpy: v0.2.0}",
         year = 2021,
        month = mar,
          eid = {10.5281/zenodo.4582723},
          doi = {10.5281/zenodo.4582723},
      version = {v0.2.0},
    publisher = {Zenodo},
       adsurl = {https://ui.adsabs.harvard.edu/abs/2021zndo...4582723J}
}

@INPROCEEDINGS{Joye2003,
       author = {{Joye}, W.~A. and {Mandel}, E.},
        title = "{New Features of SAOImage DS9}",
    booktitle = {Astronomical Data Analysis Software and Systems XII},
         year = 2003,
       editor = {{Payne}, H.~E. and {Jedrzejewski}, R.~I. and {Hook}, R.~N.},
       series = {Astron.\ Soc.\ Pac.\ Conf.\ Series},
       volume = {295},
        month = jan,
        pages = {489--492},
       adsurl = {https://ui.adsabs.harvard.edu/abs/2003ASPC..295..489J}
}

@ARTICLE{Conroy2009,
       author = {{Conroy}, Charlie and {Gunn}, James E. and {White}, Martin},
        title = "{The Propagation of Uncertainties in Stellar Population Synthesis Modeling. I. The Relevance of Uncertain Aspects of Stellar Evolution and the Initial Mass Function to the Derived Physical Properties of Galaxies}",
      journal = {\apj},
         year = 2009,
        month = jul,
       volume = {699},
       number = {1},
        pages = {486-506},
          doi = {10.1088/0004-637X/699/1/486},
archivePrefix = {arXiv},
       eprint = {0809.4261},
 primaryClass = {astro-ph},
       adsurl = {https://ui.adsabs.harvard.edu/abs/2009ApJ...699..486C}
}

@ARTICLE{Conroy2010ApJ,
       author = {{Conroy}, Charlie and {Gunn}, James E.},
        title = "{The Propagation of Uncertainties in Stellar Population Synthesis Modeling. III. Model Calibration, Comparison, and Evaluation}",
      journal = {\apj},
         year = 2010,
        month = apr,
       volume = {712},
       number = {2},
        pages = {833-857},
          doi = {10.1088/0004-637X/712/2/833},
archivePrefix = {arXiv},
       eprint = {0911.3151},
 primaryClass = {astro-ph.CO},
       adsurl = {https://ui.adsabs.harvard.edu/abs/2010ApJ...712..833C}
}

    \begin{center} 
        \begin{minipage}{\textwidth}
This paper was built using the Open Journal of Astrophysics \LaTeX\ template. The OJA is a journal which provides fast and easy peer review for new papers in the \texttt{astro-ph} section of the arXiv, making the reviewing process simpler for authors and referees alike. Learn more at \url{http://astro.theoj.org}.
        \end{minipage}
    \end{center}


\begin{appendices}
\setcounter{figure}{0} 
\renewcommand{\thefigure}{\Alph{section}.\arabic{figure}} 

\begin{figure*}[ht]
\centering
{\includegraphics[width=\textwidth,clip=true]{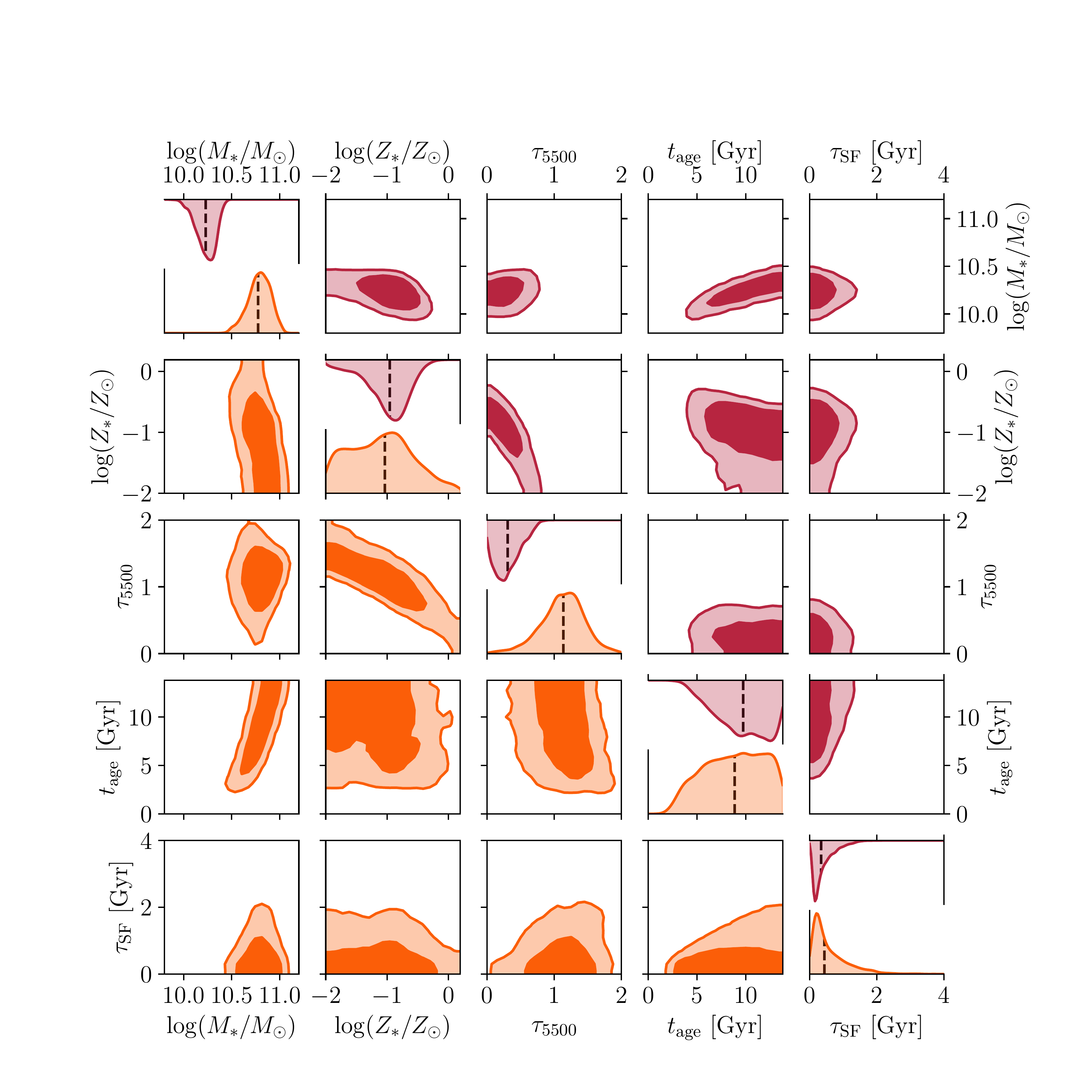}}
\caption{\small Cornerplot showing the free parameters in the \texttt{prospector} fits for \geu (lower and left panels) and SN Zwicky (upper and right panels).}
\label{fig:prospector_all_params}
\end{figure*}

\begin{figure*}[ht]
\centering
{\includegraphics[width=\textwidth,clip=true]{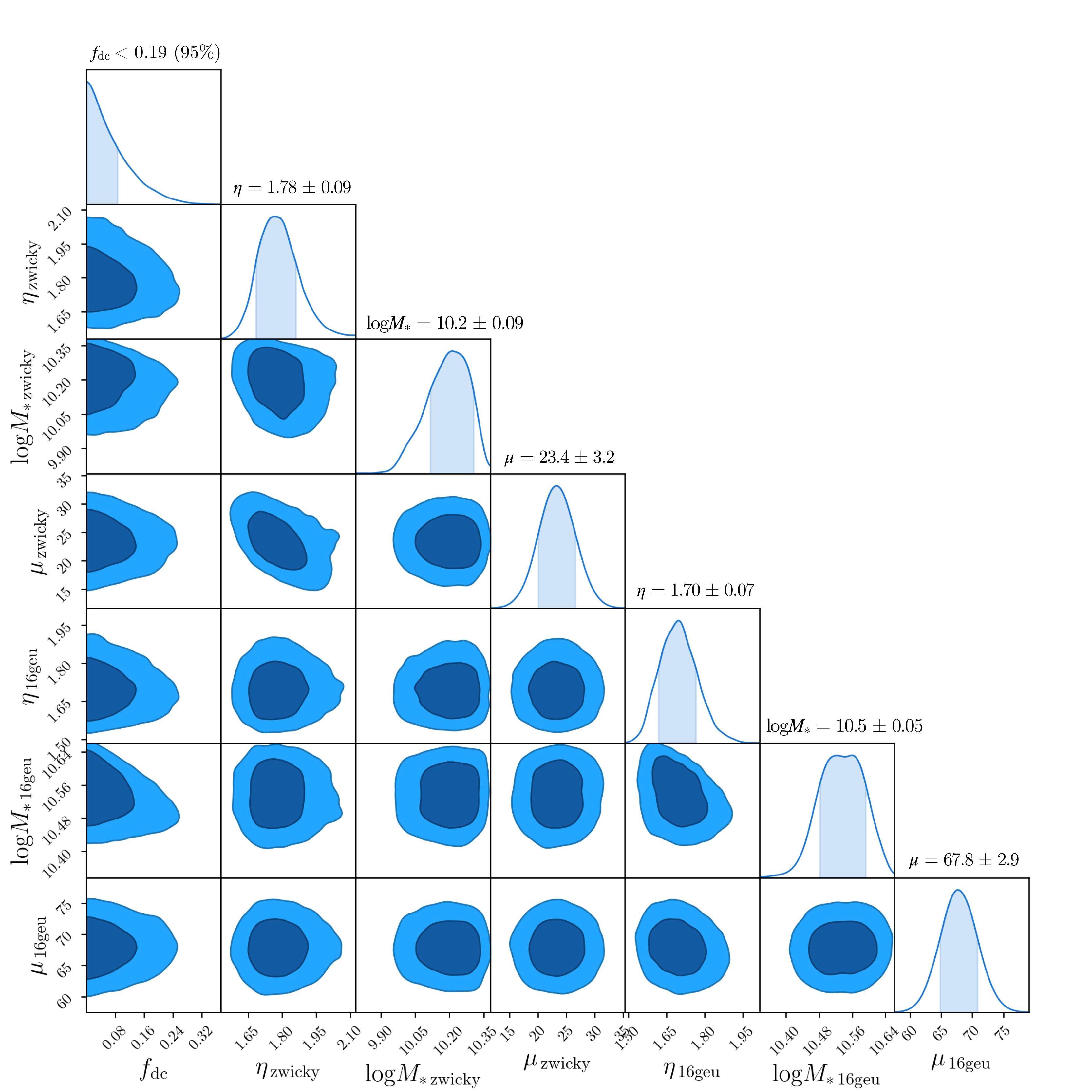}}
\caption{\small Joint fit for \geu and SN Zwicky. Each have individual parameters for their slopes ($\eta$), stellar masses ($\log M_*$), and total magnifications ($\mu$), but they share a common parameter for the dark compact object fraction (\pbh). This shows how multiple lensed SN can be combined to give a tighter upper limit on \pbh.}
\label{fig:mcmc_joint_params}
\end{figure*}

    \begin{center} 
        \begin{minipage}{\textwidth}
\section{Appendix: corner plots}\label{App:cornerplots}

Fig.~\ref{fig:prospector_all_params} shows the full corner plot of all free parameters in our \texttt{prospector} fits for the stellar mass, as described in section~\ref{subseq:stellar_mass}, both for \geu and SN Zwicky.

\medskip
Fig.~\ref{fig:mcmc_joint_params} shows the resulting parameter posteriors when SN Zwicky and \geu are combined in one single fit, where they share a common \pbh parameter. By combining the two, a tighter upper limit of \pbh $< 0.19 \ (95\%)$ is obtained. This assumes that \pbh is the same at all image positions in both galaxies.

\section*{}

\end{minipage}
\end{center}

\end{appendices}

\end{document}